\documentclass[10pt]{article}
\usepackage{fullpage,citesort,epsfig,graphics,amsbsy}
\usepackage{psfrag}
\usepackage[normal,small]{caption}
\newcommand{\beq}{\begin{equation}}
\newcommand{\eeq}{\end{equation}}
\newcommand{\bea}{\begin{eqnarray}}
\newcommand{\eea}{\end{eqnarray}}

\newcommand{\Tr}{{\rm Tr}}
\newcommand{\be}{\begin{equation}}
\newcommand{\ee}{\end{equation}}
\newcommand{\bq}{\begin{eqnarray}}
\newcommand{\eq}{\end{eqnarray}}

\newcommand{\ie}{{\it i.e.\ }}
\def\math{\mathsurround=0pt }
\def\leftrightarrowfill{$\math \mathord\leftarrow \mkern-6mu \cleaders\hbox{$\mkern-2mu \mathord- \mkern-2mu$}\hfill
 \mkern-6mu \mathord\rightarrow$}
\def\overleftrightarrow#1{\vbox{\ialign{##\crcr
     \leftrightarrowfill\crcr\noalign{\kern-1pt\nointerlineskip}
     $\hfil\displaystyle{#1}\hfil$\crcr}}}

\let\a=\alpha \let\b=\beta \let\g=\gamma \let\d=\delta 
    
\let\l=\lambda   \let\x=\xi  
   \let\f=\phi  
\let\w=\omega

\def\nn{\nonumber} \def\bd{\begin{document}} \def\ed{\end{document}}
\def\ds{\documentstyle} \let\fr=\frac \let\bl=\bigl \let\br=\bigr
\let\Br=\Bigr \let\Bl=\Bigl
\let\bm=\bibitem
\let\na=\nabla
\let\pa=\partial \let\ov=\overline
\def\ft#1#2{{\textstyle{{\scriptstyle #1}\over {\scriptstyle #2}}}}
\def\fft#1#2{{#1 \over #2}}
\def\vp{\varphi}
\def\sst#1{{\scriptscriptstyle #1}}
\def\oneone{\rlap 1\mkern4mu{\rm l}}
\def\td{\tilde}
\def\wtd{\widetilde}
\def\dalemb#1#2{{\vbox{\hrule height .#2pt
        \hbox{\vrule width.#2pt height#1pt \kern#1pt
                \vrule width.#2pt}
        \hrule height.#2pt}}}
\def\square{\mathord{\dalemb{6.8}{7}\hbox{\hskip1pt}}}
\def\wtd{\widetilde}
\def\R{\rlap{\rm I}\mkern3mu{\rm R}}
\def\im{{\rm i}}
\def\tilg{\tilde{g}}
\def\tilF{\tilde{F}}
\def\tilA{\tilde{A}}
\def\varf{\varphi}
\def\tilf{\tilde{\phi}}
\def\tilh{\tilde{h}}
\def\rme{{\rm e}}
\def\ep{\epsilon}
\def\0{{(0)}}
\def\9{{(9)}}
\def\8{{(8)}}
\def\7{{(7)}}
\def\6{{(6)}}
\def\5{{(5)}}
\def\4{{(4)}}
\def\3{{(3)}}
\def\2{{(2)}}
\def\1{{(1)}}
\newcommand{\trace}{{\rm Tr}}
\newcommand{\ub}{\overline{U}}
\newcommand{\vb}{\overline{V}}
\newcommand{\uh}{\widehat{U}}
\newcommand{\vh}{\widehat{V}}
\newcommand{\ubh}{\overline{\widehat{U}}}
\newcommand{\vbh}{\overline{\widehat{V}}}
\newcommand{\lb}{\bar{\l}}
\newcommand{\Fb}{\overline{F}}
\newcommand{\Fh}{\widehat{F}}
\newcommand{\Fbh}{\overline{\widehat{F}}}
\newcommand{\Ab}{\overline{A}}
\newcommand{\Ah}{\widehat{A}}
\newcommand{\Abh}{\overline{\widehat{A}}}
\newcommand{\Gb}{\overline{G}}
\newcommand{\Gh}{\widehat{G}}
\newcommand{\Gbh}{\overline{\widehat{G}}}
\newcommand{\Pb}{\overline{P}}
\newcommand{\Ph}{\widehat{P}}
\newcommand{\Pbh}{\overline{\widehat{P}}}
\newcommand{\Qb}{\overline{Q}}
\newcommand{\Qh}{\widehat{Q}}
\newcommand{\Qbh}{\overline{\widehat{Q}}}
\newcommand{\Bb}{\overline{B}}
\newcommand{\Bh}{\widehat{B}}
\newcommand{\Bbh}{\overline{\widehat{B}}}
\newcommand{\fhns}{\hat{F}^{\rm (NS)}}
\newcommand{\fhrr}{\hat{F}^{\rm (RR)}}
\newcommand{\ahns}{\hat{A}^{\rm (NS)}}
\newcommand{\ahrr}{\hat{A}^{\rm (RR)}}
\newcommand{\hhrr}{\hat{H}^{\rm (RR)}}
\newcommand{\hchi}{\hat{\chi}}
\newcommand{\hphi}{\hat{\phi}}
\newcommand{\htau}{\hat{\tau}}
\newcommand{\cG}{{\cal G}}
\newcommand{\cGb}{\overline{{\cal G}}}
\newcommand{\cH}{{\cal H}}
\newcommand{\cP}{{\cal P}}
\newcommand{\cPb}{\overline{{\cal P}}}
\newcommand{\cQ}{{\cal Q}}
\newcommand{\cQb}{\overline{{\cal Q}}}
\newcommand{\cM}{{\cal M}}
\newcommand{\cN}{{\cal N}}
\newcommand{\cO}{{\cal O}}
\newcommand{\cD}{{\cal D}}
\newcommand{\cL}{{\cal L}}
\newcommand{\cA}{{\cal A}}
\newcommand{\cB}{{\cal B}}
\newcommand{\hg}{\hat{g}}
\newcommand{\cE}{{\cal E}}

\newcommand{\vpp}{\mbox{$\langle{\scriptstyle++}\rangle$}}
\newcommand{\vmp}{\mbox{$\langle{\scriptstyle-+}\rangle$}}
\newcommand{\vppp}{\mbox{$\langle{\scriptstyle+++}\rangle$}}
\newcommand{\vmpp}{\mbox{$\langle{\scriptstyle-++}\rangle$}}
\newcommand{\vpmp}{\mbox{$\langle{\scriptstyle+-+}\rangle$}}

\begin{document}
\setlength{\captionmargin}{20pt}
\begin{titlepage}
\begin{flushright}
UFIFT-HEP-03-17\\
hep-th/0307203
\end{flushright}

\vskip 3cm

\begin{center}
\begin{Large}
{\bf The Fishnet as Anti-ferromagnetic Phase\\ of 
Worldsheet Ising Spins 
\footnote{Supported 
in part by the Department
of Energy under Grant No. DE-FG02-97ER-41029. 
}}
\end{Large}

\vskip 2cm
{\large 
 Charles B. Thorn\footnote{E-mail  address: {\tt thorn@phys.ufl.edu}}
and Tuan A. Tran\footnote{E-mail  address: {\tt tatran@bcm.tmc.edu}}}
\vskip0.20cm
{\it Institute for Fundamental Theory\\
Department of Physics, University of Florida,
Gainesville FL 32611}


\vskip 1.0cm
\end{center}

\begin{abstract}\noindent
We identify the strong coupling fishnet diagram 
with a certain Ising spin configuration
in the lightcone worldsheet description of planar $\Tr\Phi^3$
field theory. Then, using a mean field formalism, we
take the remaining planar diagrams into account in an average
way. Since the fishnet spin configuration is regular
but non-uniform, we introduce two mean fields $\phi,\phi^\prime$
where the fishnet diagram is the case $\phi=1,\phi^\prime=0$.
For general values of these fields, the system is
then approximated as a light-cone quantized 
string with a field dependent effective
string tension $T_{\rm eff}(\phi,\phi^\prime)$. 
We also calculate the worldsheet energy density
${\cal E}(\phi,\phi^\prime)$, and find the field
values that minimize it in the presence of a transverse
space infra-red cutoff $\epsilon>0$. The criterion
for string formation is that the tension in this
minimum energy state remains non-zero as $\epsilon\to0$.
In the most simple-minded
implementation of the mean field method, which neglects
{\it all} short range correlations of the Ising spins, we
find, in this limit, that the tension vanishes for weak and moderate
coupling, but for very large coupling does indeed stay non-zero.
However, a more elaborate treatment, 
taking temporal correlations into account
(but still neglecting spatial correlations),
removes this ``phase transition'' and the 
string tension of the minimum energy state vanishes
for all values of the coupling when $\epsilon\to0$.
Our mean field analysis thus suggests that
the ``fishnet phase'' of $\Tr\Phi^3$ theory is unstable,
and there is no string formation for any value of the coupling. 
This is probably 
a reasonable outcome given the instability of the
underlying theory. It is encouraging for our method, that an approach
designed for a string description can predict, where appropriate, the 
absence of string formation within an intuitive and simple approximation. 
\end{abstract}
\vfill
\end{titlepage}
\section{Introduction}
Ever since Maldacena proposed that IIB superstring theory on
an AdS$_5\times$S$_5$ background is equivalent to
supersymmetric Yang-Mills field theory with
extended ${\cal N}=4$ supersymmetry \cite{maldacena}, the status
of string theory in physics has dramatically changed, from a 
speculative ``theory of everything'' with no experimental
support,
to a potentially powerful
tool for analyzing the nonperturbative physics
of experimentally established quantum field theories.
The most exciting application of this new string
theory technique is, without doubt, the confinement problem of QCD.

While the vast bulk of the literature on this exciting
new program of research is dedicated to a ``top-down''
approach, seeking to recover field theoretic physics
from the string formulation \cite{klebanovs,polchinskis}, 
the only work
toward a ``bottom-up'' approach, seeking a direct
construction of a string worldsheet formalism that
sums the planar diagrams of quantum field theory, 
has been the series of articles
\cite{bardakcit,thornsheet,gudmundssontt}. These
three articles are foundational for this more
practical approach: the first sets up a worldsheet
formalism for scalar $\Phi^3$ theory, the second
for Yang-Mills theory, and the third for the whole 
range of interesting supersymmetric Yang-Mills theories,
including ${\cal N}=1,2,4$ extended supersymmetry. 

The next task is to develop an
effective and tractable framework for extracting 
nonperturbative physics from this new worldsheet
formalism. The first steps toward this goal
were taken in \cite{bardakcitmean,bardakcitimp}.
These two articles develop alternative
mean field methods on the worldsheet for approximating the
sum over planar diagrams. The first of these replaces the
Ising spins in the formalism by a homogeneous mean field, whereas the
second makes a mean field replacement of scalar 
bilinears of the target space worldsheet fields.
We continue the development
of both approaches in the present article by considering
a non-homogeneous mean field which can be used to describe
the fishnet diagram. In this way we can assess the
role of this special diagram in the context of an
average treatment of the sum of {\it all} planar diagrams.

In Section 2 we briefly review the worldsheet
formalism for the $\Tr\Phi^3$ theory. We include
a new improved way of incorporating
the mass of the scalar field. The review closes with
Eq.~(\ref{isingsumeps2}) which explicitly displays the worldsheet
system that sums the planar graphs of this theory. The
rest of the article uses mean field techniques to
approximately solve this system. Here
we want to draw attention to the essential features of
this system, including an explanation of the cutoffs
employed to completely specify the dynamics.

The worldsheet we discuss is based on light-cone
parameters, an imaginary time $\tau=ix^+=i(t+z)/\sqrt2$
in the range $0\leq\tau\leq T$
and a worldsheet spatial coordinate $0\leq\sigma\leq P^+$
chosen so that the $P^+$ density is uniform. Both
of these parameters are discretized: $\tau=ka$ and
$\sigma=lm$ with $T=Na$ and $P^+=Mm$. The limit
of a continuous worldsheet is equivalent to the
double limit $M,N\to\infty$ with $N/M=(T/P^+)(m/a)$
fixed. In this limiting process the quantity
$m/a$ which has the dimensions of force or tension
can have {\it any} value. Thus we actually have
a family of cutoff theories labeled by this
ratio. The so-called DLCQ cutoff corresponds to
$a\to0$ at fixed $m$. The continuum physics
must be independent of this ratio.

Qualitatively the worldsheet formalism maps
every planar diagram to a worldsheet with several
internal boundaries each at fixed values of $\sigma$.
The worldsheet target space field ${\boldsymbol q}(\sigma,\tau)$
satisfies Dirichlet conditions ${\boldsymbol q}={\boldsymbol k}_i$
on the $i$th boundary, and each ${\boldsymbol k}_i$
is integrated (these are the loop momenta
of the multi-loop diagram). On the discretized worldsheet
the boundaries are also discretized, occupying a
certain number of temporal links. When a boundary 
occupies a temporal link $[(i,j-1),(i,j)]$, 
Dirichlet conditions require the insertion of 
a factor $\delta({\boldsymbol q}_i^j-{\boldsymbol q}_i^{j-1})$
in the path integral. To keep track of the occurrence
of boundaries, a two valued Ising-like variable $P_i^j=0,1$
is assigned to each site. The value 1 means the site
is crossed by a boundary and 0 means it is not. Thus
each consecutive pair of sites crossed by a boundary
comes with the factor
\bea
\left[\left({2\pi m\over a}\right)^{d/2}
\delta({\boldsymbol q}_i^j-{\boldsymbol q}_i^{j-1})\right]^{P_i^jP_i^{j-1}}
\eea  
in the path integral. 
If either of the $P$'s here is zero the factor reduces to unity.
Certainly for numerical work and also for conceptual
clarity, we choose to use a Gaussian approximation to the
delta function:
\begin{eqnarray}
\left({2\pi m\over a}\right)^{d/2}\delta({\boldsymbol q}_{i}^j
-{\boldsymbol q}_{i}^{j-1})
&=&\lim_{\epsilon\to0}{1\over\epsilon^{d/2}}
\exp\left\{-{a\over2m\epsilon}({\boldsymbol q}_{i}^j-
{\boldsymbol q}_{i}^{j-1})^2\right\},
\label{deltarep}
\end{eqnarray}
If we keep $\epsilon$ finite, we see by comparing the two
right sides at ${\boldsymbol q}_{i}^j-{\boldsymbol q}_{i}^{j-1}=0$,
that $(2\pi a/m\epsilon)^{d/2}=L^d$ where $L$ is the size of the
transverse space. That is we have an infra-red cutoff $L$ in
transverse space $L^2=2\pi a/m\epsilon$. Exploring the
connection between field and string theory in the presence of an
infra-red cutoff is not unprecedented. Indeed, the work
by Berenstein, Maldacena and Nastase \cite{berensteinmn} which explores the
field/string connection in the original SUSY Yang-Mills/AdS
context defines the gauge theory on $S_3$ rather than $R_3$.
Similarly, in the present work we study the sum of planar diagrams
with $\epsilon$ finite. We must remember though that the
``field theory'' we are thereby solving is some compactified
version of the original $\Phi^3$ theory. The worldsheet formalism
for a more conventional compactification (on a
transverse torus) of $\Tr\Phi^3$ theory
is sketched in Appendix D.

Using (\ref{deltarep}) leads to the factor
\bea
\left[{1\over\epsilon}\right]^{dP_i^jP_i^{j-1}/2}
\exp\left\{-{aP_i^jP_i^{j-1}\over2m\epsilon}
({\boldsymbol q}_i^j-{\boldsymbol q}_i^{j-1})^2\right\}
\eea  
for each pair of sites crossed by a boundary. One can immediately
anticipate that the $\epsilon\to0$ limit will be very delicate
in approximate calculations. For example the mean field
approximation we shall employ replaces $P_i^j$ by a slowly varying
field $\phi$ taking values between 0 and 1. The $q$ integral
then behaves as $\epsilon^{d/2}$ whereas the prefactor is softened
and the net factor becomes $\epsilon^{(1-\phi^2)d/2}$,
which strongly suppresses the contribution as $\epsilon\to0$ 
except when $\phi=1$. We can mitigate this delicacy somewhat 
by arranging the $\epsilon$ factors as in (\ref{isingsumeps2}). 
This makes
no difference in the exact formula but makes a big difference
in the $\epsilon\to0$ limit of a mean field approximation.
Even if the $P$'s are less than 1 the singular $\epsilon$
dependence cancels between the $q$ and $b,c$ integrals.
This helps define a less singular mean field approximation,
but the need for such a finely tuned set up of
the approximation nonetheless casts
doubt on its reliability at small $\epsilon$. We can be more 
confident about our conclusions for $\epsilon$
relatively large, say of $O(1)$. It is nonetheless
important to confirm (or refute) our conclusions 
with more exact calculations such as Monte Carlo 
and block spin renormalization group studies.
In any case we think our mean
field studies give an important qualitative description
of the worldsheet dynamics
that will be crucial in interpreting these more accurate numerical
methods.

In Section 3 we give a brief description of the fishnet diagram
and its representation in terms of a specific
Ising spin pattern on the BT worldsheet. This pattern motivates the
choice of mean fields we use to illuminate the role
of the fishnet in the complete sum of planar graphs. Sections 4 and
5 explain the results of our two approaches to the 
mean field approximation based on mean fields $\phi_k^j=\langle P_k^j\rangle$
and $\varphi_k^j=\langle P_k^jP_k^{j-1}\rangle$ respectively. Section
6 gives our concluding remarks. Appendices A, B and C contain
the technical details of our calculations.

\section{Worldsheet for Summing $\Phi^3$ Planar Diagrams}
\label{sec2}

We start our review of \cite{bardakcit} by recalling the
worldsheet representation of the free gluon propagator
(we shall refer to the $\Phi$ field quanta as
``gluons'' even though their spin is 0).
Consider a gluon carrying $M$ units of $p^+=Mm$ and 
and propagating $N$ time steps. The transverse momentum 
${\boldsymbol p}\equiv {\boldsymbol q}_M-{\boldsymbol q}_0$,
where the worldsheet target space fields ${\boldsymbol q}_k^j$,
$k=0,\ldots,M$ and $j$ labeling the time slice,
live on the sites of the worldsheet lattice. ${\boldsymbol q}_k^j$
satisfies Dirichlet boundary conditions ${\boldsymbol q}_0^j
={\boldsymbol q}_0$, ${\boldsymbol q}_M^j
={\boldsymbol q}_M$.
We also must introduce Grassmann ghost variables $b_k^j,c_k^j$ on each
lattice site to provide crucial factors of $M$ when we
represent a single gluon as $M$ bits. 
$D=d+2$ will denote the space-time dimension, which
we keep general for our formal presentation.
But in our numerical work we take
the case $d=2$, four dimensional space-time. There
are $d$ components of each $q$ and $d/2$ pairs $b,c$ of ghost fields.
With this notation the action is
\bea
S&=&S_g+S_q\\
S_q&=&{a\over2m}\sum_j\sum_{i=0}^{M-1}
({\boldsymbol q}^j_{i+1}-{\boldsymbol q}^j_i)^2\\
S_g&=&-{a\over m}\sum_j
\left[{\boldsymbol b}^j_1{\boldsymbol c}^j_1
+{\boldsymbol b}^j_{M-1}{\boldsymbol c}^j_{M-1}
+\sum_{i=1}^{M-2}({\boldsymbol b}^j_{i+1}
-{\boldsymbol b}^j_i)({\boldsymbol c}^j_{i+1}-{\boldsymbol c}^j_i)\right]
\label{freeghostaction}
\eea
 Then the master formula for the worldsheet formalism is \cite{bardakcit}
\begin{eqnarray}
\exp\left\{-{T{\boldsymbol p}^2\over2p^+}\right\}
&=&\exp\left\{-N{a\over m}
{({\boldsymbol q}_M-{\boldsymbol q}_0)^2\over2M}\right\}=
\int\prod_{j=1}^N\prod_{i=1}^{M-1} 
{d{\boldsymbol c}^j_id{\boldsymbol b}^j_i\over2\pi} 
d{\boldsymbol q}^j_i\ e^{-S_g-S_q}\equiv\int DcDbD{\boldsymbol q}\ e^{-S}.
\label{nstepbits}
\end{eqnarray}

In (\ref{nstepbits}) we have imposed Dirichlet boundary conditions.
However,to dynamically 
implement momentum conservation in the
path integral we integrate over  $q^j_M$ independently,
retain the Dirichlet boundary condition at $i=0$, 
where we impose ${\boldsymbol q}_0^j={\boldsymbol q}_0$, but insert
momentum conserving delta functions at the other end $i=M$.
It is convenient but not necessary to set ${\boldsymbol q}_0=0$.

We define the dimensionless coupling 
\bea
{\hat g}^2\equiv{g^2\over64\pi^3}\left({m\over 2\pi a}\right)^{(d-4)/2}.
\eea
Note that in $D$ space-time dimensions we have all together
$d/2$ sets of $b,c$ ghosts, denoted by bold-faced letters
when referring to all such sets. 
But one of these sets, identified with non-bold type,
is singled out to handle the special factors occurring at interaction points.
When needed, we denote the remaining $(D-4)/2$ sets by checking
their symbols ${\check b},{\check c}$.

We associate factors
\bea
{\cal V}_{0i}^{j}&\equiv& {\hat g}
\exp\left\{-{a\over m}(b_{i-1}^{j}c_{i-1}^{j}
+b_{i+1}^{j}c_{i+1}^{j})\right\}\\
{\bar{\cal V}}_{0i}^{j}&\equiv& {\hat g}\exp\left\{-{a\over m}
(b_{i+1}^{j+1}-b_{i}^{j+1})
(c_{i+1}^{j+1}-c_{i}^{j+1})\right\}
\eea
with the fission and fusion vertices respectively, the different forms
due to the asymmetric way the ghost insertions assign the required 
$1/p^+$ factors to the vertices. 

Next we recall the formula that systematically
sums over all the planar diagrams on the
lattice. The worldsheet for the general planar diagram has
an arbitrary number of vertical solid lines marking the
location of the internal boundaries corresponding
to loops. Each interior link $j,j-1$ of
a solid line at spatial location $k$ requires a factor of
$\delta({\boldsymbol q}_{k}^j-{\boldsymbol q}_{k}^{j-1})$.
To supply such factors, assign an Ising spin $s_k^j
=\pm1$ to each site of the lattice. We assign $+1$ if the site $(k,j)$ 
is crossed by a vertical solid line, $-1$ otherwise. 
We also use the spin up projector $P_k^j=(1+s_k^j)/2$.

At the endpoints of each solid line we have to
supply the vertex insertions ${\cal V}_0$, ${\bar{\cal V}}_0$. 
These factors occur when there is a spin flip,
$s_k^j=-s_k^{j-1}$. In the foundational papers
we exploited the properties of Grassmann integrals
to take these factors into the exponent where they
were multiplied by bilinears in the ghosts. This was
particularly convenient for the vertices of
gauge theories, which had momentum and
spin dependent prefactors. However, for the
scalar theory treated in this article, this is
not necessary, because the prefactor
is simply ${\hat g}$ and the ghost insertions are
already in exponential form. In the general formula
we need only multiply the $bc$ term in the exponent of 
the vertex insertion by $P_i^jP_i^{j+1}(1-P_i^{j-1})$ or
$P_i^jP_i^{j-1}(1-P_i^{j+1})$ respectively. 
Note that each internal loop 
occupies at least two time steps. 

In this article we implement the Dirichlet
conditions on boundaries using the Gaussian representation of
the delta function (see the
last line of (\ref{deltarep})).
We keep $\epsilon$ finite until the end of the calculation.
Using this device, our formula for
the sum of planar diagrams becomes:
\begin{eqnarray}
T_{fi}&=&\lim_{\epsilon\to0}
\sum_{s_i^j=\pm1}\int DcDbD{\boldsymbol q}
\exp\left\{\ln{\hat g}\sum_{i,j}\left({P}_i^j{P}_i^{j+1}(1-{P}_i^{j-1})
+{P}_i^j{P}_i^{j-1}(1-{P}_i^{j+1})\right)
\phantom{\sum}\hskip-.5cm\right\}
\nonumber\\&&
\exp\left\{-{a\over2m}\sum_{i,j}
{({\boldsymbol q}_{i+1}^j-{\boldsymbol q}_{i}^{j})^2}
-{a\over2m\epsilon}\sum_{i,j}
{({\boldsymbol q}_{i}^j-{\boldsymbol q}_{i}^{j-1})^2}
P_i^jP_i^{j-1}\right\}
\nonumber\\
&&\exp\left\{{a\over m}\sum_{i,j}{\boldsymbol b}_{i}^{j}
{\boldsymbol c}_{i}^{j}\left({P}_i^j{P}_i^{j+1}(1-{P}_i^{j-1})
+{1\over\epsilon}{P}_i^j{P}_i^{j-1}\right)
\phantom{\sum}\hskip-.5cm
\right\}
\nonumber\\&&
\exp\left\{{a\over m}
\sum_{i,j}({b}_{i+1}^j-{b}_i^j)({c}_{i+1}^j-{c}_i^j)
(1-P_i^j)(1-P_{i+1}^j)(1-P_i^{j-1}P_i^{j-2})\right.\nonumber\\
&&\left.+{a\over m}\sum_{i,j}(1-P_i^j)(P_{i+1}^j
[1-P_{i+1}^{j+1}(1-P_{i+1}^{j-1})]+
P_{i-1}^j[1-P_{i-1}^{j+1}(1-P_{i-1}^{j-1})])
{b}_{i}^{j}{c}_{i}^{j}\right\}\nonumber\\
&&\exp\left\{{a\over m}
\sum_{i,j}({\check b}_{i+1}^j-{\check b}_i^j)
({\check c}_{i+1}^j-{\check c}_i^j)
(1-P_i^j)(1-P_{i+1}^j)+{a\over m}\sum_{i,j}(1-P_i^j)(P_{i+1}^j+
P_{i-1}^j){\check b}_{i}^{j}{\check c}_{i}^{j}\right\}
\label{isingsumepsilon}
\eea
The first exponent in this formula supplies a factor of
${\hat g}$ whenever a boundary is created or destroyed.
The second exponent includes the action $S_q$ for the free
propagator together with the exponent in the Gaussian representation
of the delta function that enforces Dirichlet boundary conditions
on the solid lines.
The contents of this exponent
go to the discretized action for the
light-cone quantized string, if the quantity
$a^2P_i^jP_i^{j-1}/m^2\epsilon$ is replaced by
$1/T_0^2$, with $T_0$ the string rest tension.
The third exponent incorporates the $\epsilon$ dependent
prefactor for that representation of the delta
function as a term in the ghost Lagrangian. 
The remaining exponents contain $S_g$ together with
strategically placed spin projectors that arrange the
proper boundary conditions on the Grassmann variables
and supply appropriate $1/p^+$ factors needed at the
beginning or end of solid lines.

We remark that, when Dirichlet conditions are imposed
at $i=0,M$, the expression (\ref{isingsumepsilon}) sums all the
planar multi-loop corrections to the gluon propagator. 
The evolving system is therefore in the
adjoint representation of the color group. That is, we have tacitly
assumed that the only solid lines initially and finally are
those at the boundaries of the strip. More general initial and
final states are described by allowing more solid lines initially
and finally. When the system is in a color singlet state
(a ``glueball''),
we must include diagrams in which the outer boundaries
are identified, \ie\ the diagrams should really be drawn on a
cylinder, not a strip. In this case, 
${\boldsymbol q}(p^+)
={\boldsymbol q}(0)+{\boldsymbol p}$, the variable ${\boldsymbol q}(\sigma)$
is strictly periodic only in the state of zero
total transverse momentum. 

When we regard the worldsheet as
a cylinder, we realize that the Ising spins can be in configurations
that have no correspondence to the original Feynman diagrams.
These involve time slices in which {\it all} spins are down,
signifying the complete absence of solid lines through
the slice. Several consecutive such slices represent
the evolution of a no gluon or ``nothing'' state. This
is the closest thing the formalism gets to a closed string
state in the non-interacting theory. The worldsheet formalism
associates this state with the ambiguous $0/0$: both the $bc$
and $q$ systems have a zero mode on a nothing time slice,
the ghost zero mode is responsible for the 0 in the numerator
and the $q$ zero mode for the zero in the denominator.
In the exact formula this nothing state decouples from an
initial state with gluons because the ghost insertion
associated with a terminating boundary supplies a second zero factor
making the contribution of the transition time slice $0^2/0=0$.
So the nothing state will be removed from all amplitudes that have
at least one gluon in the initial state. However if one
is considering the generic state that evolves without
regard to initial conditions, the nothing state will be included. 
In particular,
the mean field approximation employed in this article replaces 
the Ising spins by average spins, the approximate
amplitudes will not involve definite spin configurations,
and for non-zero mean fields the zero modes are removed.
The nothing state will be mixed in with multi-gluon states.
For moderate to strong coupling where the typical time slice
has many up spins this admixture should be innocuous. However, for
weak coupling when up spins are very rare, the precise
connection with straight perturbation theory
will be obscured because the mean field
method will treat the precise decoupling of the nothing
state in a rough average way. So applying the mean
field approximation at nonzero coupling and then taking
the coupling to zero will most likely lead to the
description of the nothing state rather than a state with
a small number of gluons. 

After that digression on ``nothing'' we
continue with our discussion of the
world sheet path integral.
We can simplify the first exponent with the replacement
\bea
\exp\left\{\ln{\hat g}{\sum({P}_i^j{P}_i^{j+1}
+{P}_i^j{P}_i^{j-1}-2{P}_i^j{P}_i^{j-1}{P}_i^{j+1})}\right\}
\to\prod_{i,j}\left({\hat g}\right)^{(1-s_i^js_i^{j-1})/2}.
\eea
This is valid because the left
side agrees with the right side except for spin patterns that
include one or more isolated up spin (one
preceded and followed by a down spin). But inspection shows that
the integrand is independent of the $b,c$ on the site of each such
up spin, so the ghost integral gives zero. Thus only spin patterns
in which each up spin is preceded and/or followed by an up spin
contribute to the worldsheet integral. This also explains
why the original formula vanishes at zero coupling if there
is one or more spin flip in the spin pattern. 

The formula (\ref{isingsumepsilon}) 
assumes the $\Phi$ field is massless. We now present
a new worldsheet local procedure for incorporating a mass $\mu$
for $\Phi$ that is superior to the
one offered in \cite{bardakcit}. Consider the identity 
\begin{eqnarray}
\int \prod_{i=1}^{M-1} {dc_idb_i\over2\pi} \exp\left\{{a\over m}
\left[{b_1c_1\over\alpha}+{b_{M-1}c_{M-1}\over\beta}
+\sum_{i=1}^{M-2}(b_{i+1}-b_i)(c_{i+1}-c_i)\right]
\right\}
&=&\nonumber\\
&&\hskip-1.2in{M\over\alpha\beta}\left(1+{\alpha+\beta-2\over M}\right)
\left({a\over2\pi m}\right)^{M-1}.
\label{ghostbitsab}
\end{eqnarray}
We have already exploited two special cases of this
formula: $\alpha=\beta=1$ gives 
one time slice of the worldsheet ghost 
path integral for the massless propagator, 
and the case $\alpha=\infty$, $\beta=1$ provides the
mechanism for supplying $1/p^+$ factors in the worldsheet
path integral for interacting Feynman diagrams.
 
A finite mass $\mu$ in the free propagator corresponds to
a factor
\bea
\exp\left\{-{\mu^2T\over2p^+}\right\}\to
\left(1-{\mu^2a\over dMm}\right)^{dN/2},
\label{massfactor}
\eea
where the right side is a valid discretization $T=Na, p^+=Mm$ 
of the left side, going to the latter in the limit $M,N\to\infty$.
Thus we must provide a factor $(1-\mu^2a/dmM)^{d/2}$ on each time slice.
This will be achieved with $d/2$ copies of (\ref{ghostbitsab}) if we choose
$\alpha+\beta-2=-\mu^2a/dm$ and multiply the left side
of (\ref{ghostbitsab}) by $(\alpha\beta)^{d/2}=\exp\{(d/2)\ln\alpha\beta\}$.
For example, with $\beta=1$, the choice $\alpha=1-\mu^2a/dm$ does the job.
Define the parameter $\rho\equiv (1/\alpha)-1=\mu^2a/(dm-\mu^2a)$.
Then the worldsheet action for the massive propagator
is obtained by adding  the term $-\rho b_1^jc_1^j$
to the free ghost action (\ref{freeghostaction}) and 
also multiplying the path integral by $\alpha^{Nd/2}=(1+\rho)^{-Nd/2}$. 
This latter factor
can be associated with the boundary of the world sheet, since
its exponent is proportional to the length of the boundary.
It contributes like a ``boundary cosmological constant''.
We should include one such factor for every internal boundary and
the square root of the factor for each external boundary.
(When we draw our graphs on cylinders every boundary will be
internal.) Since boundaries occur whenever $P_k^j=1$, for a
general diagram we can supply this factor by including the
term $(d/2)\ln\alpha\sum_{kj}P_k^j$ in the exponent of
the general worldsheet integrand.

In the worldsheet path integral for a general diagram, the 
added term $\rho b_k^jc_k^j$ should appear on a site with
$P_k^j=0$ when it is immediately to the right of a solid line,
that is when $P_{k-1}^j=1$ as well. Furthermore, it cannot
occur on such a site if it immediately follows the 
beginning of solid line, since the coefficient of $bc$ on such
a site must be zero (to properly include $1/p^+$ factors).
Thus we must have $P_{k-1}^{j-1}=1$. The upshot is that
the term should be multiplied by the factor 
$(1-P_k^j)P_{k-1}^{j-1}P_{k-1}^j$. Because of these restrictions,
we might wonder whether the contribution of the boundary
factor should share them. However, we think not. First of
all, two adjacent solid lines bound a gluon carrying a
single $p^+$ unit, i.e. $M=1$. According to (\ref{massfactor}) an
$M=1$ gluon should have a factor $\alpha^{d/2}$ for
each time slice. This is precisely supplied by the
boundary factor. (Since an $M=1$ gluon contains
no spin down sites, there is no $\rho bc$ term.) 
So we should include the boundary
factor for every solid line, including those adjacent to
each other. Next, what about the absence of the $\rho bc$
term at the beginning of each solid line; should we
also omit the factor $\alpha^{d/2}$ for that one time slice?
In other words, should a boundary of $N$ time steps yield
the factor $\alpha^{d(N-1)/2}$ or $\alpha^{dN/2}$? Actually this
is a matter of taste because the discrepancy between the
two choices can be absorbed in a redefinition of
the coupling constant. In
order to preserve the interpretation of these factors
as boundary cosmological constant effects we choose 
the second alternative $\alpha^{dN/2}$.

After all of these considerations we propose that the
worldsheet path integral for the sum of planar diagrams
in the massive theory be given by
\begin{eqnarray}
T_{fi}&=&\lim_{\epsilon\to0}
\sum_{s_i^j=\pm1}\int DcDbD{\boldsymbol q}
 \exp\left\{\ln{\hat g}\sum_{ij}{1-s_i^js_i^{j-1}\over2}
-{d\over2}\ln\left(1+\rho\right)\sum_{i,j}P_i^j\right\}
\nonumber\\
&&\exp\left\{-{a\over2m}\sum_{i,j}
{({\boldsymbol q}_{i+1}^j-{\boldsymbol q}_{i}^{j})^2}
-{a\over2m\epsilon}\sum_{i,j}P_i^jP_i^{j-1}
{({\boldsymbol q}_{i}^j-{\boldsymbol q}_{i}^{j-1})^2}
\right\}\label{isingsumeps2}\\
&&\exp\left\{{a\over m}
\sum_{i,j}\left[A_{ij}{\boldsymbol b}^j_{i}{\boldsymbol c}^j_{i}
-B_{ij}{b}_{i}^{j}{c}_{i}^{j}
+C_{ij}({\boldsymbol b}_{i+1}^j-{\boldsymbol b}_i^j)
({\boldsymbol c}_{i+1}^j-{\boldsymbol c}_i^j)
-D_{ij}({b}_{i+1}^j-{b}_i^j)({c}_{i+1}^j-{c}_i^j)\right]\right\}\nonumber\\
A_{ij}&=&{1\over\epsilon}{P}_i^j{P}_i^{j-1}
+{P}_i^{j+1}{P}_i^j-{P}_i^{j-1}{P}_i^j{P}_i^{j+1}
+(1-P_i^j)(P_{i+1}^j+P_{i-1}^j)+\rho(1-P_i^j)P_{i-1}^{j-1}P_{i-1}^j\\
B_{ij}&=&(1-P_i^j)\left(P_{i+1}^jP_{i+1}^{j+1}
(1-P_{i+1}^{j-1})+
P_{i-1}^j{P_{i-1}^{j+1}(1-P_{i-1}^{j-1})}
+P_i^{j-1}P_i^{j-2}P_{i+1}^j\right)\\
C_{ij}&=&(1-P_i^j)(1-P_{i+1}^j)
\\
D_{ij}&=&(1-P_i^j)(1-P_{i+1}^j)P_i^{j-1}P_i^{j-2}
\end{eqnarray}
In the rest of the article we apply mean
field methods to {\it this version} of the
worldsheet system. It is helpful to note that once
$\mu^2\neq0$, one can scale $\rho$ to $\rho=1$
by fixing an appropriate value of $m/a$. Then the
massless limit would be regained by taking ${\hat g}\to\infty$.
\section{The Fishnet Spin Pattern}
We would like to extend the mean field formalism of 
\cite{bardakcitmean,bardakcitimp} to allow for
regular spin patterns that are not necessarily
completely uniform. In spin models, such a
generalization is important for describing 
the anti-ferromagnetic phase. Similarly in
the sum of planar diagrams, the fishnet diagrams
first identified by  Nielsen and Olesen and 
of Sakita and Virasoro \cite{nielsenfishnet}
are an important subclass of 
diagrams that naturally contain a stringy interpretation.
Based on these ideas the program of
using a ``strong coupling fishnet'' diagram as the 
scaffolding for the construction of a worldsheet representation
of the sum of the planar diagrams of quantum field
theory was proposed in 1977 \cite{thornfishnet}. This work
made use of the mixed $(x^+,p^+,\boldsymbol{p})$ representation
\cite{goddardgrt,thooftlargen}
of each propagator, and further employed a discretization of
both $p^+=lm$ and $ix^+=ka$ with $k,l$ positive integers
\cite{gilest}.
In the context of this discretization, the formal strong coupling
limit singles out those diagrams in which every propagator
spans a single time unit $a$ and carries one or two units $m$
of $p^+$. Among such diagrams, the fishnet diagram is further 
restricted by the requirements (1) that it be planar, and
(2) that no pair of propagators connect the same pair of
vertices. These properties together single out a more or
less unique diagram, the strong coupling fishnet. For $\Tr\Phi^3$
field theory, this diagram, calculated
in \cite{beringrt} is shown on the left in Fig.~\ref{fishnet}.
It was shown in \cite{thornfishnet,beringrt} that the diagram describes
the propagator of the lightcone quantized string \cite{goddardgrt},
discretized as in \cite{gilest}.

\begin{figure}[htb]
\psfrag{'M'}{$$}
\psfrag{'N'}{$$}
\begin{center}
\includegraphics[width=15cm]{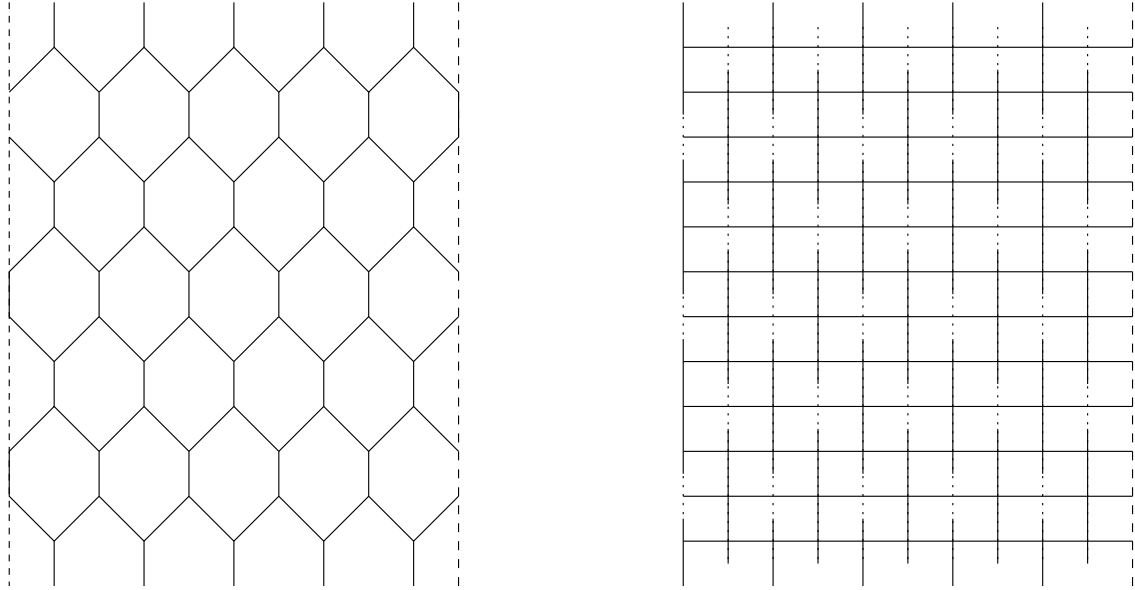}
\caption{The spin pattern on the right corresponds
to  the $\Tr\Phi^3$ fishnet diagram on the left. Both
diagrams are drawn on a cylinder, with opposite
sides identified.}
\label{fishnet}
\end{center}
\end{figure}

Of course there are many other diagrams that are important
at strong coupling: the fishnet diagram was to be
regarded as a worldsheet template for the inclusion, in
principle, of all the rest of the planar diagrams. The proposal
amounted to a nontrivial reorganization of the summation
of planar diagrams in a fashion in which a string
worldsheet description occupied center stage. However, in the
references cited above, no approach to the inclusion of
``the rest of the planar diagrams'', beyond a brute force
calculation, was offered.

The worldsheet formalism \cite{bardakcit},
reviewed in the previous section,
offers powerful techniques for an approximate evaluation of
the contribution of all planar diagrams. We begin by
identifying which pattern of Ising spins corresponds to the
fishnet diagram just described. We have drawn the corresponding 
Bardakci-Thorn (BT) worldsheet on the right
in Fig.~\ref{fishnet}. The Ising spin at a given
site is $+1$ or $-1$ depending on whether the site is crossed
by a solid or dotted line respectively. Following the sites
for a fixed spatial value, we find the 
repeated pattern $-+++-+++-+++\cdots$. The corresponding 
pattern at the neighboring spatial sites is $++-+++-+++-+\cdots$.
This has an anti-ferromagnetic character, though not quite the
strictly alternating pattern of the classic Ising
anti-ferromagnet.

Our calculations are based on the mean field method, developed
for this system in \cite{bardakcitmean,bardakcitimp}.
The first of these articles assigned the mean
field $\phi_k^j=\langle P_k^j\rangle$, and studied the
energy for a completely homogeneous pattern $\phi_k^j=\phi$.
In Section 4 we generalize this work to the fishnet spin pattern
by introducing {\it two} mean fields $\phi,\phi^\prime$,
with $\phi$ the average spin at the sites marked by $+$ in the
previous paragraph and with $\phi^\prime$ the average spin
at the sites marked by $-$. Note that with $\phi\neq\phi^\prime$
the translational symmetry of the lattice is 
broken by the assignment. But the homogeneous spin pattern
can still be accessed as the special case $\phi^\prime=\phi$. 
We shall calculate the
bulk energy per site ${\cal E}(\phi,\phi^\prime)$ of the system,
as a function of $\phi$ and $\phi^\prime$. The fishnet
diagram corresponds precisely to the case $\phi=1$ and $\phi^\prime=0$.
Configurations away from these values take into account
all other diagrams in an average way. Our work therefore
provides a first look at how the complete sum of planar
diagrams alters the string interpretation given by the
fishnet.

The second mean field article \cite{bardakcitimp} applied the
mean field method to the matter fields rather than to the
Ising spins. However, this approach also leads to a mean
field description of the spins in which the mean field
$\varphi_k^j= \langle P_k^{j-1}P_k^j\rangle$, and is
associated with temporal links rather than sites.
For a link joining two $+$ spins, the field has the
value $1$, whereas it is zero for all other possibilities.
For the fishnet pattern we then find the link field
values $01100110011\cdots$ alternating with 
$10011001100\cdots$ on spatial slices. Although this
second approach is in some ways superior to the first, 
its application to the
full-fledged theory is considerably more complicated than
the first and won't be attempted here. Instead, we
explore the role of the spin order parameter 
$\varphi_k^j=\langle P_k^jP_k^{j-1}\rangle$ singled out
by this second approach. In section 5 we simply repeat
the calculations of Section 4 using this alternate
choice for mean field.

\section{Mean Field Treatment of the Ising Spins}
In this section we work directly with the Ising spin system
and approximate the $P_i^j$'s in certain terms of
the action by a mean field as was done in
\cite{bardakcitmean}. We do this by first
setting up the standard effective action formalism,
by adding a source 
term $\sum_{kj}\kappa_k^jP_k^j$ for the $P$ variables to the action.
Then we write the path integral in the presence of
$\kappa$ as $e^{W(\kappa)}$ and define the effective
field $\phi_k^j\equiv\langle P_k^j\rangle_\kappa 
=\partial W/\partial\kappa_k^j$. The effective action is then
obtained by Legendre transformation
\bea
A(\phi)&\equiv& W(\kappa)-\sum_{kj}\kappa_k^j\phi_k^j\\
{\partial A\over\partial\phi_k^j}&=&-\kappa_k^j.
\eea
Then it follows that, when the source is removed by setting
it to zero, the spin expectation $\phi$ is a stationary point
of the effective action.

So far everything is exact. The mean field approximation
comes into play in attempting an approximate calculation
of $W(\kappa)$. A straight perturbation theory would treat
the terms coupling the matter fields and the spin fields
as a perturbation. The way we implement the mean field
approximation is to improve on perturbation
theory by replacing the spin dependent coefficients
of the matter terms in the action by their expectation
values, regarded as fixed for the purposes of the spin sum.
This decouples the spin sum from the matter integrals and
makes it tractable. This decoupled spin sum then
yields an approximation to the $\kappa$ dependence of
$W$, which in turn determines the expectations $\langle P\rangle,
\langle PP^\prime\rangle,\langle PP^\prime P^{\prime\prime}\rangle$,
which are used as the zeroth approximations to the 
coefficients of the matter terms. One then looks for
stationary points of the approximate total effective
action. This zeroth order mean field approximation can
then be improved by treating the differences between
the actual spin dependent coefficients and their
assumed expectation values as a perturbation.

The extent of the approximation depends on how much of the
$P$ dependence is replaced by its expectation value. 
The simplest option is to
make this replacement for {\it all} terms in the action
(excluding the source term itself of course).
But we shall also (as in \cite{bardakcitmean}) consider replacing
only the spatially coupled $P$'s and those
multiplying matter fields by their expectations, since the
remaining terms can be explicitly included in the 
spin sums \footnote{In the standard mean field treatment of the Ising model
every spin in the action is replaced with a mean field. But
replacing say only the spatially coupled spins by mean fields leads
to an improved approximation to the critical temperature. In the
two dimensional Ising model the usual mean field treatment
gives $T^{mf}_{c}\approx 1.76T_{c}^{exact}$. In contrast replacing
only the spatial spins by mean fields leads to the improved
estimate $T^{partial}_{c}\approx 1.55T_{c}^{exact}$. As is
well known, neither version does a good job with critical exponents.}.
The first option neglects all short range correlations between
spins, whereas the second takes at least some of the temporal
correlations into account.

In all cases the matter ($q$ and $bc$) integrals are evaluated in the
presence of the mean fields $\phi$. These integrals are 
done in Appendix C for the fishnet pattern involving the
two mean fields $\phi,\phi^\prime$. In our version of the
mean field approximation the monomials of $P$'s in the
matter action are replaced by their expectations. The
fishnet spin pattern requires only five 
such quantities: $\vpp,\vmp,\vppp,\vmpp,\vpmp$, all
defined in Appendix A. The result of the
$q$ integration 
(Eq.~\ref{qint}) is repeated here for convenience.
\bea
\left({2\pi m/a}\right)^{-dMN/2}
e^{W_q}=
\prod_{l<M/2} 
\left[(\alpha^2-\beta^2)^{N-2}{\sinh N\xi^l_+
\over\sinh 2\xi^l_+}{\sinh N\xi^l_-\over\sinh 2\xi^l_-}
\right]^{-d/2}.\nonumber
\eea
In view of the $N=T/a$ dependence in this formula it is clear that
the $\xi^l_\pm$ represent excitation energies of the system:
\bea
\Delta E^l_\pm={1\over a}\xi^l_\pm .
\eea
Finite energy excitations correspond to $\xi^l=O(1/M)$, which entails
$\sqrt{R_l}\sinh\kappa_l=O(1/M)$ or $\sqrt{R_l}\cosh\kappa_l=1+O(1/M^2)$.
Since $\omega_l\sim (l\pi/M)$ for $l\ll M$ we read off in this limit
\bea
\sqrt{R_l}\sinh\kappa_l\sim{l\pi\over M}
\sqrt{2+\alpha\over2\alpha+\alpha^2-\beta^2}.
\eea
Since $\Delta{\rm mass}_l^2=2Mm\Delta E\equiv 2\pi T_{\rm eff} l$ we read
off the effective string tension in terms of the background fields:
\bea
T_{\rm eff}={m\over a}\sqrt{2+\alpha\over2\alpha+\alpha^2-\beta^2}
={m\over a}\sqrt{\vpp+\vmp+4\epsilon
\over2(\vpp+\vmp+\vpp\vmp/\epsilon)}\; ,
\eea
where $\alpha=(\vpp+\vmp)/2\epsilon$ and
$\beta=(\vpp-\vmp)/2\epsilon$.
\begin{figure}[htb]
\includegraphics[width=8cm]{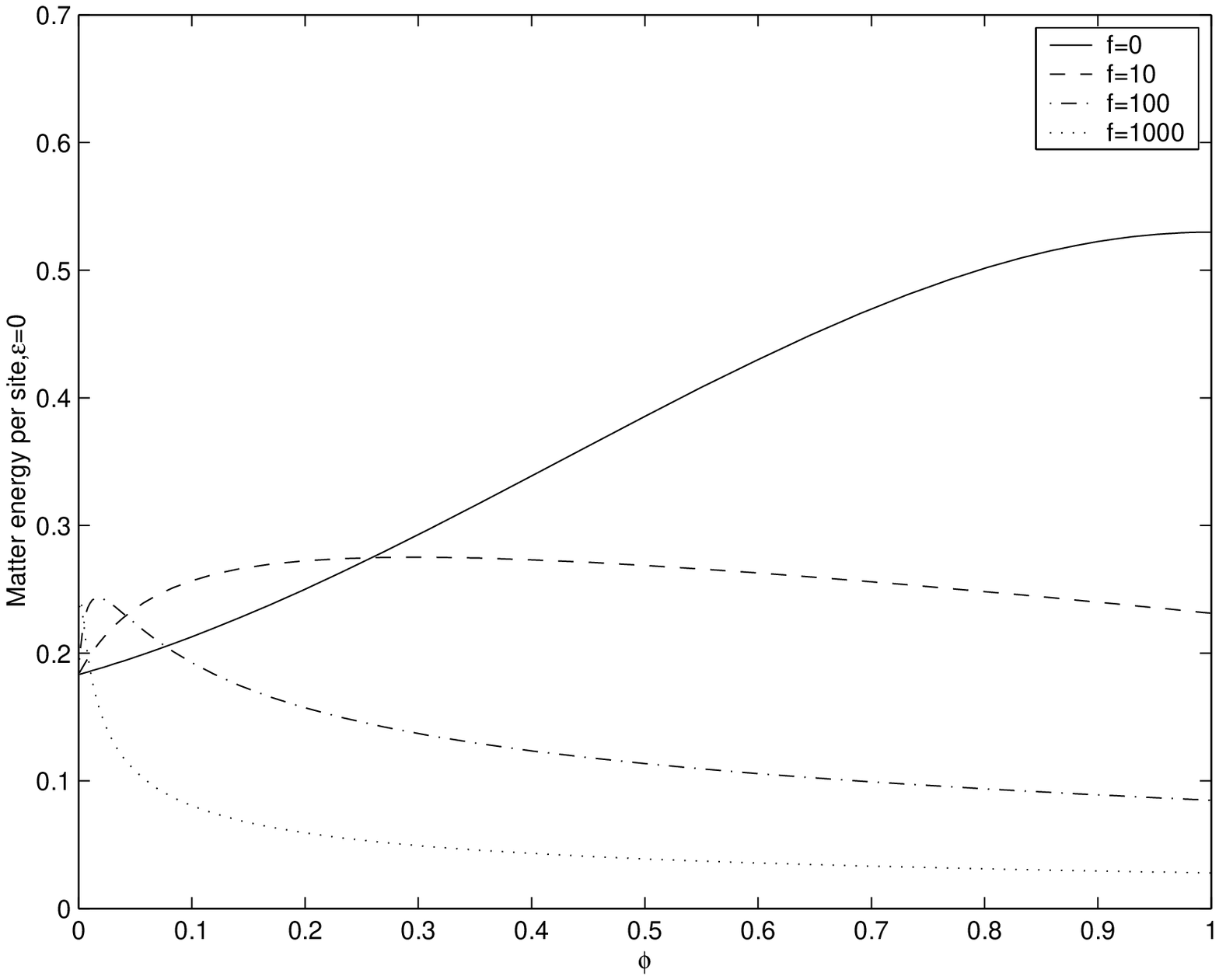}\quad
\includegraphics[width=8cm]{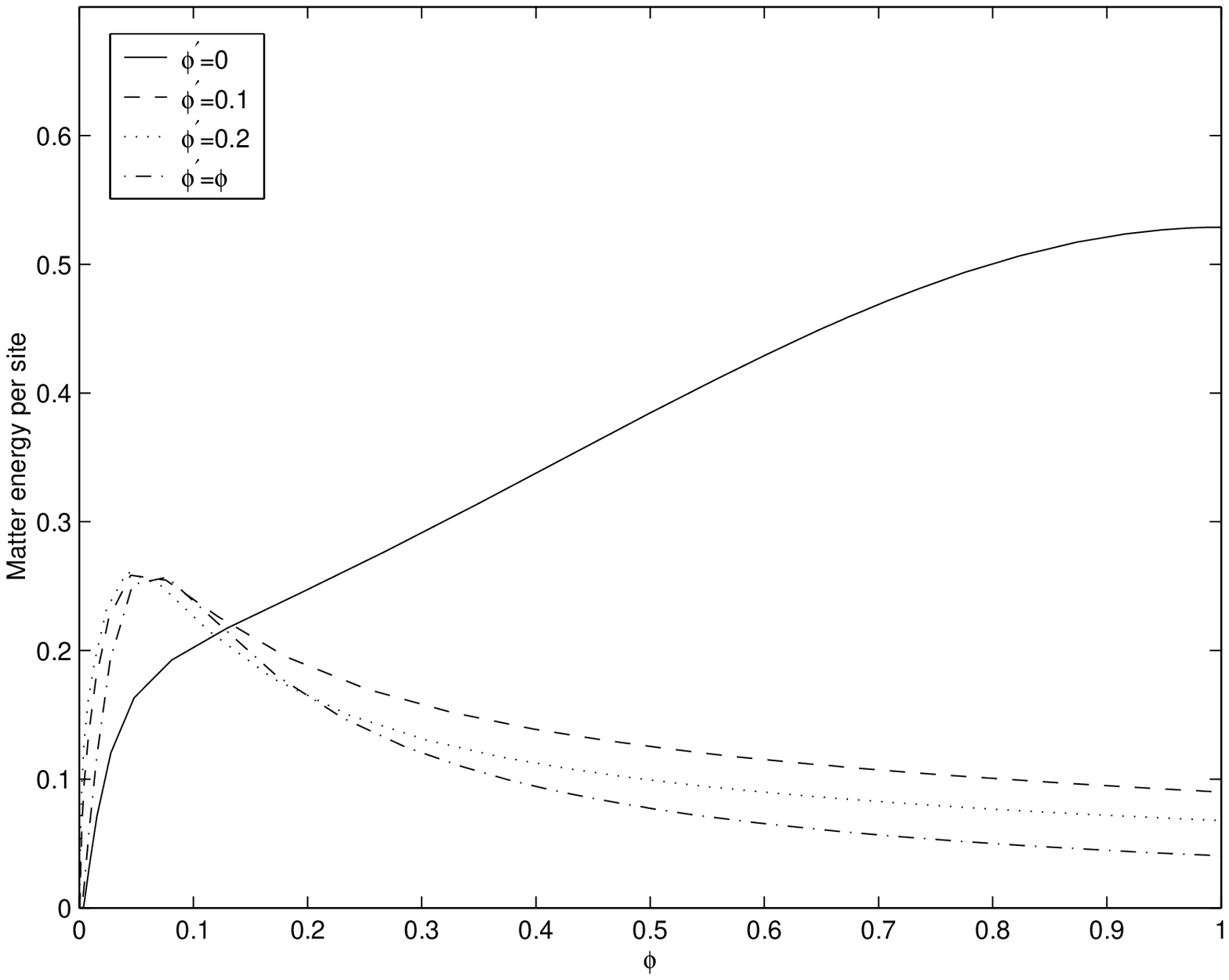}
\caption{The contribution of $q$ plus ghost fields to the bulk
energy. In the left graph we have taken $\epsilon\to0$ holding
$\phi^\prime=\epsilon f$ fixed. The right graph is plotted for
fixed $\epsilon=0.001$. Note that the effect of finite $\epsilon$
is concentrated near $\phi=0$.}
\label{matenergy}
\end{figure}

For fixed $\phi,\phi^\prime$, both $\alpha$ and $\beta$ are 
of $O(1/\epsilon)$ as $\epsilon\to0$. 
With no relation between them, it is easily seen
that the tension is of $O(\sqrt{\epsilon})$ and hence vanishes. If, however
the limit can be taken in a way 
that $(\alpha^2-\beta^2)/\alpha$ is finite, the
tension will also be finite. This can be achieved either with both
$\vpp,\vmp=O({\epsilon})$ or $\vpp=O(1)$ while $\vmp
=O(\epsilon)$. In the second case, writing $\phi^\prime=
\epsilon f$ and $\vmp=\epsilon fG(\phi)$,
we find
\bea
T_{\rm eff}^{\epsilon=0}(\phi,f)={m\over a\sqrt{2(1+fG(\phi))}}\; .
\eea
For generic fields the mean field approximation
amounts to a string description of the system, with
a field dependent tension. But 
only if $fG(\phi)$ is finite for the field
values that minimize the energy, will the system behave
like an actual physical string.

The matter contribution to the ground state bulk energy is
(see Eq.~\ref{bulkenergy}):
\bea
{\cal E}_m(\phi,\phi^\prime)=-{d\over4}(\xi_1+\xi_2)+{d\over2M}\sum_{l<M/2}
[\x^l_++\xi^l_-+\ln(\alpha^2-\beta^2)]
+{d\over2}{3\phi+\phi^\prime\over4}\ln(1+\rho)
\nonumber
\eea
Since the explicit expression is quite complicated, we display this
function graphically.
In Fig.~\ref{matenergy} we show the matter energy as a
function of $\phi$ for a sample of $\phi^\prime$ values. 
In these graphs we are assuming the
factorization $\vpp=\phi^2,\vmp=\phi\phi^\prime,
\vppp=\phi^3,\vmpp=\vpmp=\phi^2\phi^\prime$.
The effect of the infra-red cutoff $\epsilon$ is seen by comparing
the case with $\epsilon=0$, $f=\phi^\prime/\epsilon$ fixed, to
the case with $\epsilon=0.001$, $\phi^\prime$ fixed. It
clearly has the most dramatic effect in the region of small $\phi$.

\begin{figure}[htb]
\psfrag{Spin energy per site, g}[l][l][.46]{{Spin energy per site, ${\hat g}$}}
\includegraphics[width=8cm]{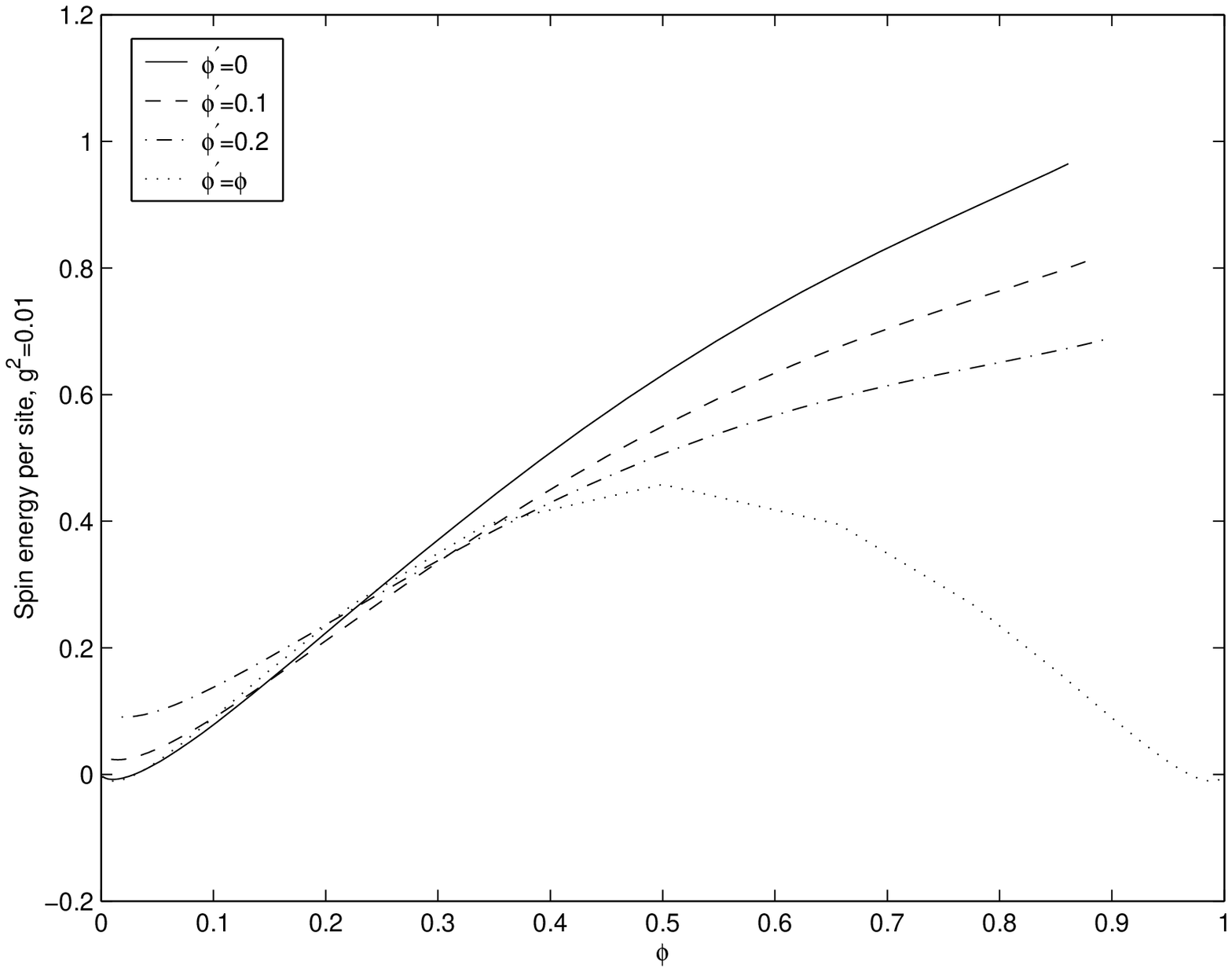}\quad
\includegraphics[width=8cm]{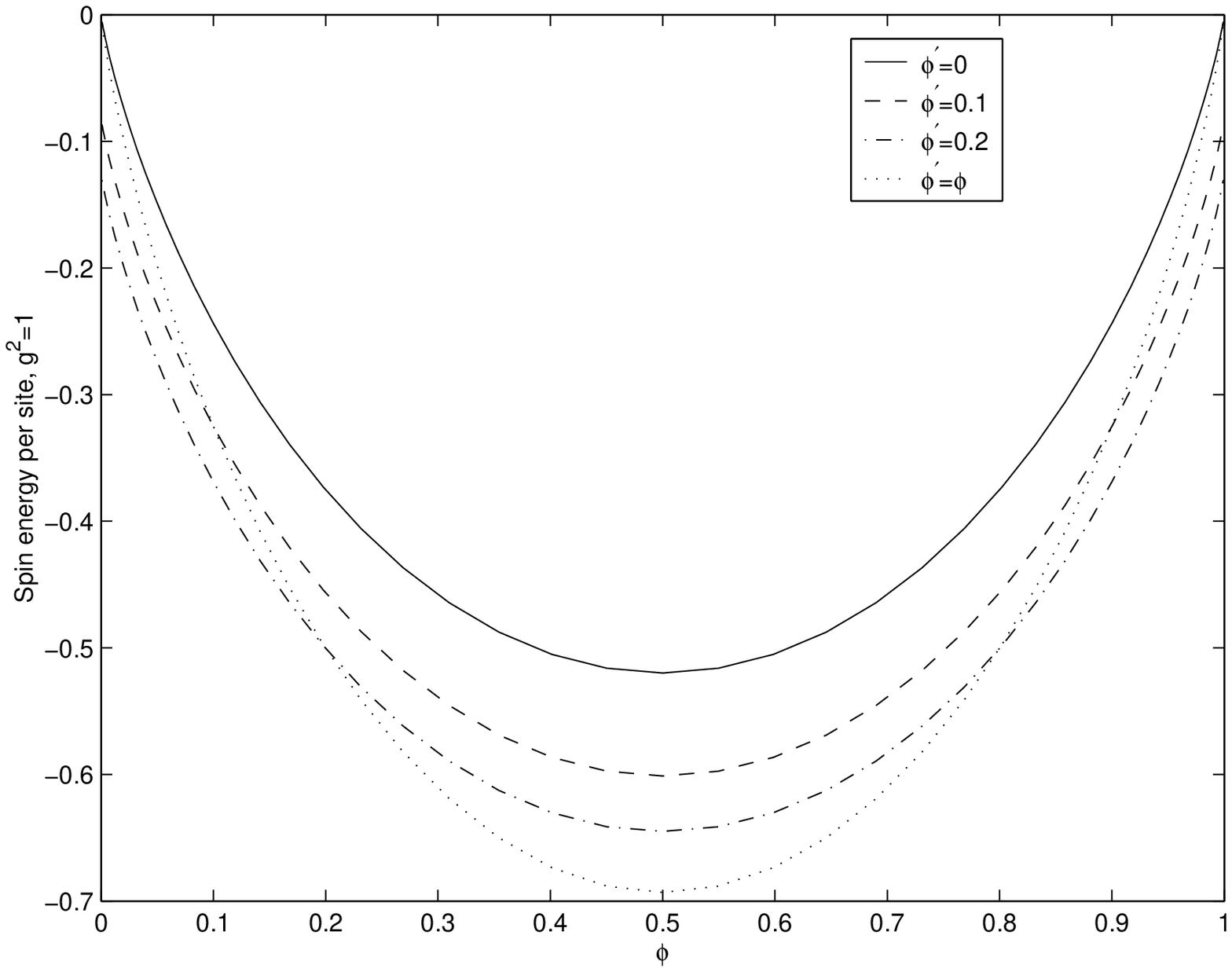}
\includegraphics[width=8cm]{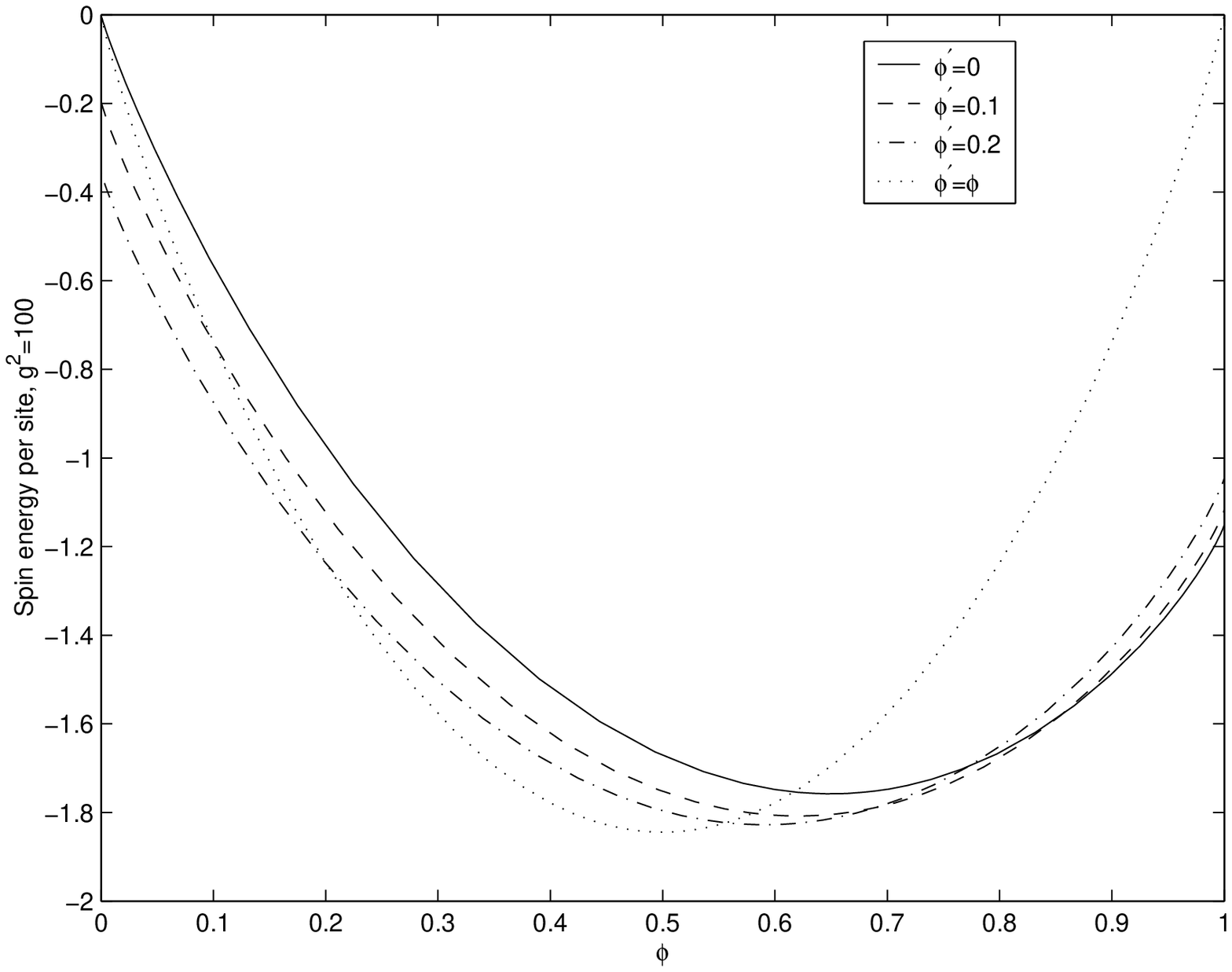}\quad
\includegraphics[width=8cm]{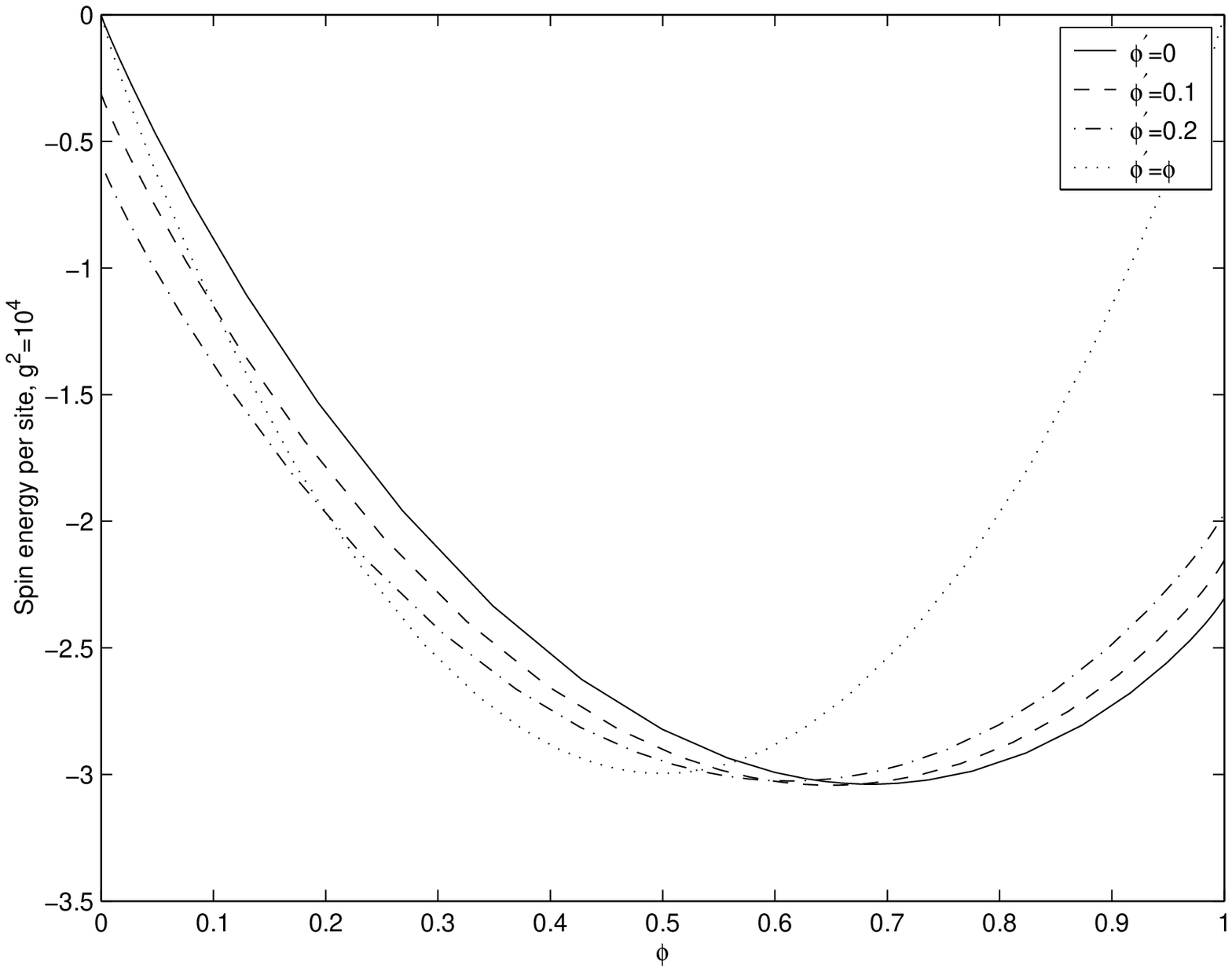}
\caption{The contribution of the spin sum to the bulk
energy per site using (\ref{spinenergynocorr}), which
neglects all short range correlations.}
\label{spinenergypolyg}
\end{figure}

As already mentioned, we present two treatments of the Ising
spin contribution to the energy.
In the first (simplest but most drastic) 
option all spin dependence in the action is replaced by
its corresponding expectation value, and 
the spin sum reduces to a trivial independent sum on
each site:
\bea
\prod_{kj}\sum_{P_k^j=0,1}e^{\kappa_k^jP_i^j}
=\exp\left\{\sum_{kj}\ln(1+e^{\kappa_k^j})\right\}.
\eea 
This version completely neglects short range correlations.
Since this represents the entire $\kappa$ dependence
of the approximate $W$ the relation of $\phi$ to $\kappa$
is independent on each site:
\bea
\phi_k^j={e^{\kappa_k^j}\over1+e^{\kappa_k^j}}, \qquad{\rm or}\qquad
e^{-\kappa_k^j}={1\over\phi_k^j}-1.
\eea
Because the spin sum is independent on each site all 
expectations of the monomials of $P$'s appearing in the
action factorize into the corresponding monomials of $\phi$.

We denote by ${\cal A}_s$ the spin part of the effective action,
including the $-\kappa\phi$ term of the Legendre transform
and also the terms involving the coupling constant:
\bea
{\cal A}_s&=&\sum_{kj}\left[-\kappa_k^j\phi_k^j+\ln(1+e^{\kappa_k^j})\right]
+2\ln{\hat g}\sum_{kj}\langle(P_k^j(1-P_k^{j+1})\rangle\\
&\to&\sum_{kj}\left[-\phi_k^j\ln\phi_k^j-(1-\phi_k^j)\ln(1-\phi_k^j)\right]
+\ln{\hat g}^2\sum_{kj}\phi_k^j(1-\phi_k^{j+1})\\
&\to&-MN\bigg\{
{3\over4}[\phi\ln{\phi}+(1-\phi)\ln(1-\phi)]
+{1\over4}[\phi^\prime\ln{\phi^\prime}+(1-\phi^\prime)\ln(1-\phi^\prime)]
\nonumber\\
&&-{1\over4}\left({3}\phi+\phi^{\prime}
-2\phi\phi^\prime-2\phi^2\right)\ln{\hat g}^2\bigg\}
\equiv-MN{\cal E}_s(\phi,\phi^\prime),
\label{spinenergynocorr}
\eea
where the last line specializes to the fishnet spin pattern.
Since the pattern is regular in the temporal direction the coefficient
of $-N$ is just the energy in units of $1/a$ and the coefficient
of $-MN$ is the energy per site ${\cal E}_s$.

We display the spin energy per site for various couplings in 
Fig.~\ref{spinenergypolyg}. These graphs display
three distinct phases for the spin system. At weak
coupling the $\phi^\prime=\phi$ curve indicates 
a degenerate ground state with $\phi\approx0,1$. These
values correspond to the Ising spins aligned with values
$-1,+1$ respectively, a ferromagnetic phase. The
curves with $\phi^\prime$ constrained to $0,0.1,0.2$
approximate the $\phi^\prime=\phi$ curve more closely as $\phi^\prime\to0$.
At intermediate coupling the minimum of the $\phi^\prime=\phi$
curve is fixed at $\phi=1/2$, corresponding to 
$\langle s_k^j\rangle=0$ for the Ising spins. This is a
disordered phase. The constrained $\phi^\prime$ curves
give more information about this phase. At ${\hat g}^2=1$
they all show a minimum for $\phi=1/2$. For ${\hat g}^2>1$
the minima of these curves are above $1/2$, approaching $1/2$
as $\phi^\prime$ increases. In other words the
$\phi$ sites are anti-correlated with the
$\phi^\prime$ sites. This anti-correlation will
tend to reduce the paramagnetic susceptibility. 
For ${\hat g}^2<1$ (not shown)
the curves would have minima less than $1/2$ that all lie on the
$\phi^\prime=\phi$ curves. In other words the
$\phi$ sites are positively correlated with the
$\phi^\prime$ sites, and this will tend
to increase the paramagnetic susceptibility. 
In summary, although the phase is
not ordered, it shows a paramagnetic susceptibility that increases
as ${\hat g}^2$ decreases. 
Finally for strong coupling the $\phi^\prime=\phi$
curve no longer contains the minimum energy. The
minimum energy is on the curve with $\phi^\prime=0$ at $\phi\approx0.7$,
an anti-ferromagnetic phase. 
This phase structure is not surprising because the
spin energy is essentially the mean field free energy
of an Ising model, and the phases we have described are
well understood properties of the Ising model provided
its dimensionality $d>1$. The mean field description fails
to predict the critical exponents for $d=2$.

\begin{figure}[htb]
\psfrag{Spin energy per site, g}[l][l][.46]{{Spin energy per site, ${\hat g}$}}
\includegraphics[width=8cm]{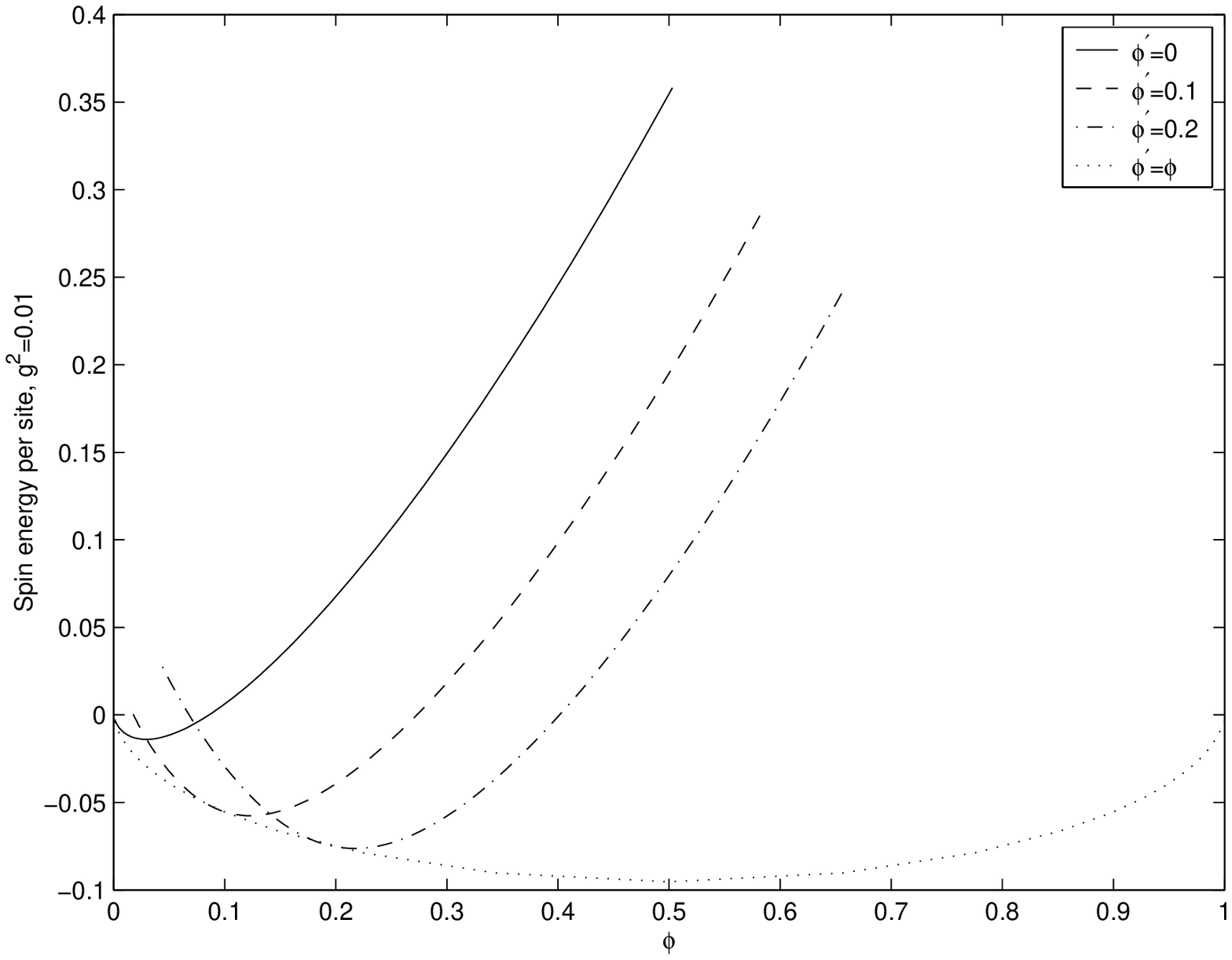}\quad
\includegraphics[width=8cm]{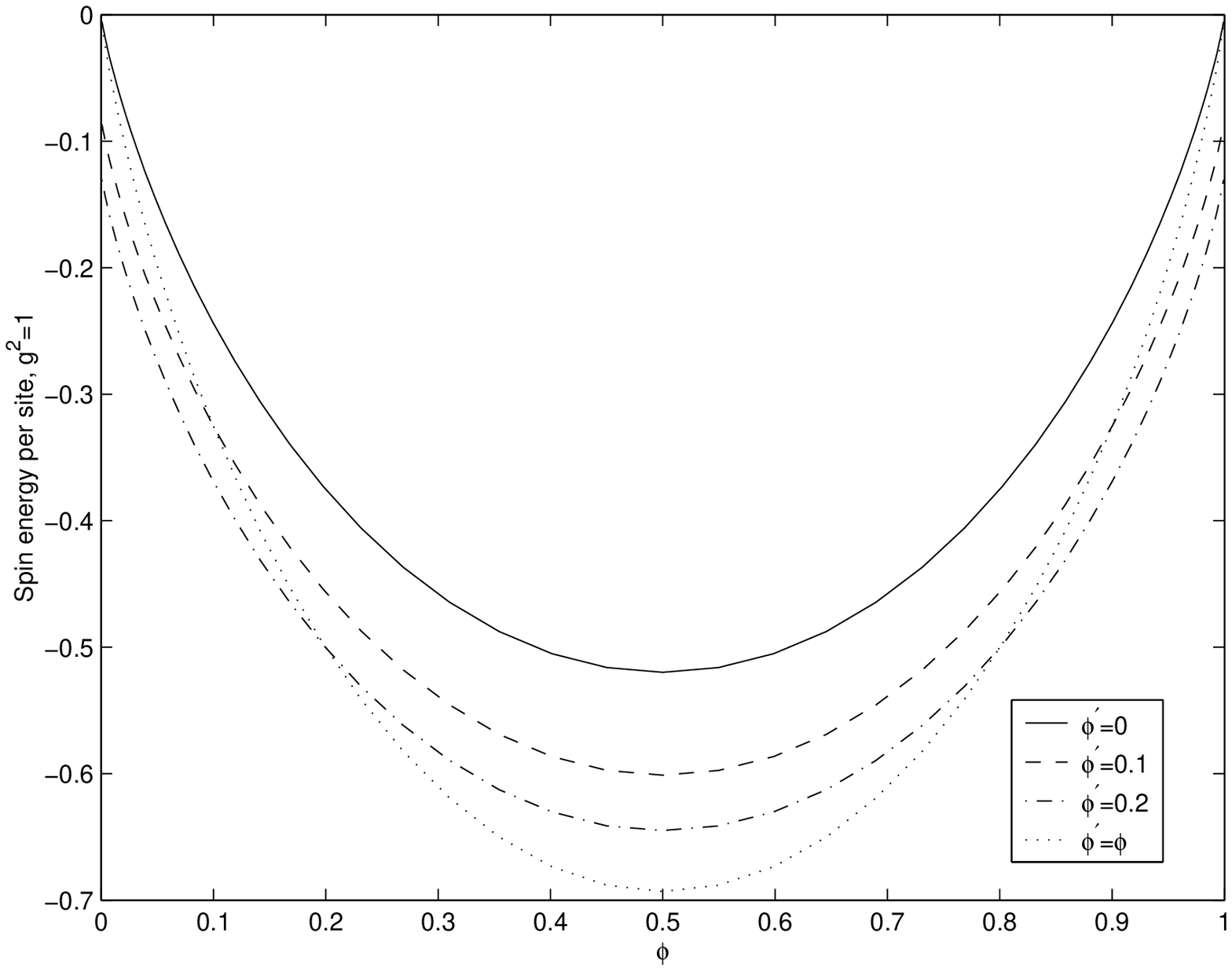}
\includegraphics[width=8cm]{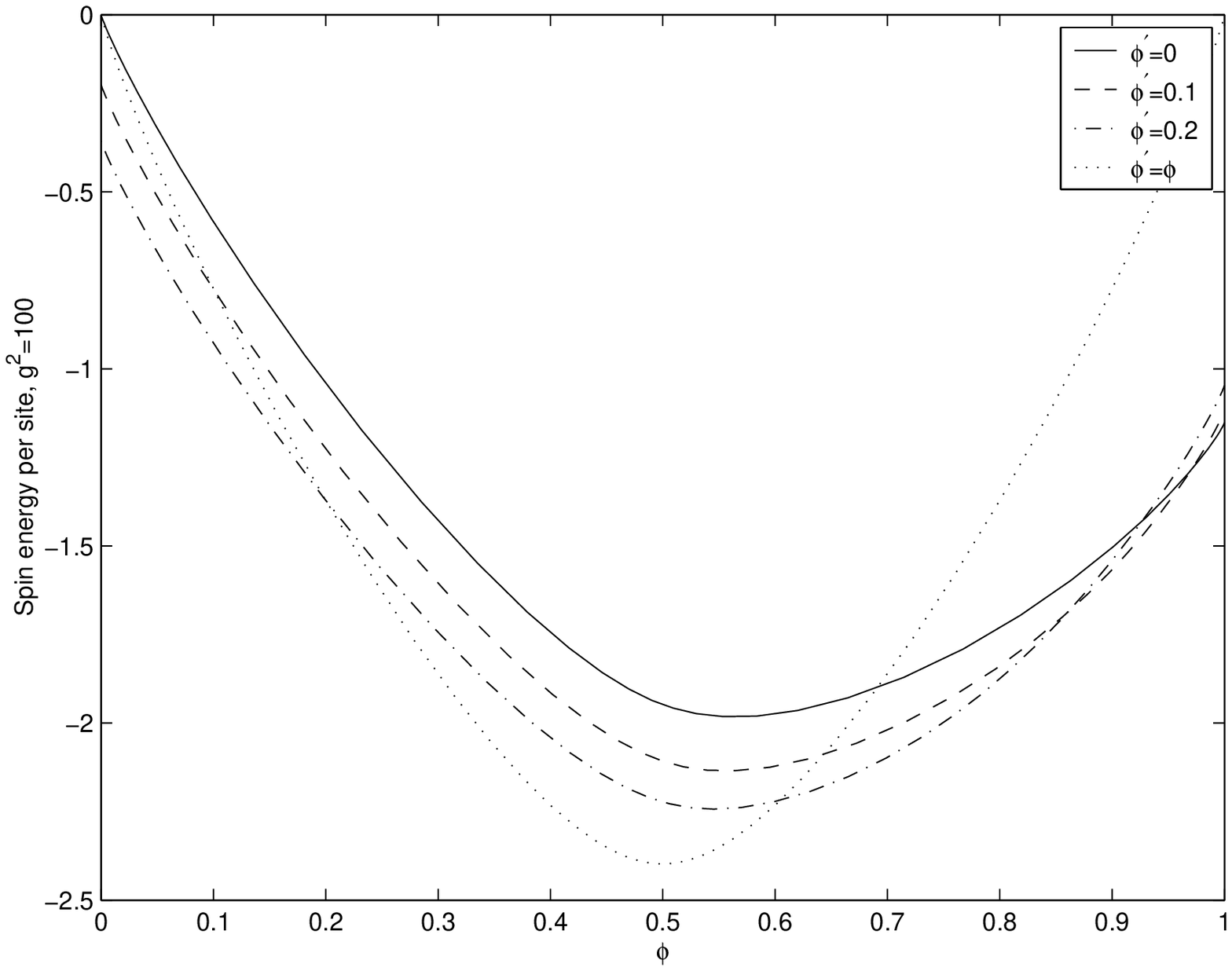}\quad
\includegraphics[width=8cm]{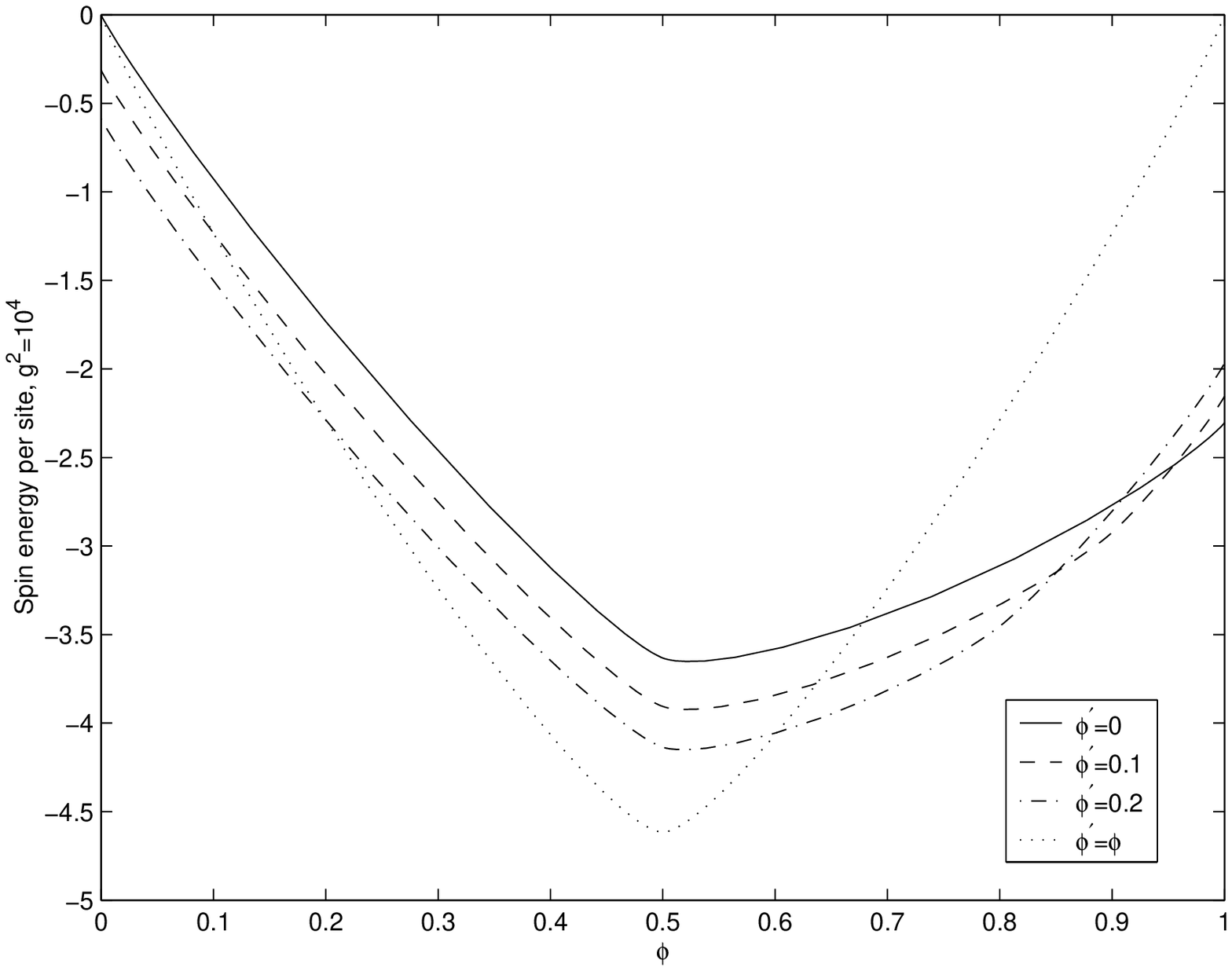}
\caption{The contribution of the spin sum to the bulk
energy, using (\ref{spinenergycorr}) which includes coupling dependence,
and hence some temporal correlations,
in the spin sum. Note the disappearance of the phase transitions.}
\label{spinenergynonpolyg}
\end{figure}

However, the spin sums we have actually
done correspond to Ising couplings in one
dimension only (the temporal direction), and it is well-known
that the 1$d$ Ising model displays no phase transitions:
there is only a single disordered phase, albeit with
paramagnetic properties. In our worldsheet system there
are some direct spatial couplings
of the spins, and there are also indirect spatial couplings
induced through the spin-matter coupling terms. When these
spatial couplings are brought into play, a 
phase transition is no longer ruled out,
but it still behooves us to do a better job on the one-dimensional
spin sums that removes this artificial phase transition. 
For that reason we now turn to 
the more sophisticated treatment of the spin sums
sketched in Appendix A. In this better approach the
spins multiplying $\ln{\hat g}$ are {\it not} replaced by
mean fields but rather incorporated exactly in the
spin sum. However this sum is tractable only after one specializes
to a regular spin pattern, here the fishnet spin pattern.
Then the problem is solved by finding the eigenvalues
of the $2\times2$ transfer matrix and choosing the largest
eigenvalue. In the notation of Appendix A the spin energy 
per site now becomes
\bea
{\cal E}_s=-{1\over4}\ln t_+^4(\kappa,\kappa^\prime)+{1\over2}\phi\kappa
+{1\over4}K\phi+{1\over4}\phi^\prime\kappa^\prime.
\label{spinenergycorr}
\eea
Again we display the spin energy
in this case graphically in
Fig.~\ref{spinenergynonpolyg}.
We see that the indications of a phase transition
seen in Fig.~\ref{spinenergypolyg}  have now
disappeared, since we now have an essentially exact
treatment of the 1$d$ Ising model reflected in the spin sum.
Now the minimum energy is always on the $\phi^\prime=\phi$
curve and located at $\phi=1/2$, the disordered phase.
But the $\phi^\prime\neq\phi$ curves still show
that this phase is paramagnetic: for ${\hat g}^2<1$
the susceptibility is greater than in the disordered phase
and it is less for ${\hat g}^2>1$. We shall see that
these qualitative features are maintained when we
add in the energy of the matter sector. The $\phi^\prime=0$
curve follows the fate of the fishnet diagram, corresponding
to the endpoint $\phi=1$, when the rest of the
planar diagrams are included. In the following
analysis we shall always display
plots for the total energy per site
in this second more accurate treatment of the spin sum.
\begin{figure}[htb]
\psfrag{Total energy per site, g}[l][l][.45]{{Total energy per site, ${\hat g}$}}
\includegraphics[width=8cm]{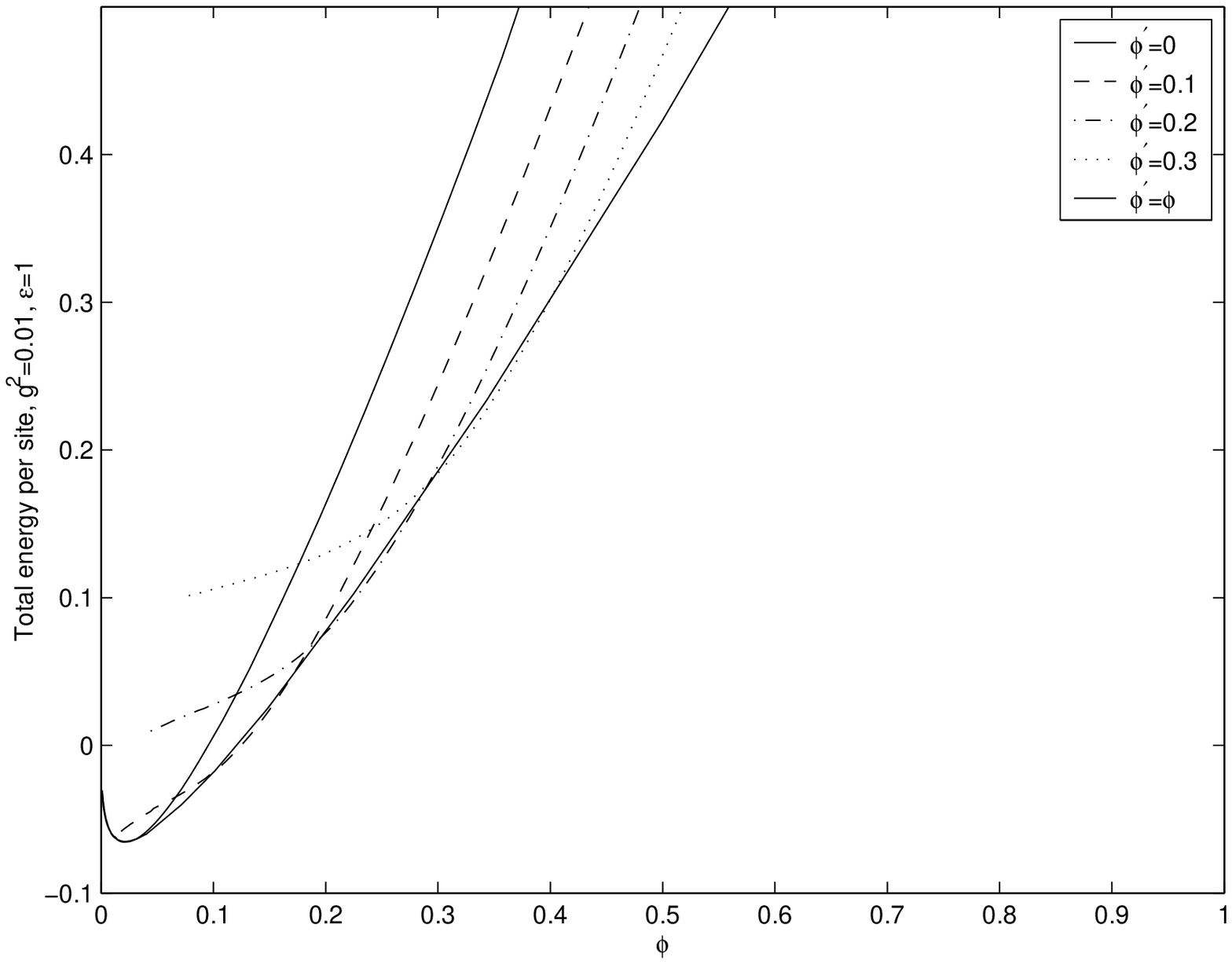}\quad
\includegraphics[width=8cm]{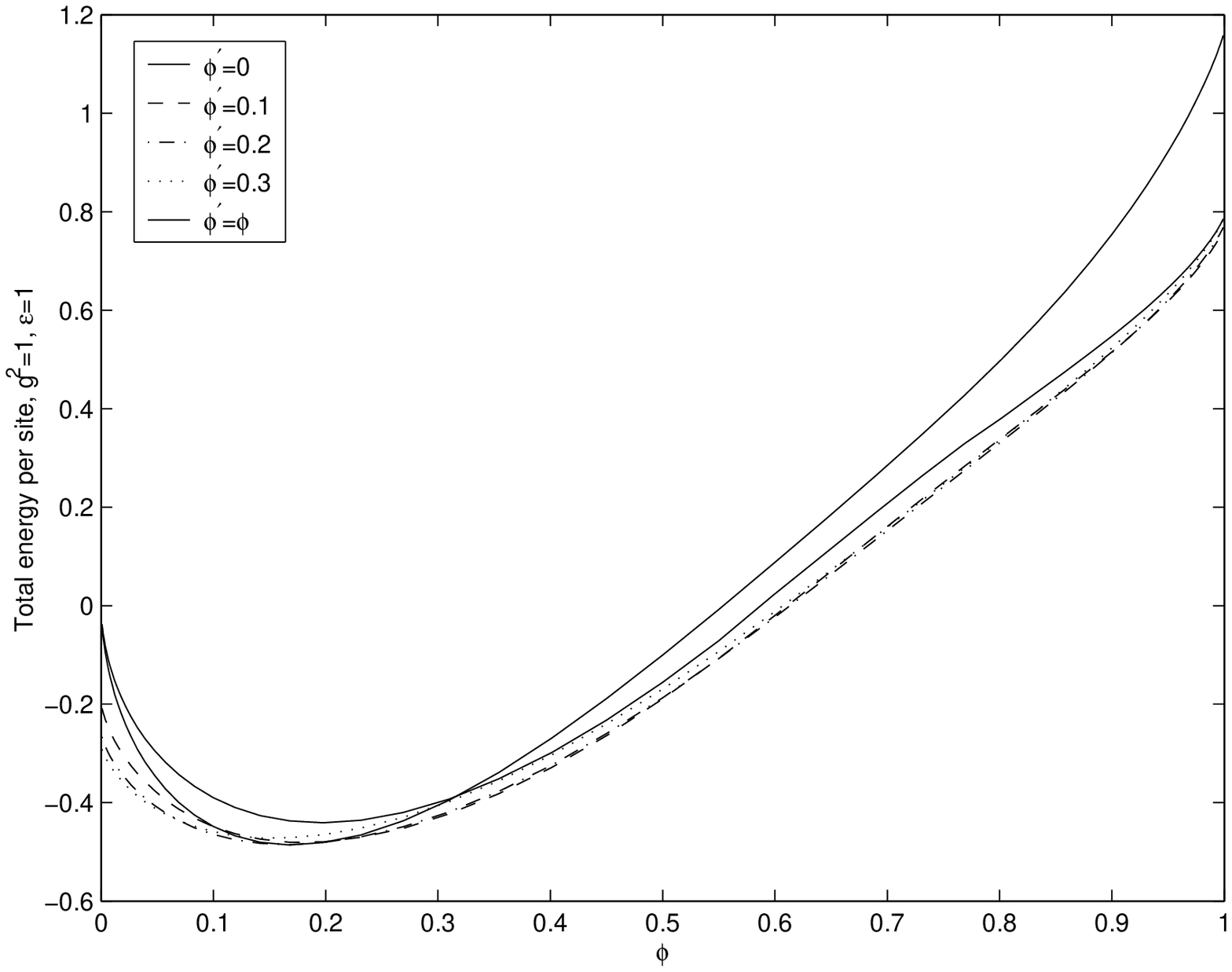}
\includegraphics[width=8cm]{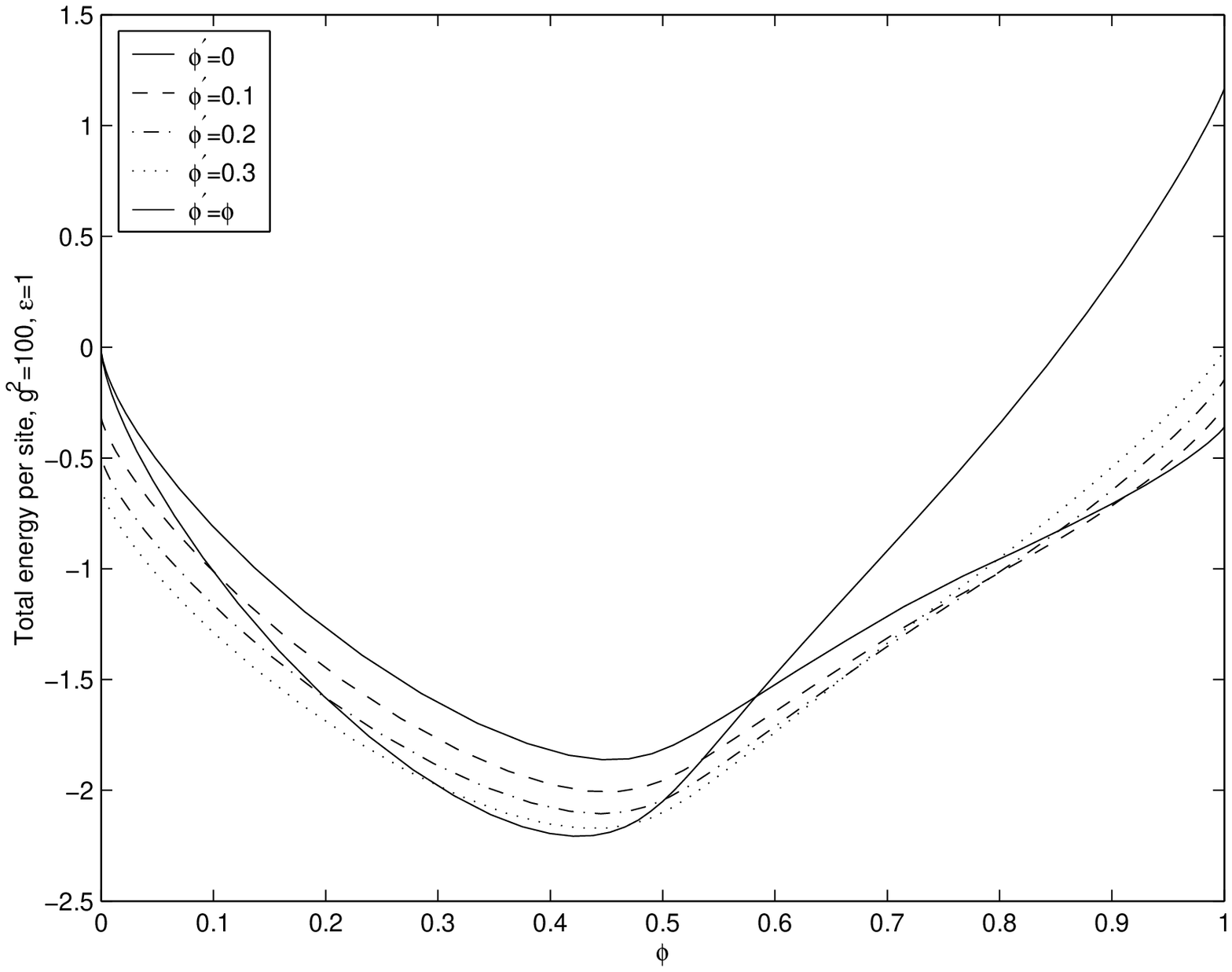}\quad
\includegraphics[width=8cm]{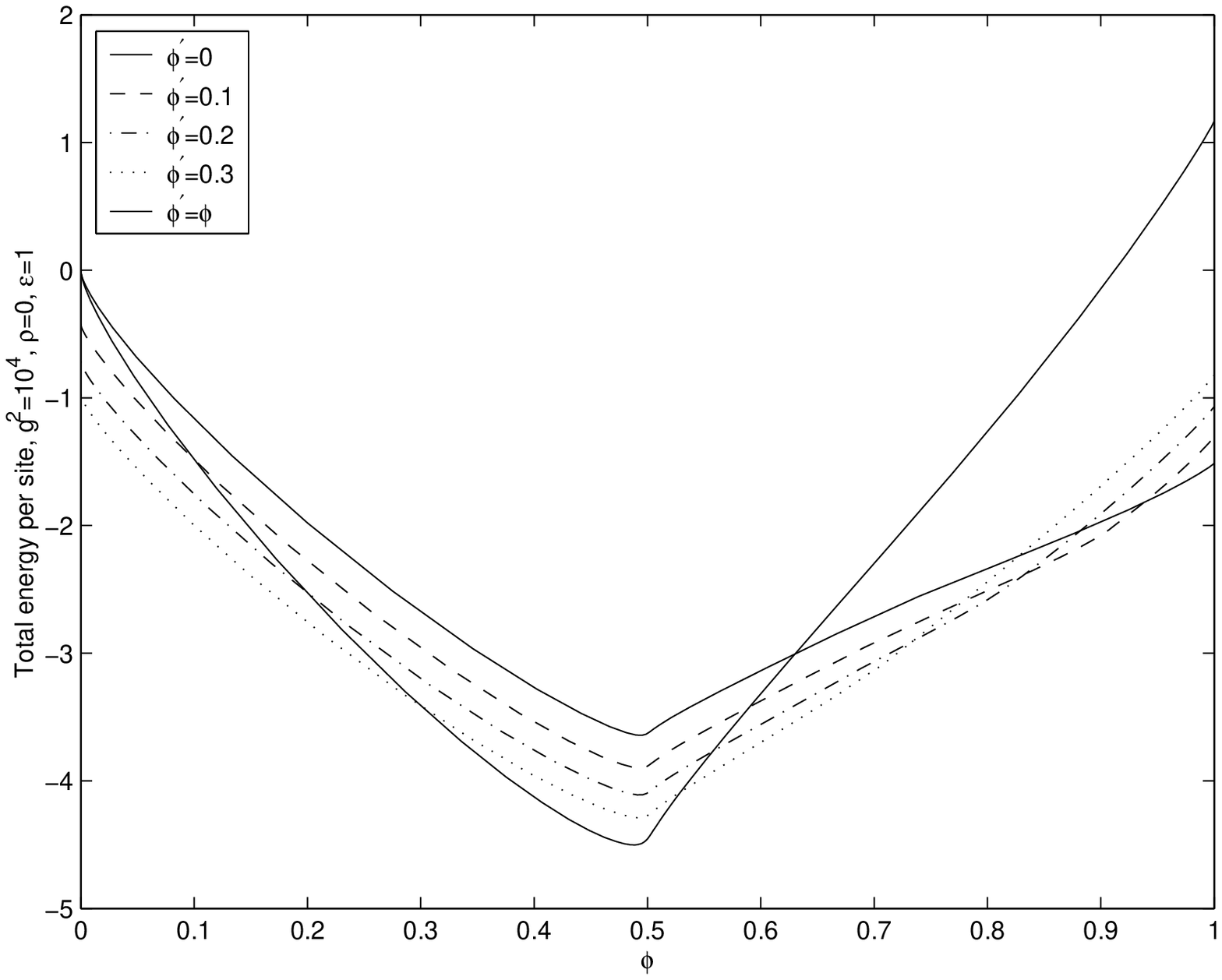}
\caption{The total energy for $\epsilon=1$. The location of the
minimum depends dramatically on the coupling.}
\label{totalenergynonpolyg0}
\end{figure}

\begin{figure}[htb]
\psfrag{Total energy per site, g}[l][l][.45]{{Total 
energy per site, ${\hat g}$}}
\includegraphics[width=8cm]{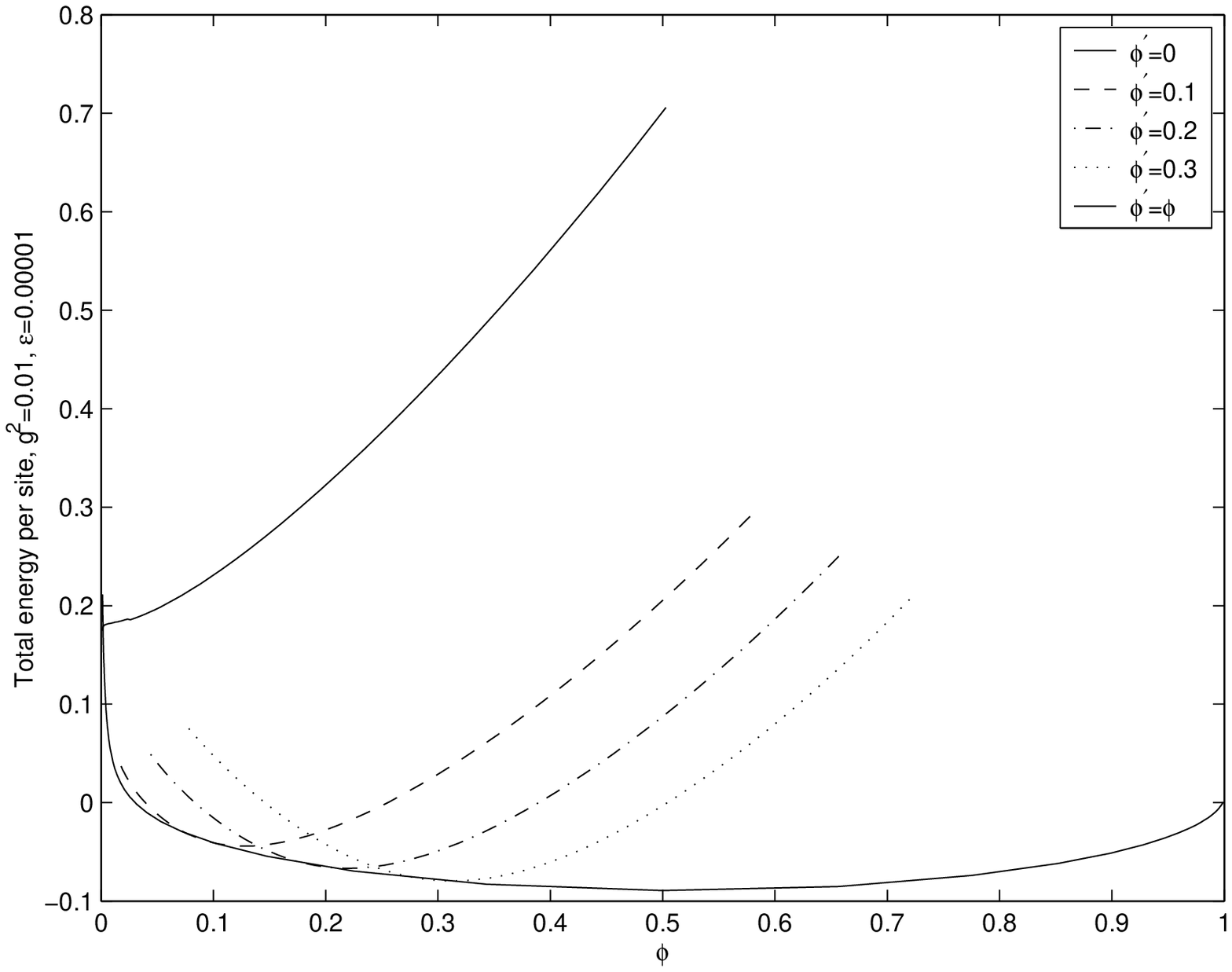}\quad
\includegraphics[width=8cm]{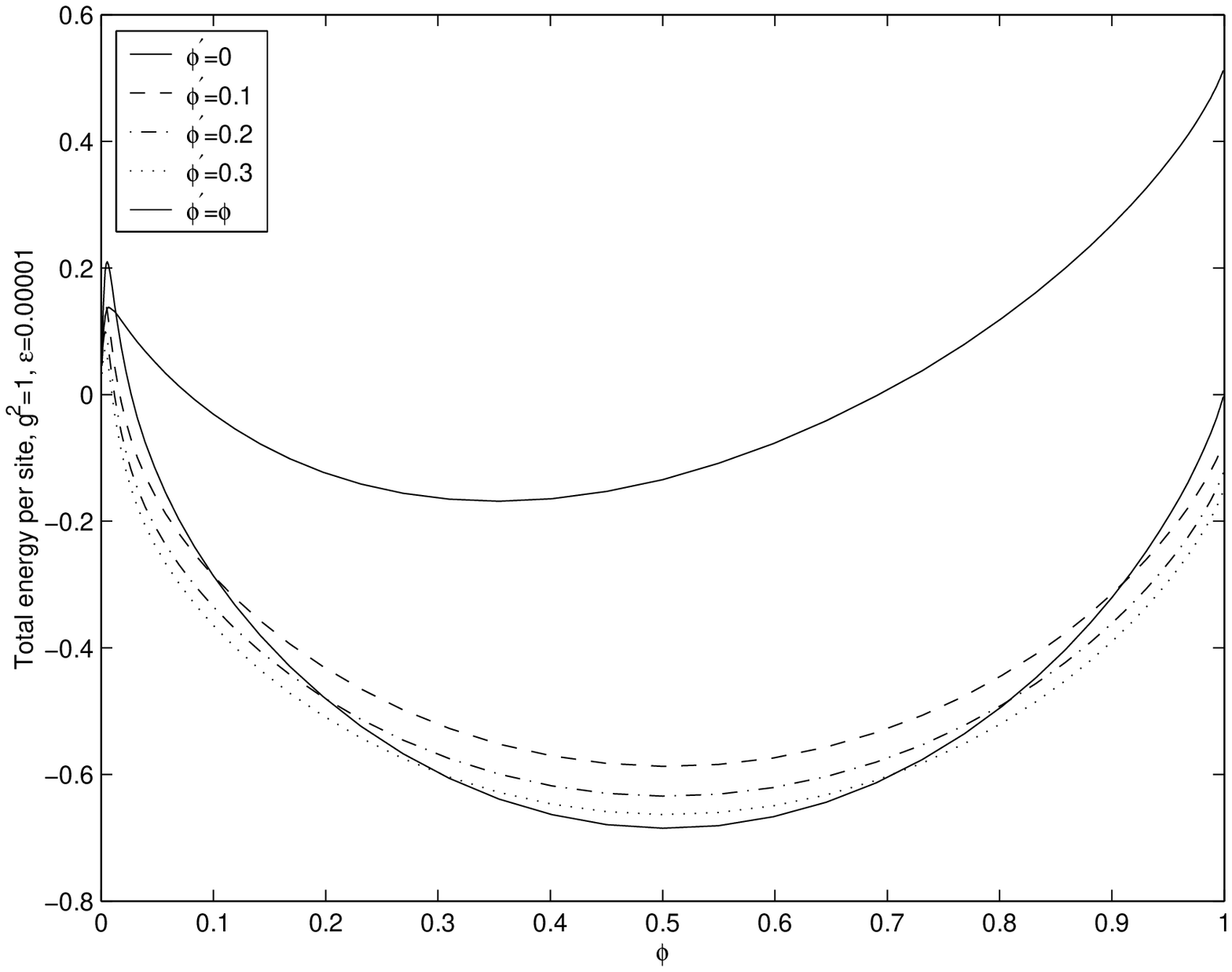}
\includegraphics[width=8cm]{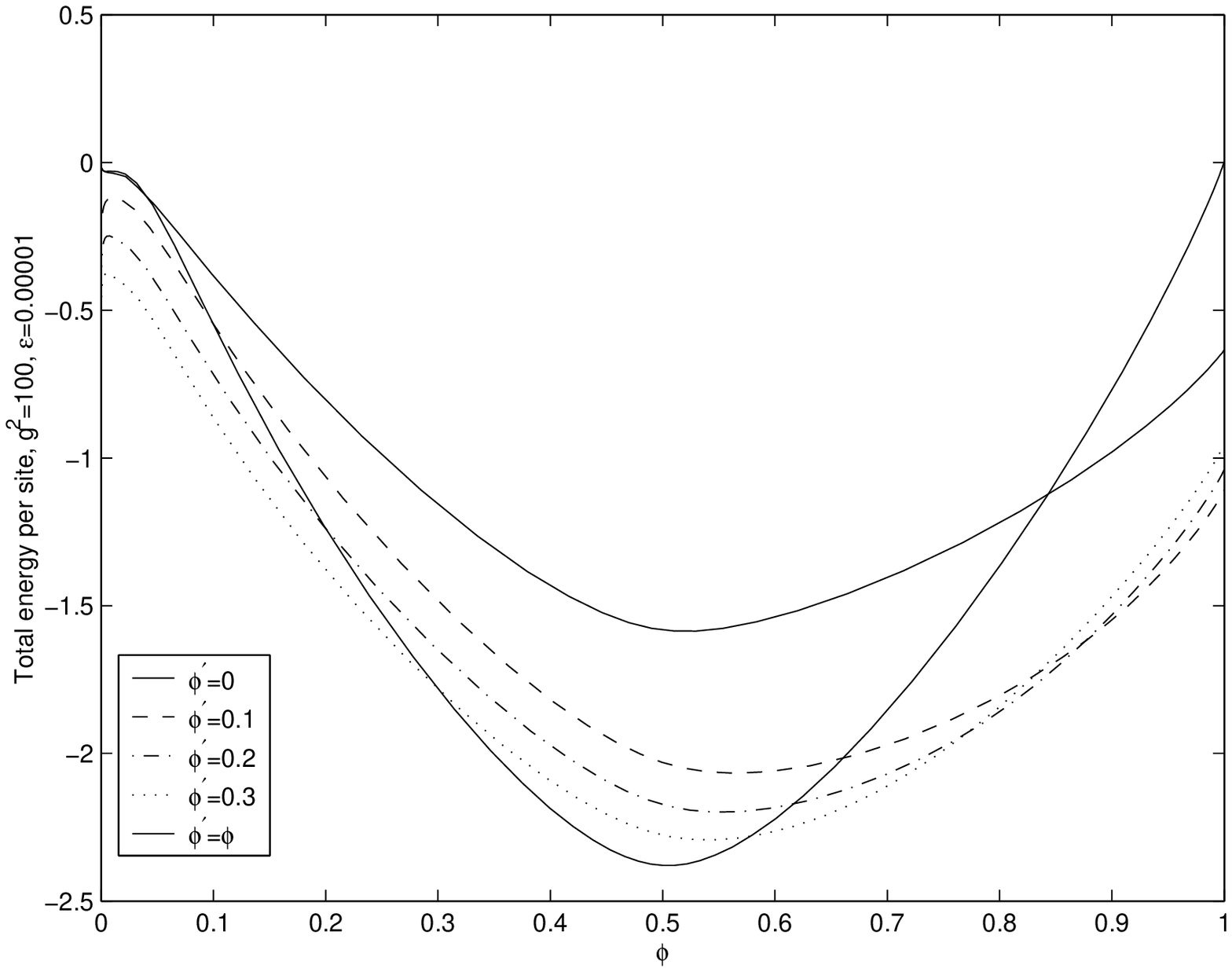}\quad
\includegraphics[width=8cm]{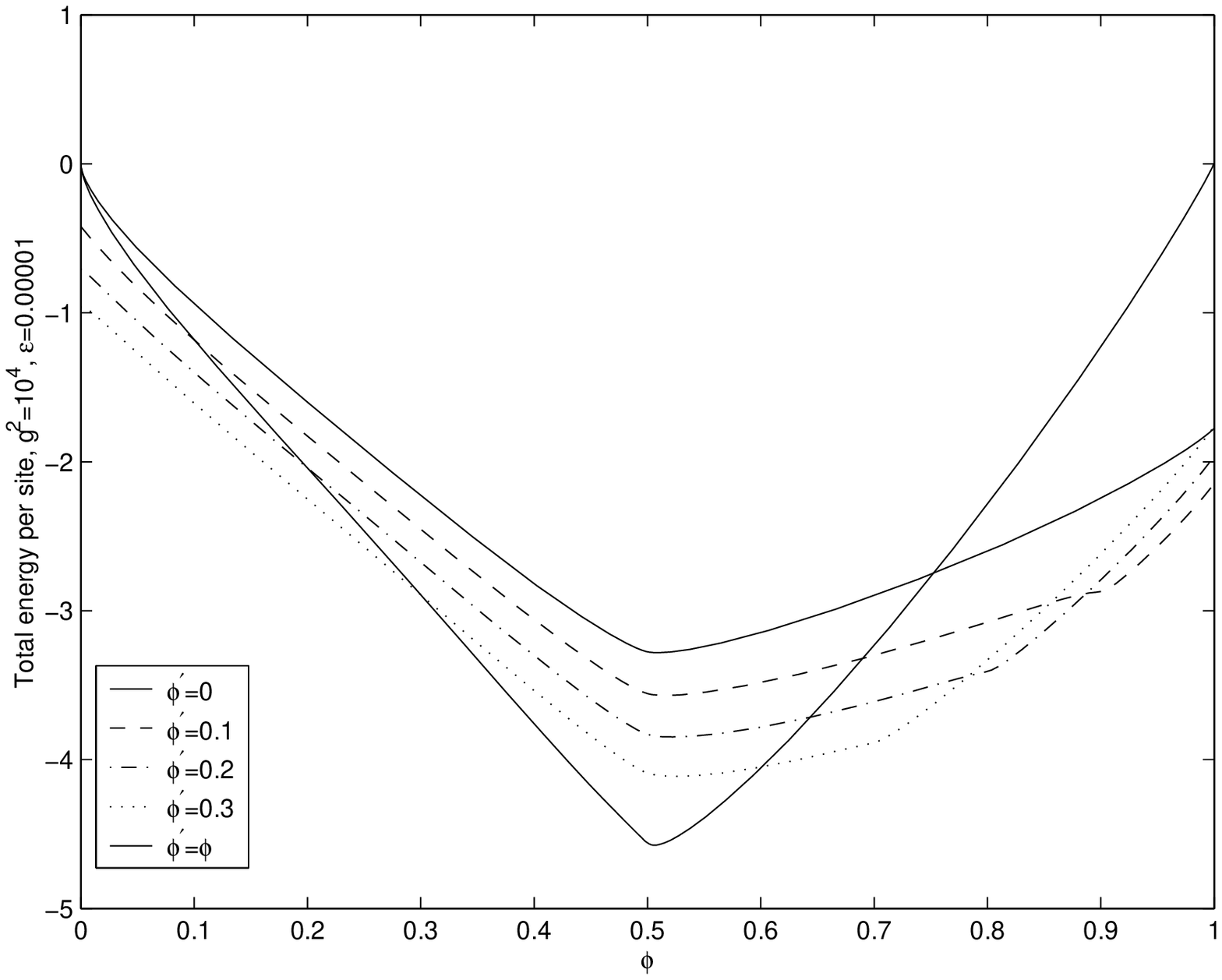}
\caption{The total energy for $\epsilon=0.00001$, corresponding
to a very small infra-red cutoff (very large transverse space).
The location of the absolute minimum is at $\phi=0.5$
for all coupling, but the paramagnetic properties of the
spins show a clear dependence on the coupling.}
\label{totalenergynonpolyg5}
\end{figure}
\begin{figure}[htb]
\psfrag{Total energy per site, g}[l][l][.45]{{Total energy per site, ${\hat g}$}}
\includegraphics[width=8cm]{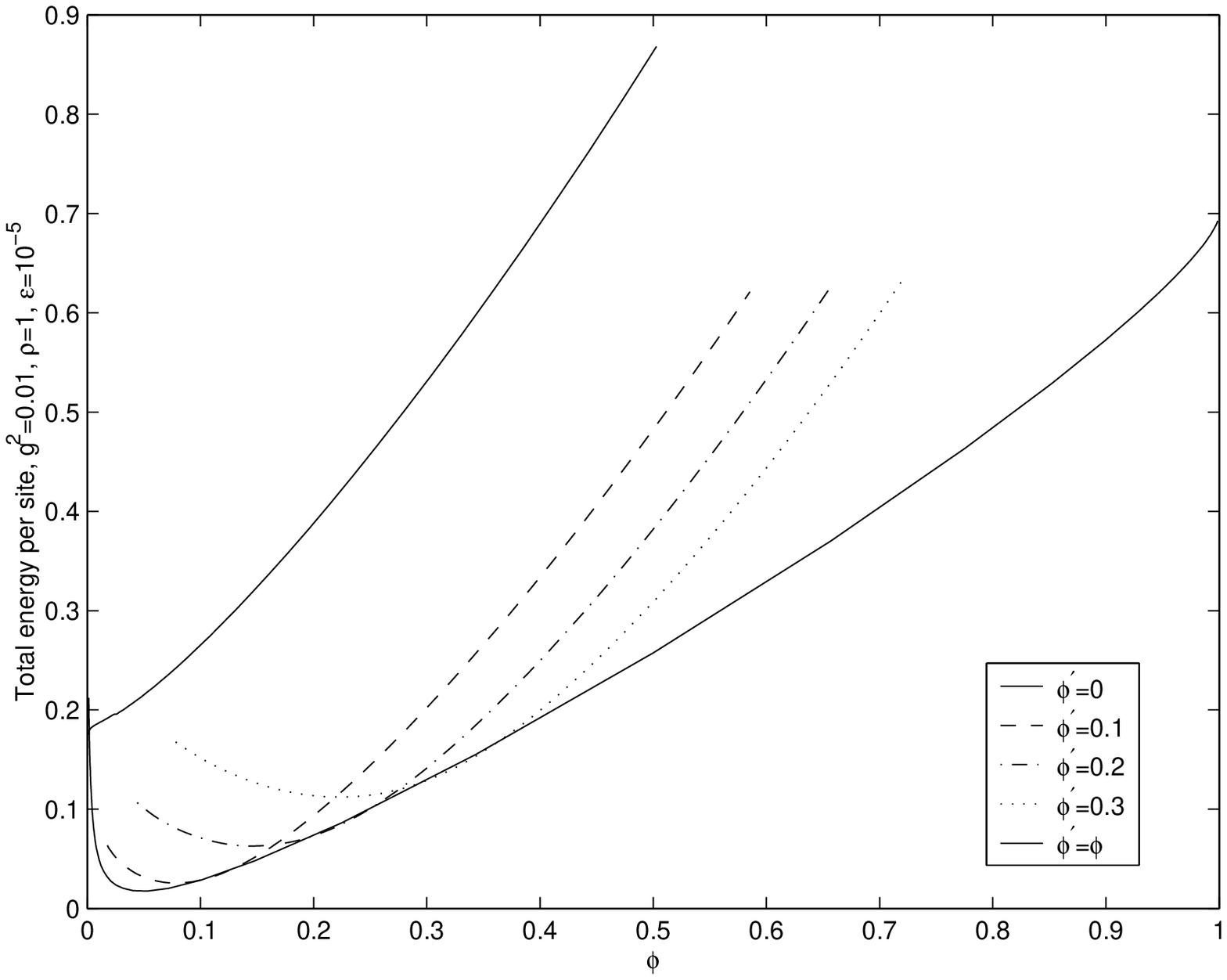}\quad
\includegraphics[width=8cm]{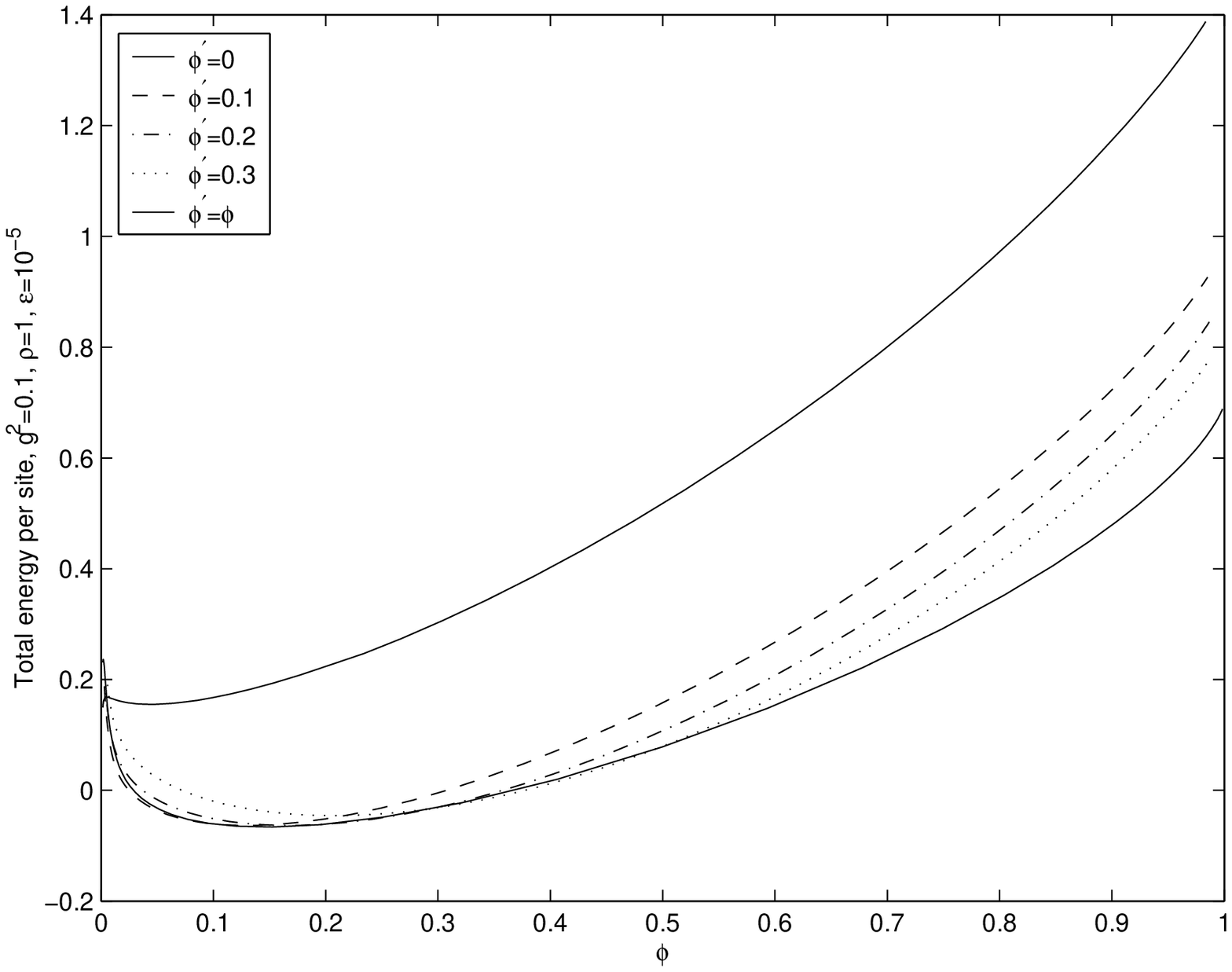}
\includegraphics[width=8cm]{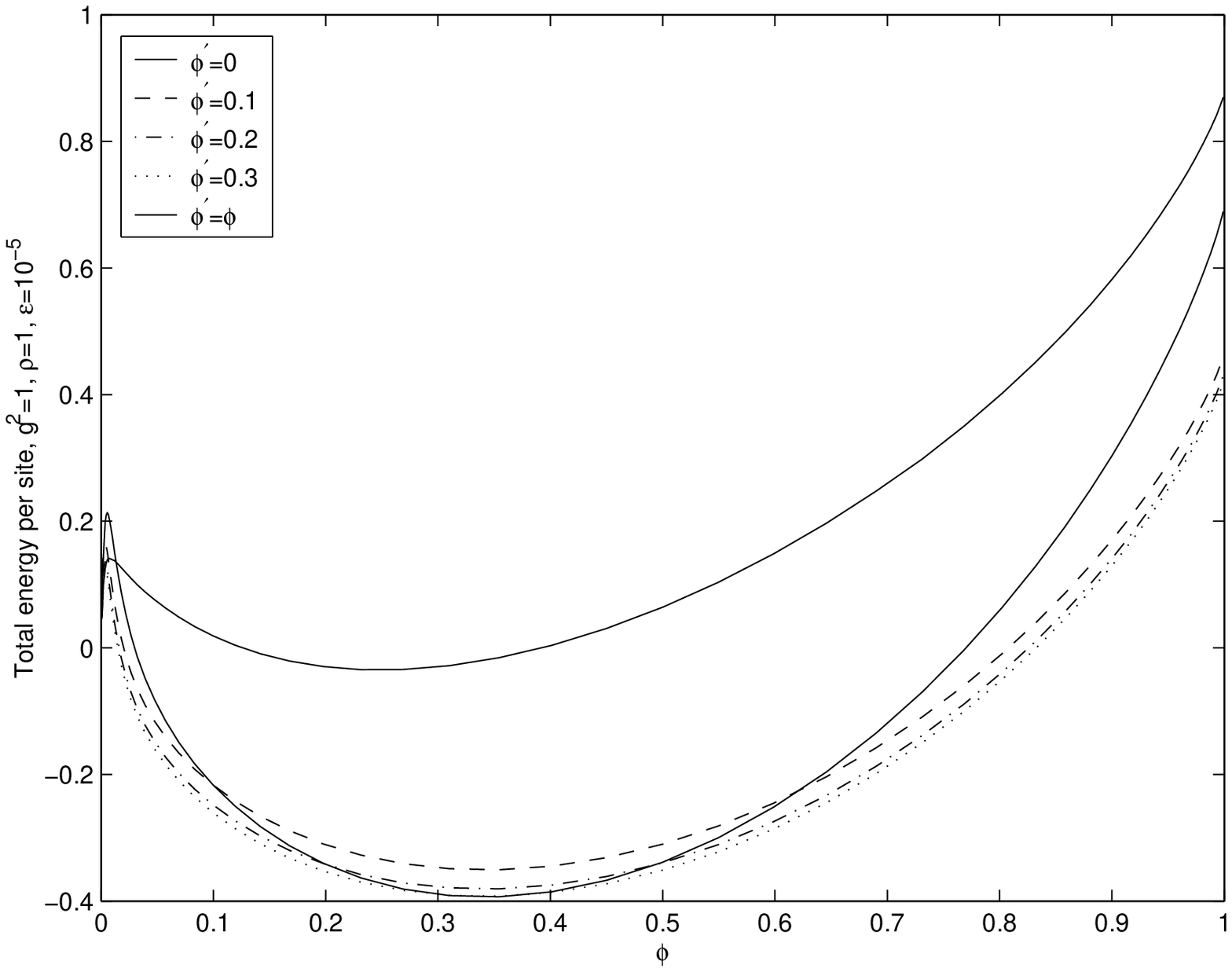}\quad
\includegraphics[width=8cm]{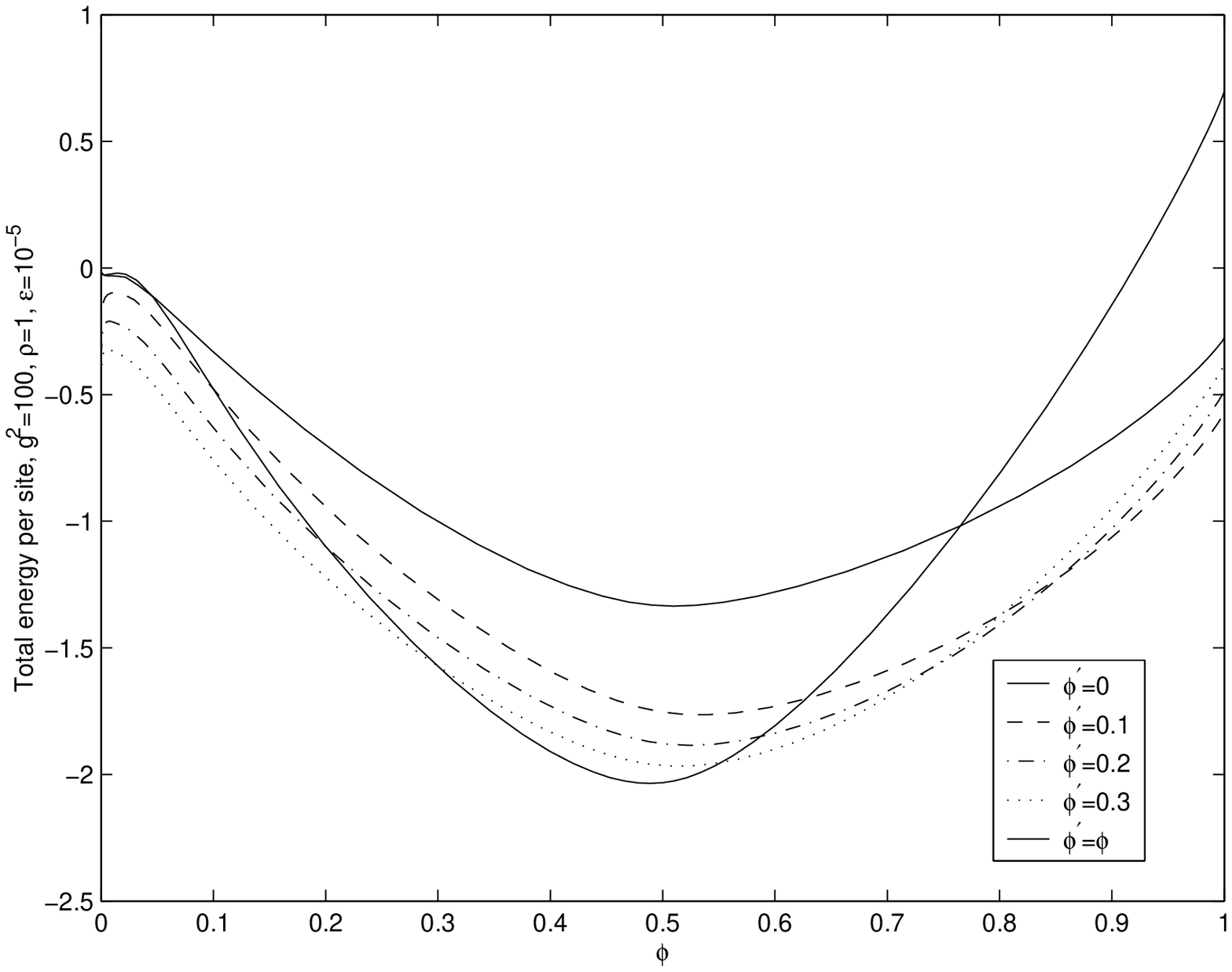}
\caption{The total energy for the massive case, $\rho=1$, $\epsilon=0.00001$.
There is evidence for ferromagnetic behavior at weak coupling
but none for anti-ferromagnetic behavior at
strong coupling. Note the qualitative similarity of these
curves to those of Fig.~\ref{totalenergynonpolyg0}.}
\label{totalenergynonpolyg5r1}
\end{figure}
\begin{figure}[htbp]
\psfrag{Total energy per site, g}[l][l][.45]{{Total 
energy per site, ${\hat g}$}}
\includegraphics[width=8cm]{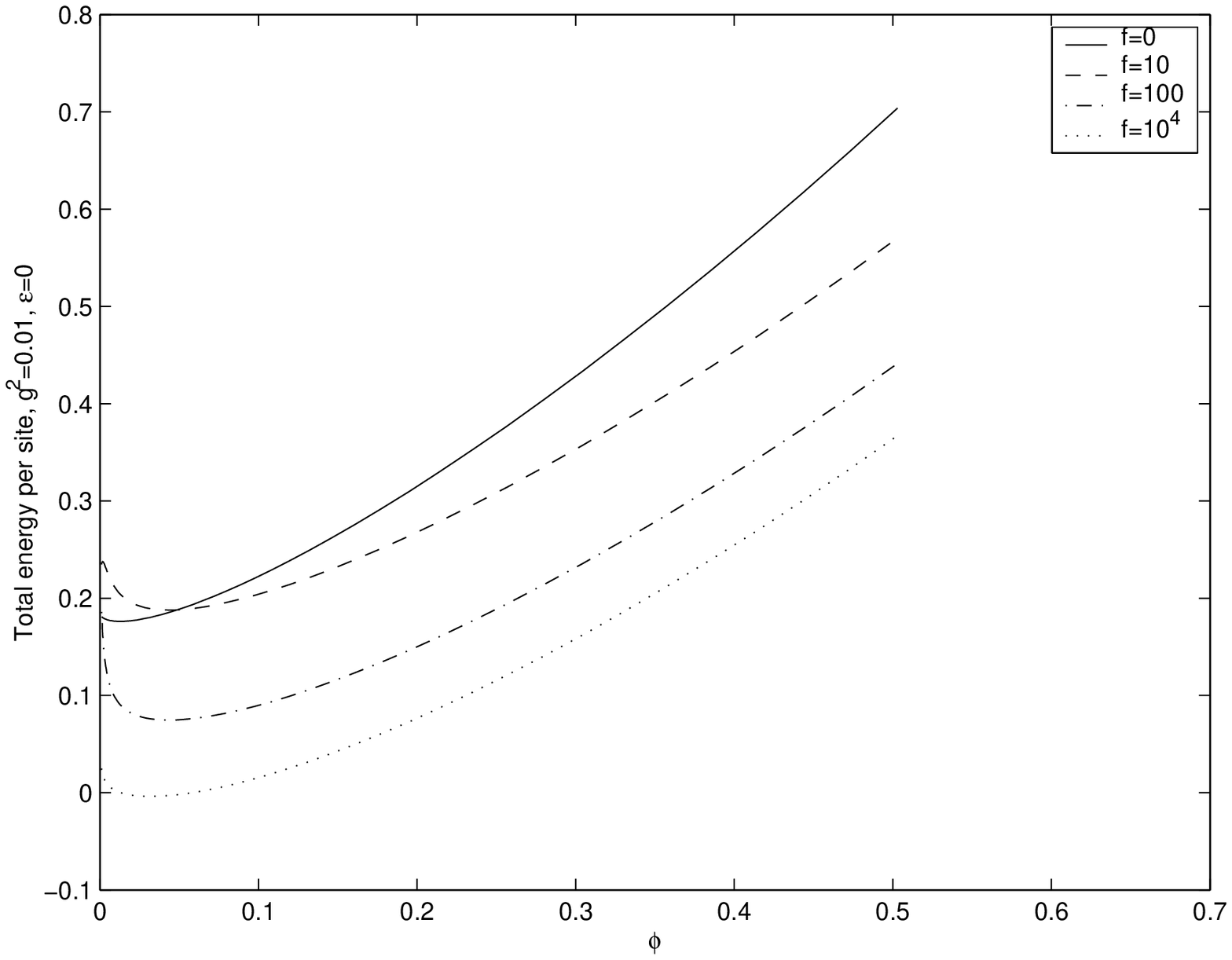}\quad
\includegraphics[width=8cm]{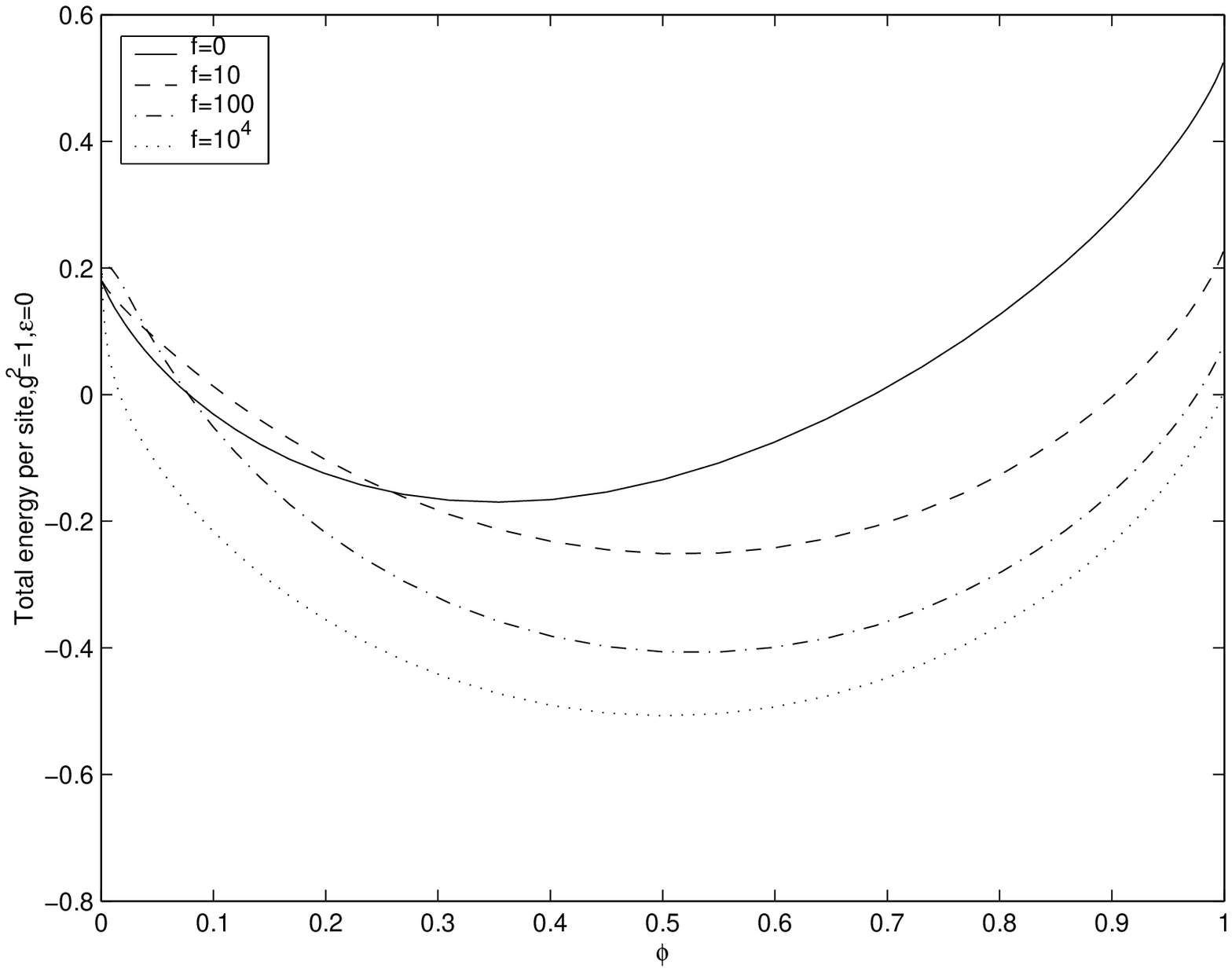}
\includegraphics[width=8cm]{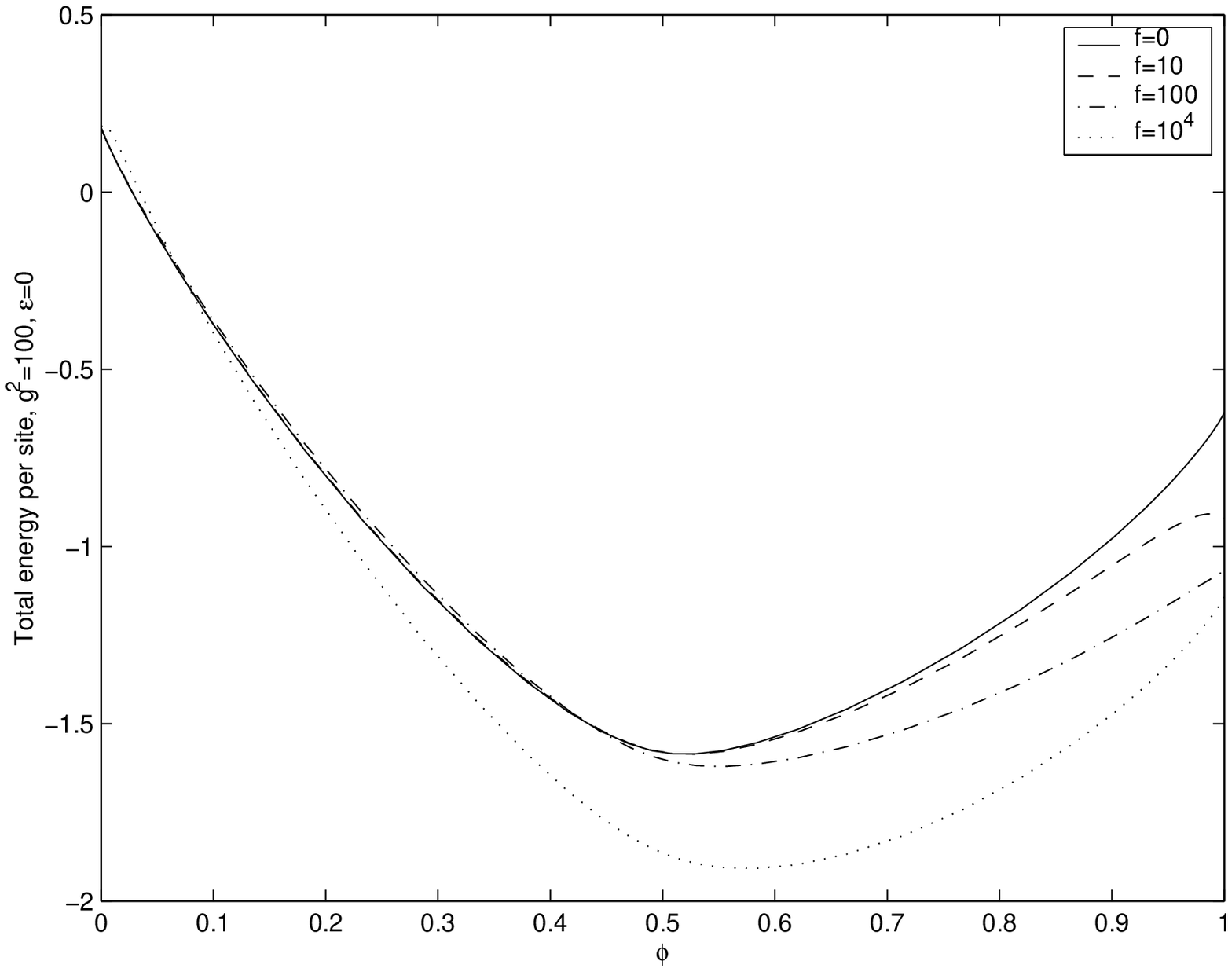}\quad
\includegraphics[width=8cm]{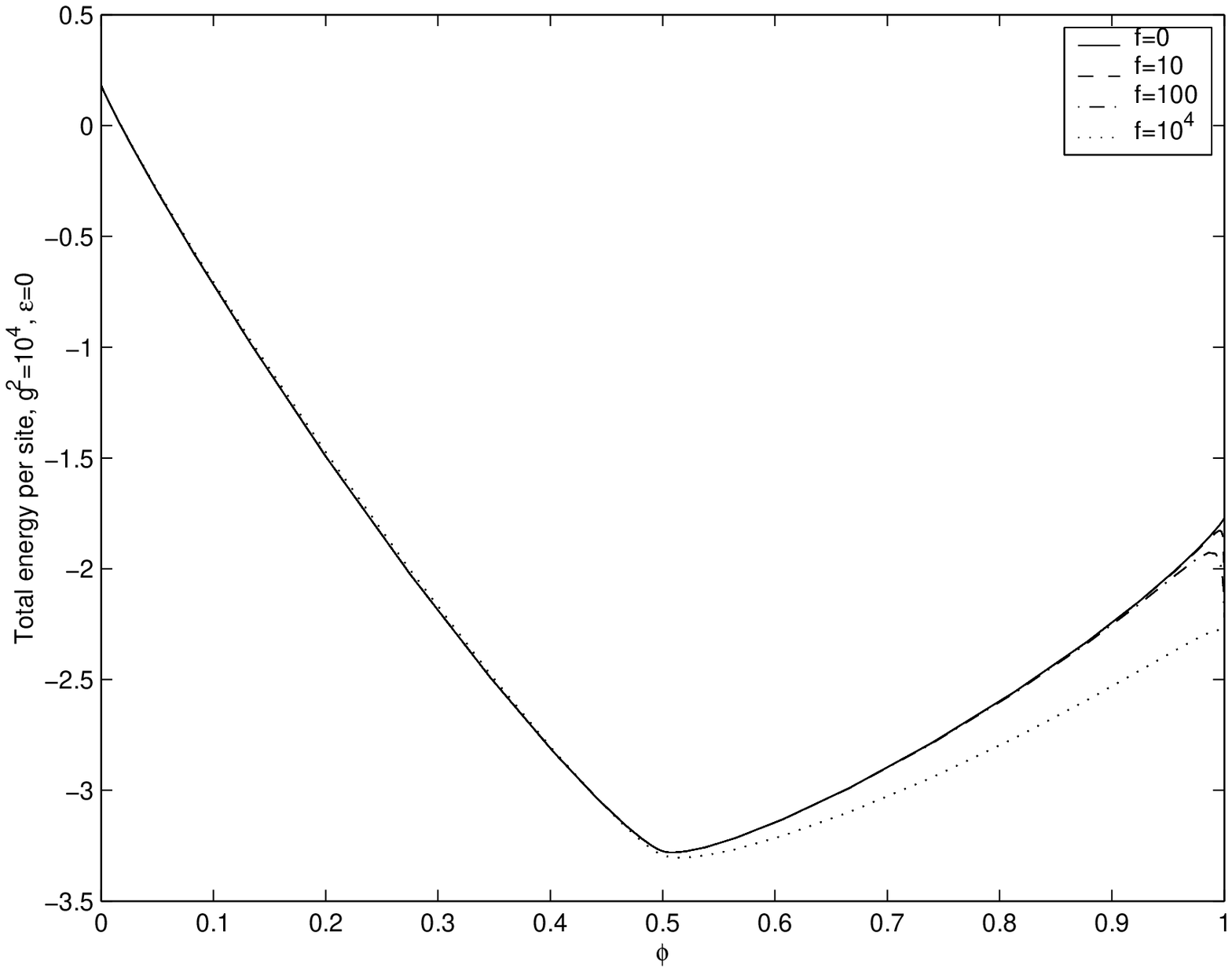}
\caption{The total energy for $\epsilon=0$ for 
fixed $f=\phi^\prime/\epsilon$.}
\label{totalenergye0}
\end{figure}
\begin{figure}[htbp]
\psfrag{g}[l][l][.5]{${\hat g}$}
\includegraphics[width=8cm]{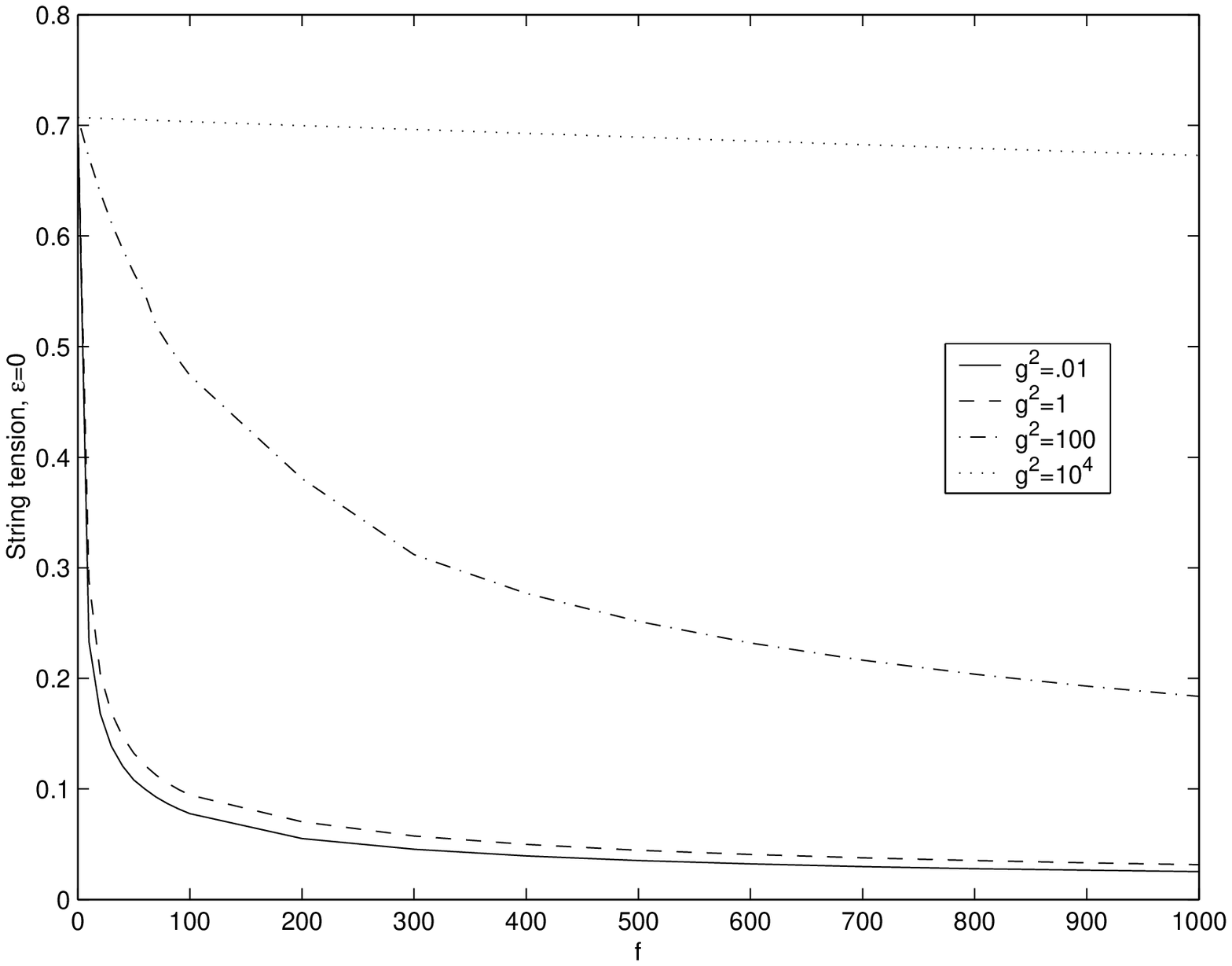}\quad
\includegraphics[width=8cm]{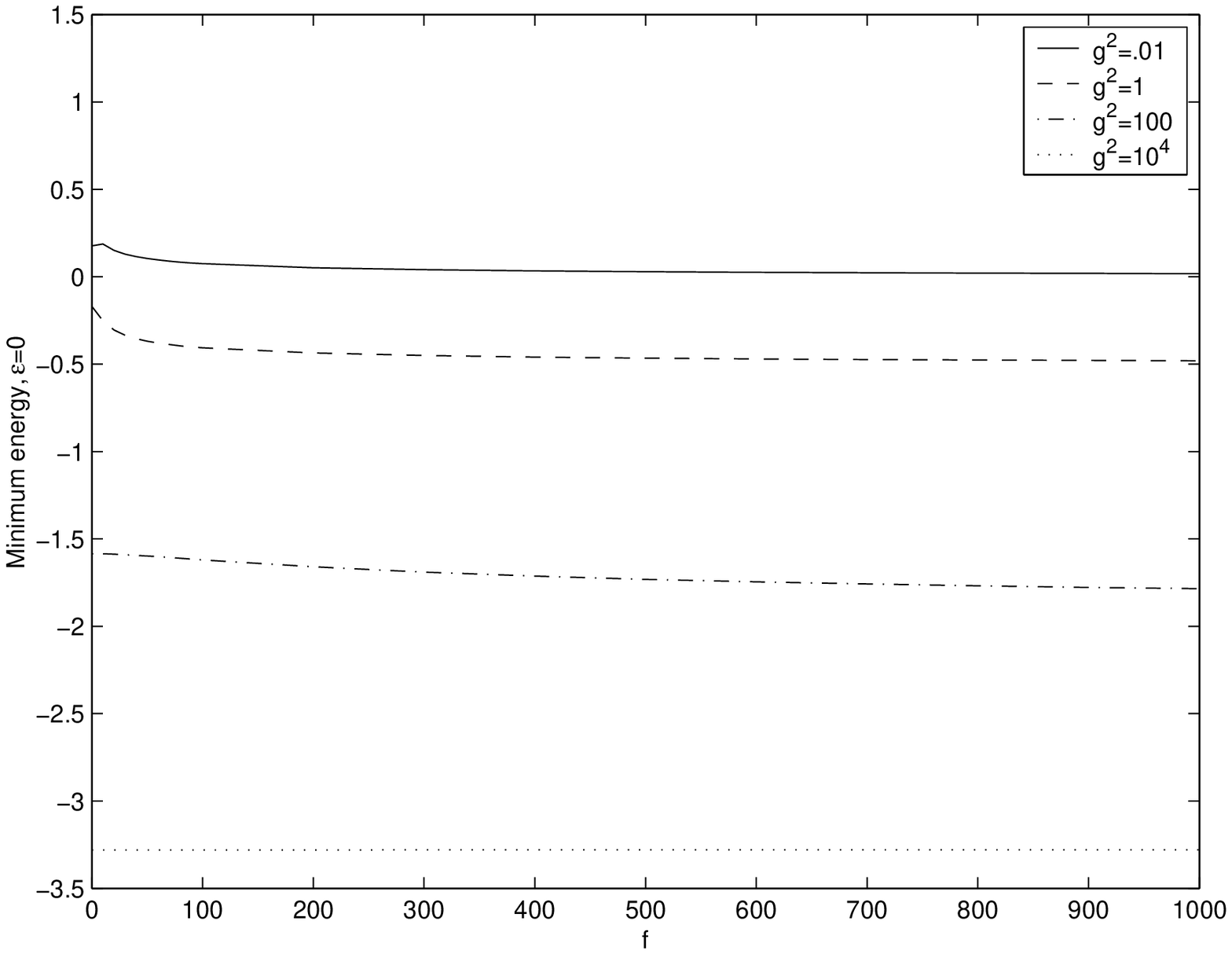}\quad
\caption{String tension on the left and
minimum energy on the right as functions of $f=\phi^\prime/\epsilon$
with $\epsilon\to0$. The monotonic decrease
in both curves indicates the vanishing of string tension in the
ground state.}
\label{tensione0}
\end{figure}

We now come to the total energy. The details of the
matter contribution to the energy are
contained in Appendix C. For ease of comparison
we shall display graphs for the same parameters
we chose in the spin energy. We start by
displaying the total energy curves for
$\epsilon=1$ so that the infra-red
cutoff in transverse space is $L\sim O(\sqrt{a/m})$. 
These results are shown in Fig.~\ref{totalenergynonpolyg0}.
We first note that as with the spin energy, it is
still true that the lowest energy solution always
falls on the $\phi=\phi^\prime$ curve,
and there is no indication of a phase transition. 
However, the
location of the minimum now depends 
dramatically on the value of ${\hat g}^2$. It
evidently approaches $\phi=0$ for weak coupling and $\phi=1/2$
for strong coupling. Looking at the $\phi^\prime=0$
curves we see first that the fishnet diagram $\phi=1$
on this curve predicts positive energy for weak and intermediate
coupling. By ${\hat g}^2=100$ this diagram predicts negative
energy. But in all cases the minimum of the $\phi^\prime=0$
curve is negative but higher in energy than the uniform $\phi^\prime=\phi$
case.

Next we display graphs
with $\epsilon=0.00001$ and the same range of couplings.
We see from Fig~\ref{totalenergynonpolyg5} that 
qualitatively the energy curves resemble more closely
those of the spin energy alone. The minimum energy again lies
on the $\phi^\prime=\phi$ curve and is located at
$\phi=1/2$ for all couplings. And again we see the
same paramagnetic tendencies in the fixed $\phi^\prime$
curves for strong and weak coupling. The string tension
for this lowest energy state can be found explicitly.
Since $\phi=\phi^\prime=1/2$ at this point, it follows that
$\kappa=0$ so $\vpp=\vmp=\phi/t_+(0)=1/(2+2{\hat g})$.
Thus
\bea
T_{\rm eff}={m\over a}\sqrt{2\epsilon(1+{\hat g})},
\eea
which vanishes for $\epsilon\to0$ at fixed ${\hat g}$. 
Vanishing tension implies  among other things a gapless 
energy spectrum. 

In Fig.~\ref{totalenergynonpolyg5r1} we display 
the energy curves for a massive scalar field. It is
interesting that, for weak coupling, there are indications
of a ferromagnetic behavior. (We were able to do calculations at zero
mass because of our cutoffs.) Notice though that
for ${\hat g}^2>1$ the massive case is very similar to
the massless case. In particular, at strong coupling we see a continuous
approach to the disordered phase $\phi=1/2$ and no anti-ferromagnetic
behavior. Notice that although there are quantitative
differences, the energy curves for the massive case at $\epsilon=.00001$
are qualitatively similar to those of the massless case at $\epsilon=1$.
This is perhaps reasonable because both $\rho=1$ and $\epsilon=1$
set an infra-red cutoff of the same order of magnitude $L^2=O(a/m)$.
\begin{table}[htb]
\begin{center}
\begin{tabular}{|c|c|c|c|c|c|c|c|c|}
\hline&&&&&&&&\\
${\hat g}^2$&$\rho$&$\epsilon$ &${\cal E}_{min}$ & 
$\phi_{min}$ & $\kappa_{min}$ & 
$\langle{\scriptstyle ++}\rangle$ &
$\langle{\scriptstyle +++}\rangle$ & ${a\over m}T_{\rm eff}$\\
\hline&&&&&&&&\\
.01&0&1&-.0651&0.0208&-.6598&.0106&.0054&9.713\\
.01&0&$10^{-5}$&-.0893&.5&0&.4545&.4132&.00492\\
.01&1&$10^{-5}$&.0176&0.0531&-.3959&.0351&.0232&.0169\\
.1&1&$10^{-5}$&-0.0663&0.1527&-.6015&.0761&.0379&.0115\\
\hline&&&&&&&&\\
1&0&1&-.4860&.1680&-1.6000&.0282&.0047&5.955\\
1&0&$10^{-5}$&-.6850&.5&0&.250&.125&.00632\\
1&1&$10^{-5}$&-.3931&0.3543&-.6000&.1256&.0445&.00892\\
\hline&&&&&&&&\\
100&0&1&-2.208&.4380&-2.049&.0132&3.95$\cdot10^{-4}$&8.704\\
100&0&$10^{-5}$&-2.3792&.5&0&.0455&.0041&.01482\\
100&1&$10^{-5}$&-2.0353&0.4895&-.4189&.0360&.0026&.0167\\
\hline&&&&&&&&\\
$10^4$&0&1&-4.501&.4892&-3.031&.001&2.25$\cdot10^{-6}$&
31.62\\
$10^4$&0&$10^{-5}$&-4.574&.505&1.8189&.0124&3.03$\cdot10^{-4}$&.0284\\
\hline
\end{tabular}
\end{center}
\caption{The worldsheet parameters for $\Tr\Phi^3$
theory, fixed by minimizing the
energy in the mean field approximation.}
\label{parameters}
\end{table}

Let us now consider what these calculations 
tell us about the role of the fishnet diagram.
We can follow the fate of the fishnet diagram in two
stages. Starting at $\phi^\prime=0,\phi=1$ we first
let $\phi$ relax at fixed $\phi^\prime=0$ to the
minimum energy at $\phi^\prime=0$. Since
$\vmp=0$, the string tension
along this curve is $(m/a)\sqrt{(1+4\epsilon/\vpp)/2}$,
which is simply $m/(a\sqrt2)$ as $\epsilon\to0$.
Then one can begin to relax $\phi^\prime$ putting
$\phi^\prime=\epsilon f$. Then one can smoothly send
$\epsilon$ to zero at fixed $f$, and study how the
physics evolves as $f$ increases. The field dependence of the
energy is shown in Fig.~\ref{totalenergye0}. One should
keep in mind that because of our limiting procedure
all these curves are infinitesimally near the
$\phi^\prime=0$ curves of the previous graphs. 
In Fig.~\ref{tensione0} we show the string tension
corresponding to the minimal point on each fixed
$f$ curve as a function of $f$. Next to it
we display the corresponding minimal energy as a function
of $f$. We see that both
string tension and energy generally decrease as $f$ increases.
Note, however that a small energy barrier between
$f=0$ and $f>10$ is visible on the ${\hat g}^2=.01$
curve. Although not visible there are similar
tinier barriers at higher coupling. 
It is clear from the previous graphs that the
general decreasing trend
continues till infinite $f$ where the tension goes to zero and
the energy reaches its minimum. Our conclusion is that,
for $\Tr\Phi^3$ theory, the fishnet phase is not
stable and though there is a stringy description for
generic values of the fields, the system relaxes to a
ground state with no mass gap and zero string tension.
In the table we have compiled a list of the relevant
parameters characterizing the physics of this global
minimum for various $\hg, \rho, \epsilon$.

What would we expect to happen in a theory like QCD
which ought to confine quarks by forming 
stringy flux tubes? The failure of the $\Tr\Phi^3$
theory to form string isn't just signified by the
monotonic decrease of effective tension with $f$, but
rather with the correlation of this behavior with
the monotonic decrease of the associated system energy.
If the system energy had a minimum at finite $f$, its
ground state would support a non-vanishing string tension
at the corresponding $f$. This is what we should
expect to happen for QCD and other confining theories.
Moreover, for QCD this should occur for couplings of $O(1)$.

\section{Mean Field Treatment of the Matter Fields (Large $d=D-2$)}

In the previous section we built the mean field approximation
on the mean field $\phi=\langle P\rangle$ representing the
average value of the Ising spin, following \cite{bardakcitmean}.
But another, perhaps superior,  approach is to instead focus on a mean
field defined as the average value of an $SO(d)$ scalar 
bilinear of the matter fields, as suggested in \cite{bardakcitimp}.
(Such a description should be accurate in the parametric
limit $d\to\infty$. The existence of this parametric limit
in which the approximation is good is a significant
virtue of this approach, although the underlying
field theories we are describing are only meaningful
in sufficiently low dimensions.)
In this approach we want our mean field to be an order parameter
which measures the presence of an effective string tension,
so it is natural to define
\bea
\chi_k^j\equiv -{a\over m\epsilon}\left\langle{1\over2}
({\boldsymbol q}_k^j-{\boldsymbol q}_k^{j-1})^2
-{\boldsymbol b}_k^j{\boldsymbol c}_k^j\right\rangle.
\eea
Accordingly, we add a source term to the worldsheet
action
\bea
-{a\over m\epsilon}\sum_{kj}J_k^j\left[{1\over2}
({\boldsymbol q}_k^j-{\boldsymbol q}_k^{j-1})^2
-{\boldsymbol b}_k^j{\boldsymbol c}_k^j\right].
\eea
The path integral in the presence of $J$ is defined
to be $e^{W(J)}$, the field is then 
$\chi_k^j=\partial W/\partial J_k^j$, and the effective action
is given by
\bea
{\cal A}\equiv W-\sum_{kj} \chi_k^jJ_k^j.
\eea

The mean field approximation is then implemented by
establishing the relation between $\chi$ and $J$ 
as that arising from the path integral over
matter fields neglecting
the terms in the action that couple the matter to the
Ising spins. In other words we take the $J$ dependence of
$W$ to be that of $W_m$ defined by
\bea
e^{W_m(J)}&\equiv&
\int DcDbD{\boldsymbol q}
\exp\left\{-{a\over2m}\sum_{k,j}
{({\boldsymbol q}_{k+1}^j-{\boldsymbol q}_{k}^{j})^2}
-{a\over2m\epsilon}\sum_{k,j}
{J}_k^j{({\boldsymbol q}_{k}^j-{\boldsymbol q}_{k}^{j-1})^2}
+{a\over m\epsilon}
\sum_{k,j}{J}_k^j{\boldsymbol b}^j_{k}{\boldsymbol c}^j_{k}\right\}
\nonumber\\
&&\exp\left\{{a\over m}
\sum_{k,j}({\boldsymbol b}_{k+1}^j-{\boldsymbol b}_k^j)
({\boldsymbol c}_{k+1}^j-{\boldsymbol c}_k^j)\right\}
\eea
We then use this approximate path integral to approximate the
matter dependence of the terms in the action coupling to the
Ising spins by their average values. In practice we assume
that $J$ (and hence also $\chi$) is uniform in $k,j$. Then
by consulting Appendix C, we find that
\bea
W_m(J)&=&-MN\left[{2d\over\pi}{\rm Re}\left\{i{\rm Li}_2\left({1\over i}
\sqrt{\epsilon\over J}\right)\right\}-d\ln\left({1\over2}+\sqrt{
{1\over4}+{\epsilon\over J}}\right)\right]\\
-{a\over m\epsilon}\left\langle{1\over2}
({\boldsymbol q}_k^j-{\boldsymbol q}_k^{j-1})^2
-{\boldsymbol b}_k^j{\boldsymbol c}_k^j\right\rangle
&=&{1\over MN}{\partial W_m\over\partial J}=\chi(J)\\
{a\over m}\left\langle{\boldsymbol b}_k^j{\boldsymbol c}_k^j\right\rangle
&=&{d\over2}{a\over m}\left\langle{b}_k^j{c}_k^j\right\rangle
={d\epsilon\over2\sqrt{J^2+4\epsilon J}}\\
{a\over m}\left\langle({\boldsymbol b}_{k+1}^j-{\boldsymbol b}_k^j)
({\boldsymbol c}_{k+1}^j-{\boldsymbol c}_k^j)\right\rangle
&=&{d\over2}{a\over m}\left\langle({b}_{k+1}^j-{b}_k^j)
({c}_{k+1}^j-{c}_k^j)\right\rangle
={d\over2}\left[1-{J\over\sqrt{J^2+4\epsilon J}}\right].
\eea
Inserting these results into the summand for the spin sum,
we find that the spin sum is just a two dimensional
generalized Ising model with Boltzmann factor $e^{A(P)}$ with
\bea
A(P)&=&\ln{\hat g}\sum_{kj}{1-s_k^js_k^{j-1}\over2}
-{d\over2}\ln\left(1+\rho\right)\sum_{k,j}P_k^j
+ {\chi(J)}\sum_{kj}P_k^jP_k^{j-1}
\nonumber\\
&&+\sum_{kj}\left\{{\epsilon\over\sqrt{J^2+4\epsilon J}}
\left[{d\over2}A^\prime_{kj}-B_k^j\right]
+\left[1-{J\over\sqrt{J^2+4\epsilon J}}\right]
\left[{d\over2}C^\prime_{kj}-D_k^j\right]\right\}\label{2dising-2pcase}
 \\
A^\prime_{kj}&=&{P}_k^{j+1}{P}_k^j-{P}_k^{j-1}{P}_k^j{P}_k^{j+1}
+(1-P_k^j)(P_{k+1}^j+P_{k-1}^j)+\rho(1-P_k^j)P_{k-1}^{j-1}P_{k-1}^j\\
C^\prime_{kj}&=&(1-P_k^j)(1-P_{k+1}^j)-1=P_k^jP_{k+1}^j-P_k^j-P_{k+1}^j
\eea
and $B$ and $D$ are unchanged. Defining $e^{W_s(J)}=\sum_{P=0,1}e^A$,
the total effective action is
\bea
{\cal A}\approx W_m(J)+W_s(J)-MNJ\chi(J).
\eea
and the value of $J$ (or of $\chi(J)$) is determined as a
stationary point of ${\cal A}$ that minimizes the energy
per site ${\cal E}=-{\cal A}/MN$. To calculate $W_s$ with
no further approximation we would need the exact solution
of the two dimensional spin system defined
by (\ref{2dising-2pcase}). This system differs from the soluble
two dimensional Ising model by linear terms in the
spins (a constant magnetic field) and also higher than
quadratic terms in the spin. For generic $J$
and coupling parameters this model certainly
has no analytic solution. A numerical solution can
be attempted, but we leave this problem for
future research.

Instead we note that, in this setup, the mean field $\chi$
acts as a source for the spin bilinear 
$\langle P_k^jP_k^{j-1}\rangle$, and so its
conjugate $J$ is in this sense a mean field for this bilinear.
Thus a natural mean field approach to the two dimensional
Ising system encountered here is to use 
$\varphi_k^j\equiv\langle P_k^jP_k^{j-1}\rangle$ as
the mean field. But once we treat the Ising spins in this
way, nothing is really gained by introducing a source for the 
matter bilinears, and so we will only source
the spin bilinear, repeating the mean field
treatment of the previous section using this
new order parameter. In the rest of this section
we describe the results of doing this, and we compare them
to those of the more traditional mean field treatment of the
previous section. Although this second study will not
be different in spirit from that of the previous
section, it will give us some idea of the robustness
of the mean field method.

We introduce the source by inserting $e^{\sum \chi_k^jP_k^jP_k^{j-1}}$
into the original path integral (\ref{isingsumeps2}), and we
define $W(\chi)$ by calling the resulting path integral
$e^W$. Then our effective field is $\varphi_k^j=\partial W/\partial\chi_k^j$
and the effective action is
\bea
{\cal A}(\varphi)=W(\chi)-\sum_{kj}\varphi_k^j\chi_k^j.
\eea
As in Section 4 we define the mean field approximation by
doing the spin sum with the spin dependent terms in the
matter action fixed to their expectation values which are then 
self-consistently determined. In contrast to section 4,
the mean fields here are the bilinears 
$\langle P_k^{j-1}P_k^{j}\rangle$ and
the derived expectations are now the linear and cubic
ones $\langle P_k^j\rangle$ and $\langle P_k^{j-1}P_k^{j}P_k^{j+1}\rangle$.

For the fishnet spin pattern 
$\varphi_k^j=\varphi,\varphi,\varphi^\prime,\varphi^\prime,\varphi,\varphi,
\ldots$, there are three distinct values each for the linear and
cubic cases. As explained in
Appendix A.2, where these
expectations are evaluated, we denote the linear ones by $\phi,\phi^\prime$, 
and ${\tilde \phi}$ and the cubic ones by $\langle{\scriptstyle+++}\rangle$,
$\langle {\scriptstyle-++}\rangle$, 
and $\langle{\scriptstyle+-+}\rangle$. The matter integrals
are evaluated in Appendix C.2, basically
by substituting a new set of coefficients for the $\alpha,\beta,\gamma,
\delta $ coefficients of C.1.
Now the effective string tension
depends directly on the background fields
\bea
T_{\rm eff} 
=\fr{m}{a}\sqrt{\fr{(4\ep + \vp +\vp^\prime)}{2(\vp+\vp^\prime)
    +\vp\vp^\prime/\ep)}}. 
\label{eff-tension-2pcase}
\eea
As before the tension goes to zero as $\ep\to0$ unless $\vp^\prime=\cO(\ep)$.

\begin{figure}[htb]
\psfrag{varphi}[l][l][.55]{$\varphi$}
\psfrag{phi^prime = phi}[l][l][.55]{$\vp^\prime=\vp$}
\psfrag{phi^prime = 0.1}[l][l][.55]{$\vp^\prime=0.1$}
\psfrag{phi^prime = 0.2}[l][l][.55]{$\vp^\prime=0.2$}
\psfrag{phi^prime = 0.3}[l][l][.55]{$\vp^\prime=0.3$}
\psfrag{phi^prime = 0}[l][l][.55]{$\hskip-6.7pt\vp^\prime=0$}
\psfrag{Spin energy per site}[l][l][.55]{Spin energy per site}
\psfrag{g^2=100, \epsilon=10^(-5), \rho=0}[l][l][.55]
{$\hg^2=0.01, \epsilon=10^{-5}, \rho=0$}
\psfrag{g}[l][l][.46]{${\hat g}$}
\includegraphics[width=8cm,height=5.5cm]{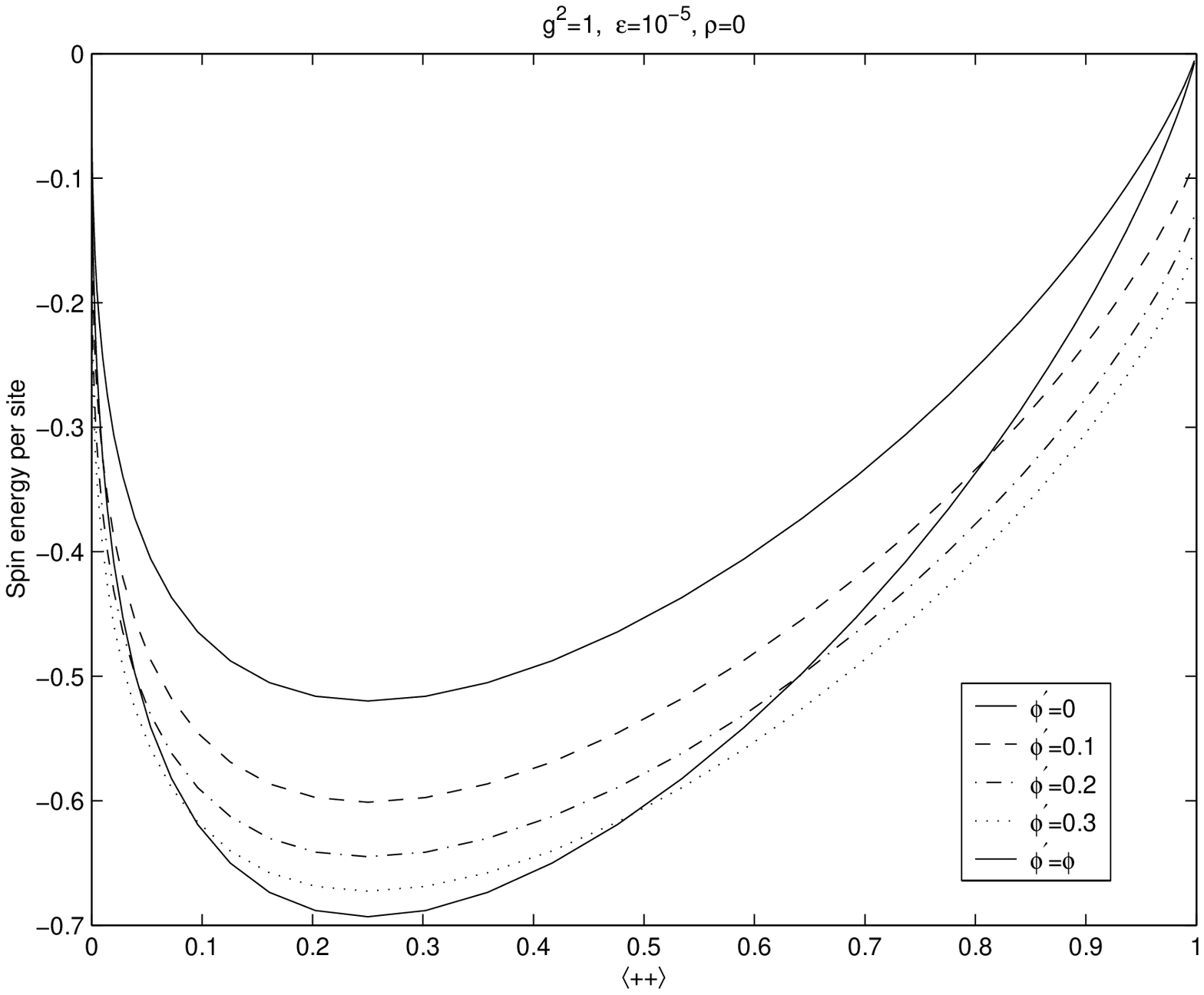}\quad
\psfrag{g^2=1,epsilon=10^(-5),rho=0}[l][l][.55]
{$\hg^2=1, \epsilon=10^{-5}, \rho=0$}
\includegraphics[width=8cm]{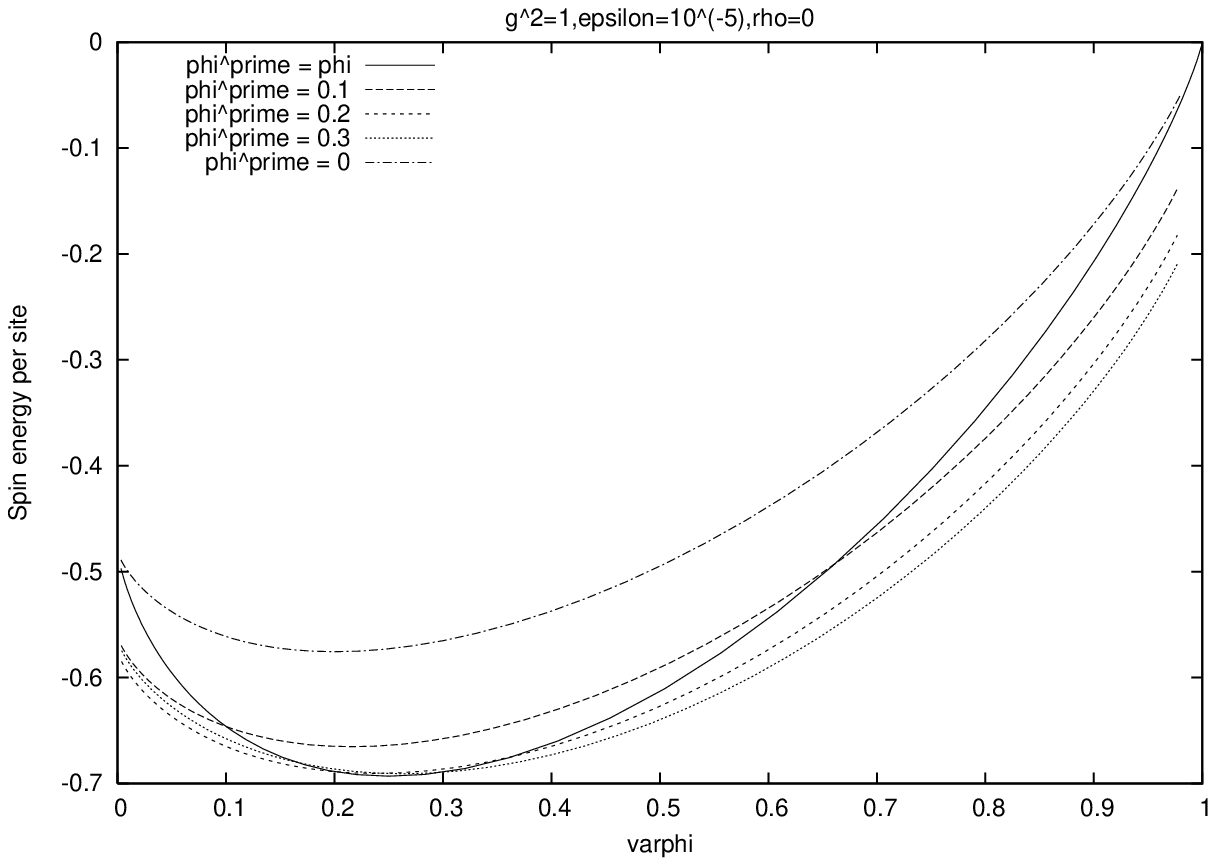}
\caption{The spin energy per site for the massless case, 
$\rho=0$, $\epsilon=10^{-5}$. On the left are the
$\hg^2 =1$ spin energy curves from Fig.~\ref{spinenergynonpolyg}
plotted versus $\vpp$. On the right are the analogous curves
from the mean field treatment of Section 5.}
\label{spinenergycompare}
\end{figure}

The energy function ${\cal E}(\vp,\vp^\prime)$ is certainly a different
function of its two variables than the energy function of Section
4, both because the variables are differently defined and
because the different structure of the source terms leads to an
effectively different treatment of fluctuation effects in 
the corresponding mean field approximations. To display these
differences graphically we choose a typical coupling ${\hat g}^2=1$
for which they are evident. In Fig.~\ref{spinenergycompare}
we show on the left the spin energy curves for this
case from Fig.~\ref{spinenergynonpolyg} in Section 4, replotted as 
a function of $\vpp$, the physical quantity measured by $\varphi$. 
On the right we show the corresponding curves from the treatment
of this section as functions of $\varphi$. The most dramatic
difference compared to Fig.~\ref{spinenergynonpolyg} is
the move of the location of the minimum
energy toward the origin. This effect is
clearly accounted for by the change of independent variables
from $\phi$ to $\vpp$. Next the shape of the curves on
the left is slightly different in detail than those on the
right, but the ordering of the curves is the same in both
figures in the region near the minimum. Note that the $\phi=\phi^\prime$
($\vp=\vp^\prime$) 
curves in the left and right are essentially identical. However there
is a significant lowering in energy of the constrained
curves on the right compared to those on the left. This 
quantitative difference in the mean field treatments is
reasonable considering that the constraints imposed
on bilinears $\vp, \vp^\prime$ in the present section are less severe
than the constraints imposed on $\phi^\prime$ in Section 4. 
Hence the constrained system in the former case
should be closer to the unconstrained ground state 
and the corresponding energies should be lower. Also the
slicing of the two dimensional surfaces is slightly
different in the two cases. This interpretation
is supported by the fact that the ground state curve is
not significantly altered. 
These two differences, the move toward the origin,
and the lowering in energy of the constrained curves but not
of the ground state curves are present in the spin
energy for other couplings, but there are no other significant
differences, so we don't display all those cases.

Finally we turn our attention to the total energy
per site. In Fig.~\ref{totennomass-2pcase}
we display the total energy curves for the massless
case and the same set of couplings used in Section 4.
These graphs should be compared to Fig.~\ref{totalenergynonpolyg5}.
The most dramatic apparent difference is the
movement of the minimum toward the origin as
the coupling increases. As discussed in the previous
paragraph, this is explained by the change of independent
variables from $\phi$ to $\vpp$. For the strongest coupling
displayed the minimum is for $\varphi\approx.01$.
We also see the trend that the constrained curves generally lie
lower in energy in Fig.~\ref{totennomass-2pcase} than
in Fig.~\ref{totalenergynonpolyg5}. But note that these
constrained curves are not really comparable at strong coupling
because in Fig.~\ref{totennomass-2pcase} 
most of the $\vp^\prime$ choices are larger than
the location of the minimum, in contrast to Fig.~\ref{totalenergynonpolyg5}
where {\it all} of the $\phi^\prime$ choices are smaller than the
location of the minimum.
From Fig.~\ref{totennomass-2pcase}, 
we see that the minimum energy is on the $\vp=\vp^\prime$ curve and
the location of this minimum point depends upon the value of
$\hg^2$. For weak coupling, the minimum point tends to $\vp=0.5$ and it
approaches $\vp=0$ for strong coupling. 
\begin{figure}[htb]
\psfrag{varphi}[l][l][.55]{$\varphi$}
\psfrag{phi^prime = phi}[l][l][.55]{$\vp^\prime=\vp$}
\psfrag{phi^prime = 0.1}[l][l][.55]{$\vp^\prime=0.1$}
\psfrag{phi^prime = 0.2}[l][l][.55]{$\vp^\prime=0.2$}
\psfrag{phi^prime = 0.3}[l][l][.55]{$\vp^\prime=0.3$}
\psfrag{phi^prime = 0}[l][l][.55]{$\hskip-6.7pt\vp^\prime=0$}
\psfrag{Total energy per site}[l][l][.55]{Total energy per site}
\psfrag{g^2=0.01,epsilon=10^(-5),rho = 0}[l][l][.55]{$\hg^2=0.01, 
\epsilon=10^{-5}, \rho=0$}
\includegraphics[width=8cm]{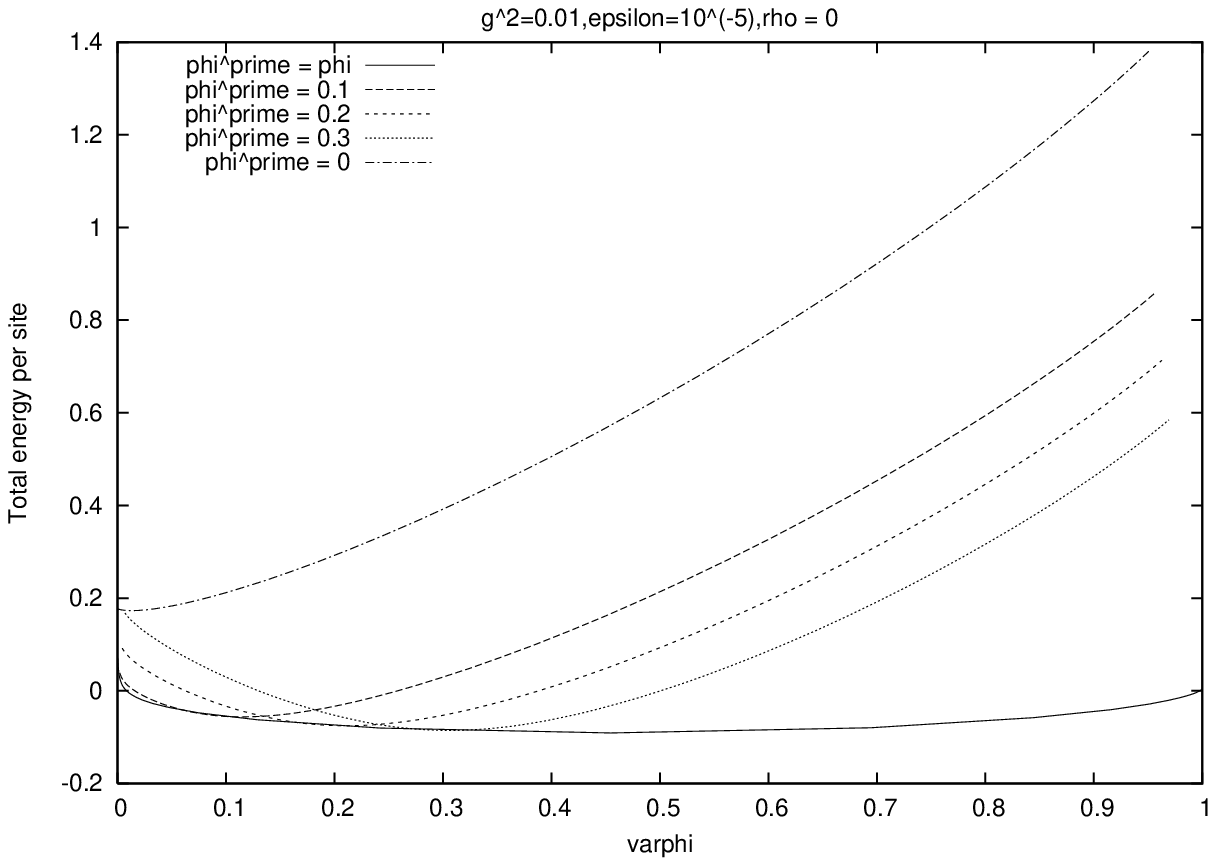}\quad
\psfrag{varphi}[l][l][.55]{$\varphi$}
\psfrag{Total energy per site}[l][l][.55]{Total energy per site}
\psfrag{g^2=1,epsilon=10^(-5),rho = 0}[l][l][.55]
{$\hg^2=1, \epsilon=10^{-5}, \rho=0$}
\includegraphics[width=8cm]{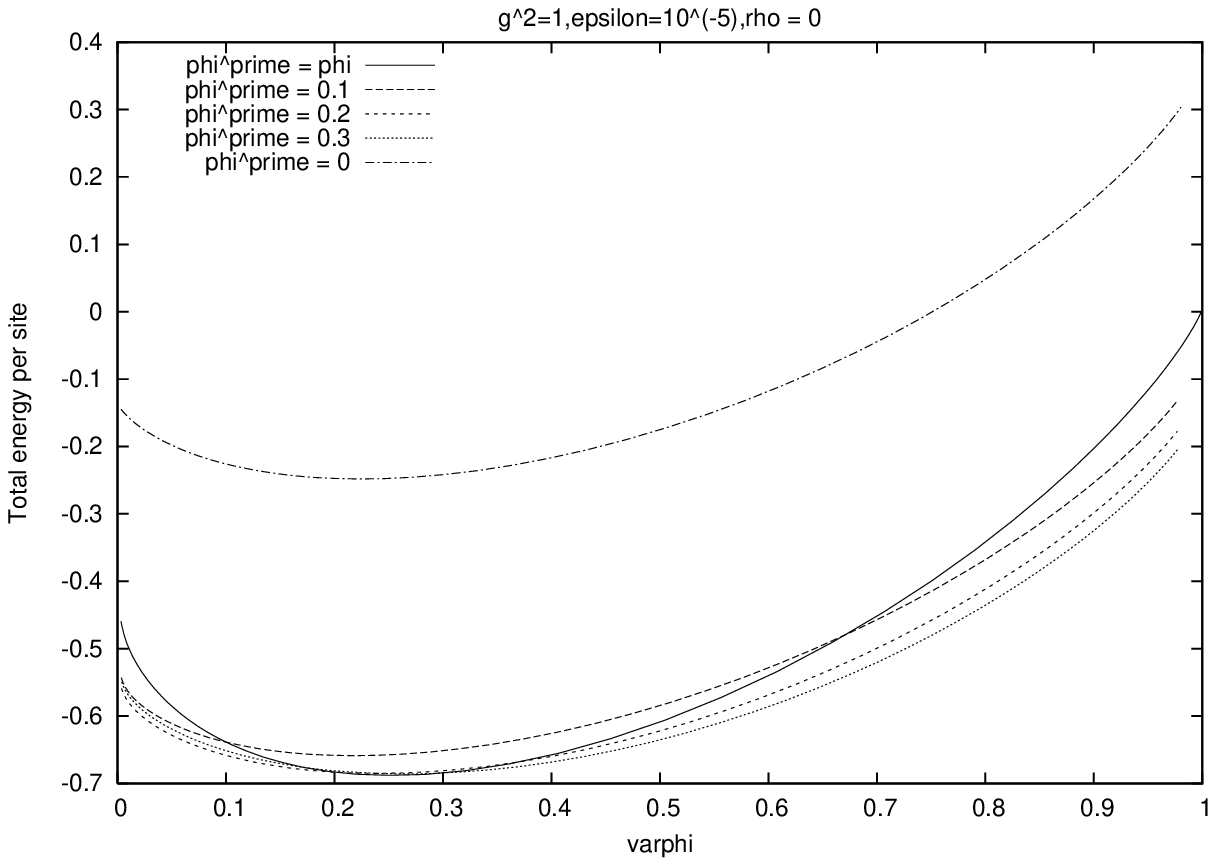}
\psfrag{Total energy per site}[l][l][.55]{Total energy per site}
\psfrag{g^2=100,epsilon=10^(-5),rho = 0}[l][l][.55]
{$\hg^2=100, \epsilon=10^{-5}, \rho=0$}
\includegraphics[width=8cm]{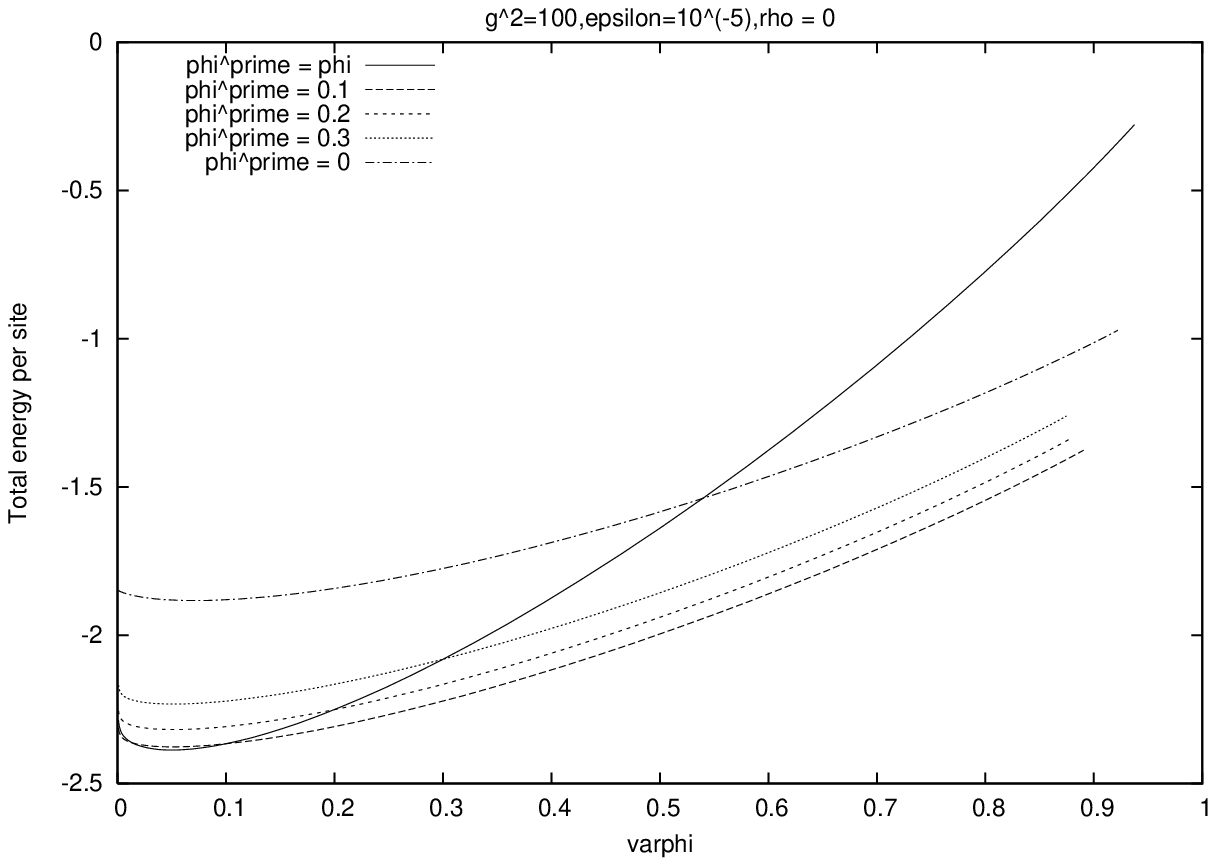}\quad
\psfrag{varphi}[l][l][.55]{$\varphi$}
\psfrag{Total energy per site}[l][l][.55]{Total energy per site}
\psfrag{g^2=10^4,epsilon=10^(-5),rho = 0}[l][l][.55]{$\hg^2=10^4, 
\epsilon=10^{-5}, \rho=0$}
\includegraphics[width=8cm]{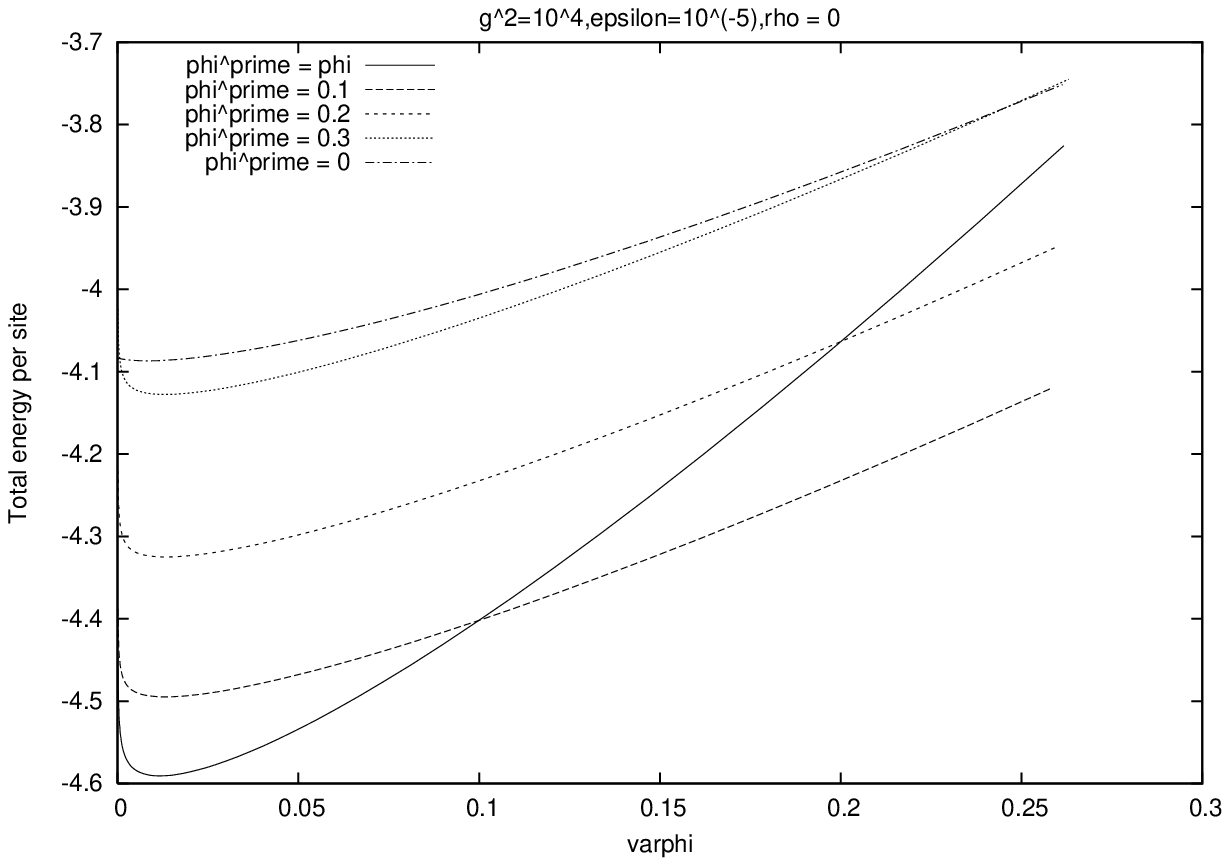}
\caption{The total energy per site for the massless case, 
$\rho=0$, $\epsilon=10^{-5}$.}
\label{totennomass-2pcase}
\end{figure}
The massive case $\rho=1$
is presented in Fig.~\ref{totenmass-2pcase} and should be compared
to Fig.~\ref{totalenergynonpolyg5r1}. In the latter the
location of the minimum is near $\phi=0$ at weak coupling and
approaches $\phi=0.5$ at strong coupling. In contrast 
in Fig.~\ref{totenmass-2pcase}, again
because of the different variables chosen, the minimum approaches
$\vp=0$ in the strong coupling limit in addition to the
weak coupling limit. But the behavior of the constrained curves
reflects the qualitatively different physics of the two limits.
There is ferromagnetic behavior at weak coupling, with $\vp$
strongly correlated with $\vp^\prime$. At strong coupling
there is very little if any such correlation.  
\begin{figure}[htb]
\psfrag{varphi}[l][l][.55]{$\varphi$}
\psfrag{phi^prime = phi}[l][l][.55]{$\vp^\prime=\vp$}
\psfrag{phi^prime = 0.1}[l][l][.55]{$\vp^\prime=0.1$}
\psfrag{phi^prime = 0.2}[l][l][.55]{$\vp^\prime=0.2$}
\psfrag{phi^prime = 0.3}[l][l][.55]{$\vp^\prime=0.3$}
\psfrag{phi^prime = 0}[l][l][.55]{$\hskip-6.7pt\phi^\prime=0$}
\psfrag{Total energy per site}[l][l][.55]{Total energy per site}
\psfrag{g^2=0.01,epsilon=10^(-5),rho = 1}[l][l][.55]
{$\hg^2=0.01, \epsilon=10^{-5}, \rho=1$}
\includegraphics[width=8cm]{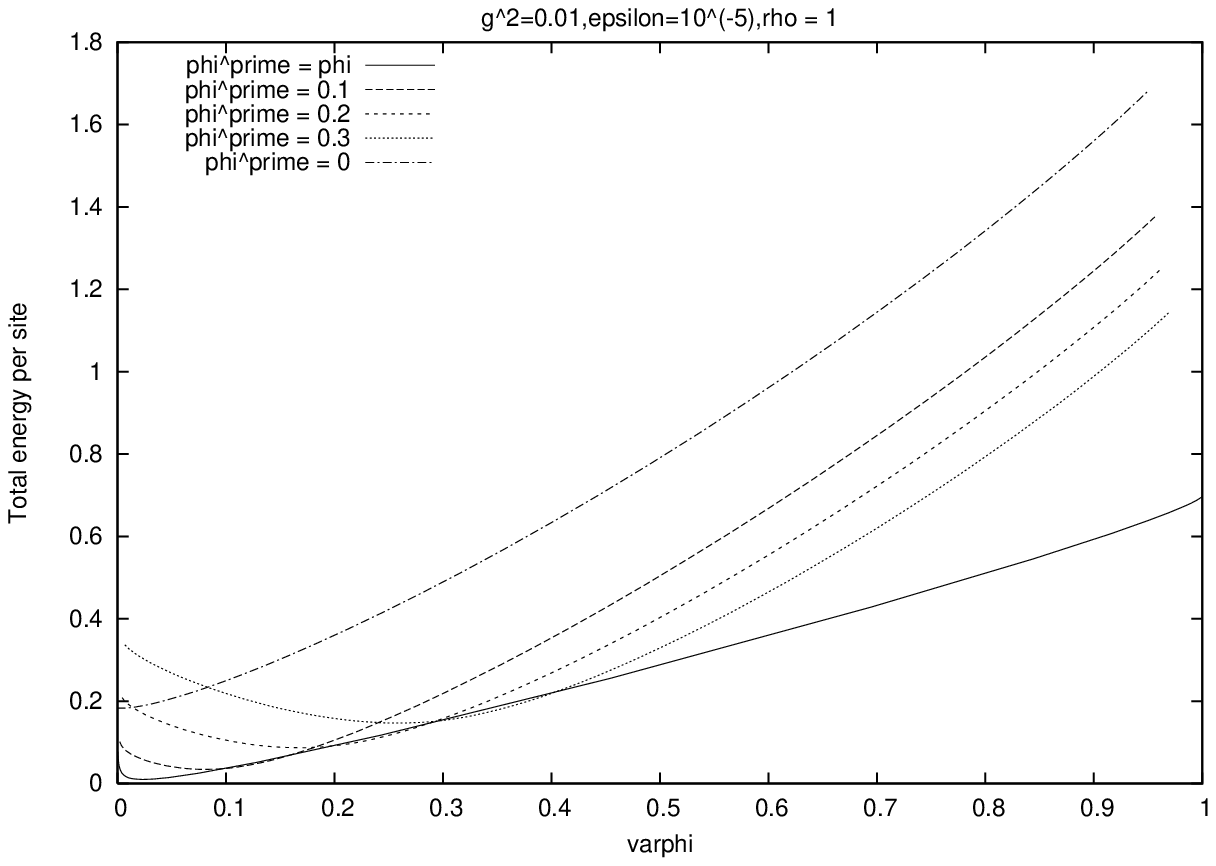}\quad
\psfrag{Total energy per site}[l][l][.55]{Total energy per site}
\psfrag{g^2=1,epsilon=10^(-5),rho = 1}[l][l][.55]
{$\hg^2=1, \epsilon=10^{-5}, \rho=1$}
\includegraphics[width=8cm]{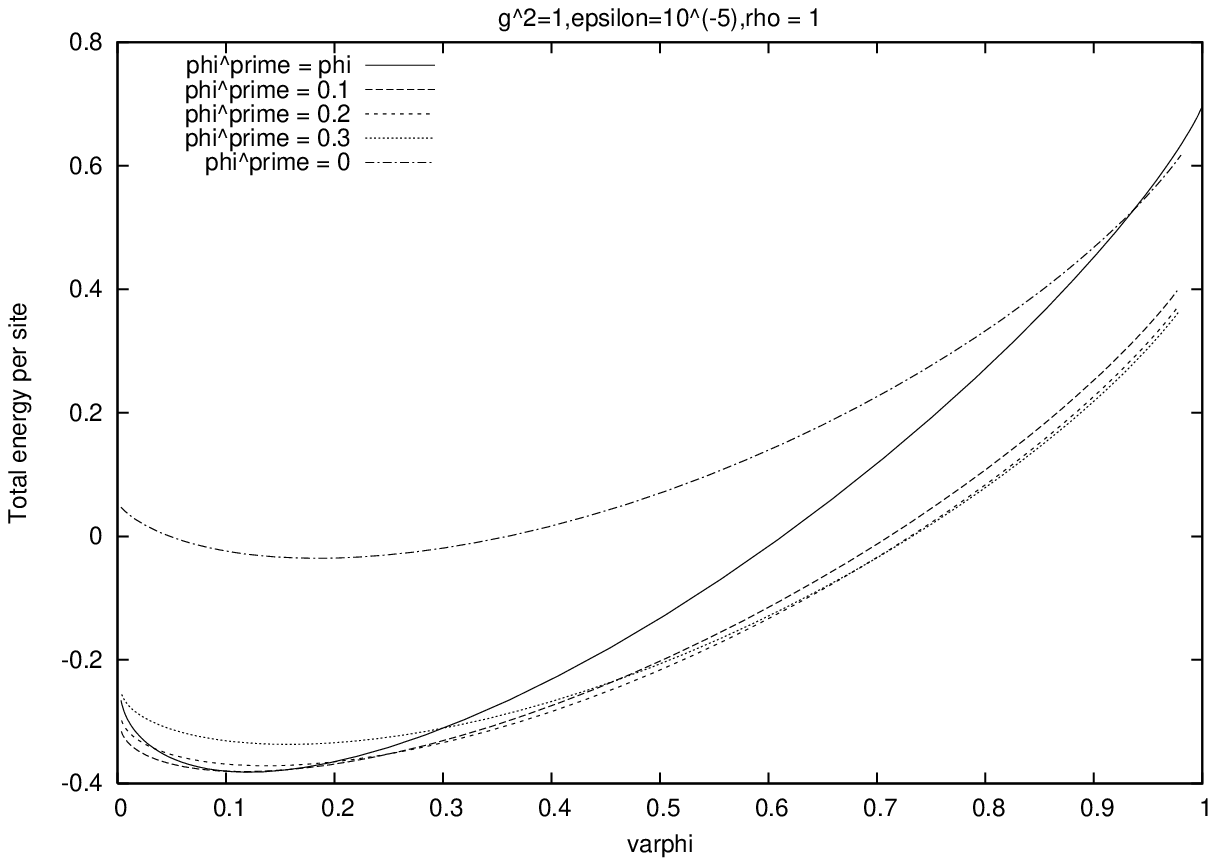}
\psfrag{Total energy per site}[l][l][.55]{Total energy per site}
\psfrag{g^2=100,epsilon=10^(-5),rho = 1}[l][l][.55]
{$\hg^2=100, \epsilon=10^{-5}, \rho=1$}
\includegraphics[width=8cm]{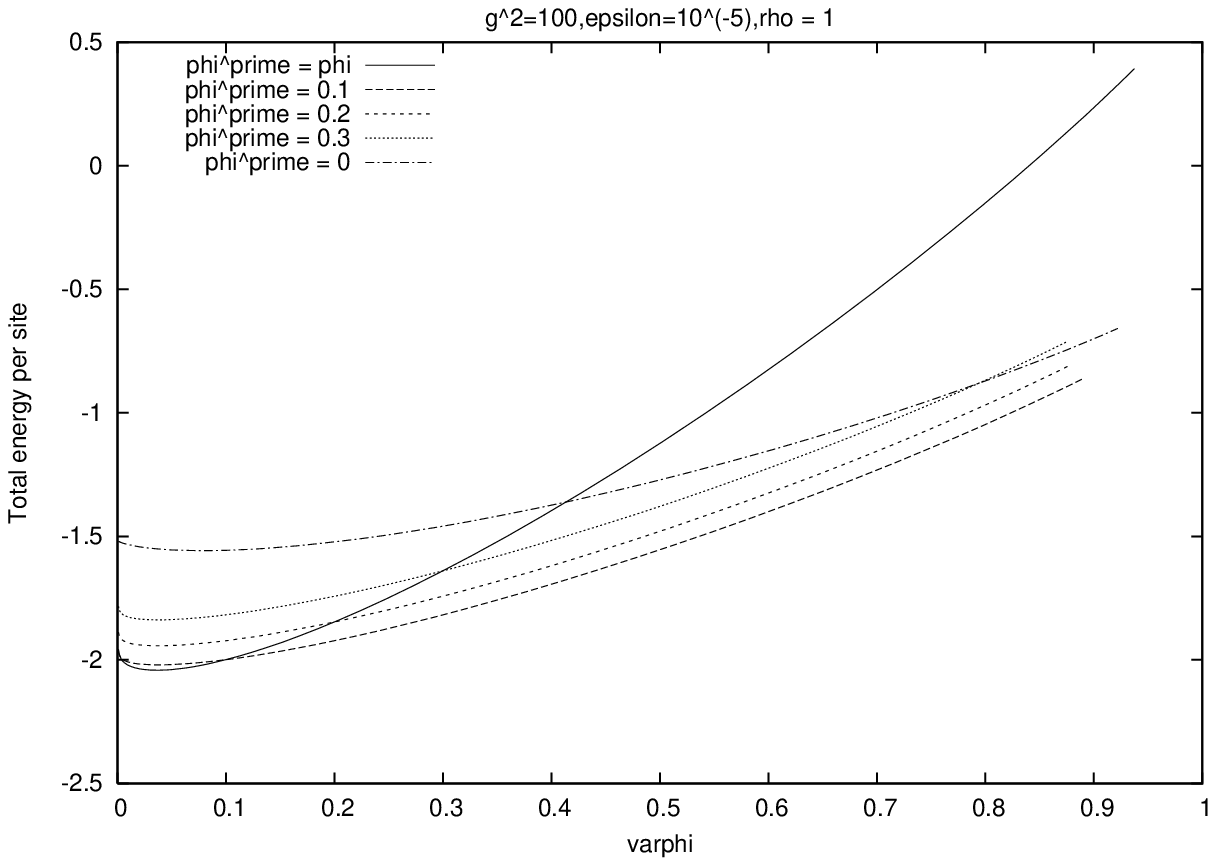}\quad
\psfrag{Total energy per site}[l][l][.55]{Total energy per site}
\psfrag{g^2=10^4,epsilon=10^(-5),rho = 1}[l][l][.55]
{$\hg^2=10^4, \epsilon=10^{-5}, \rho=1$}
\includegraphics[width=8cm]{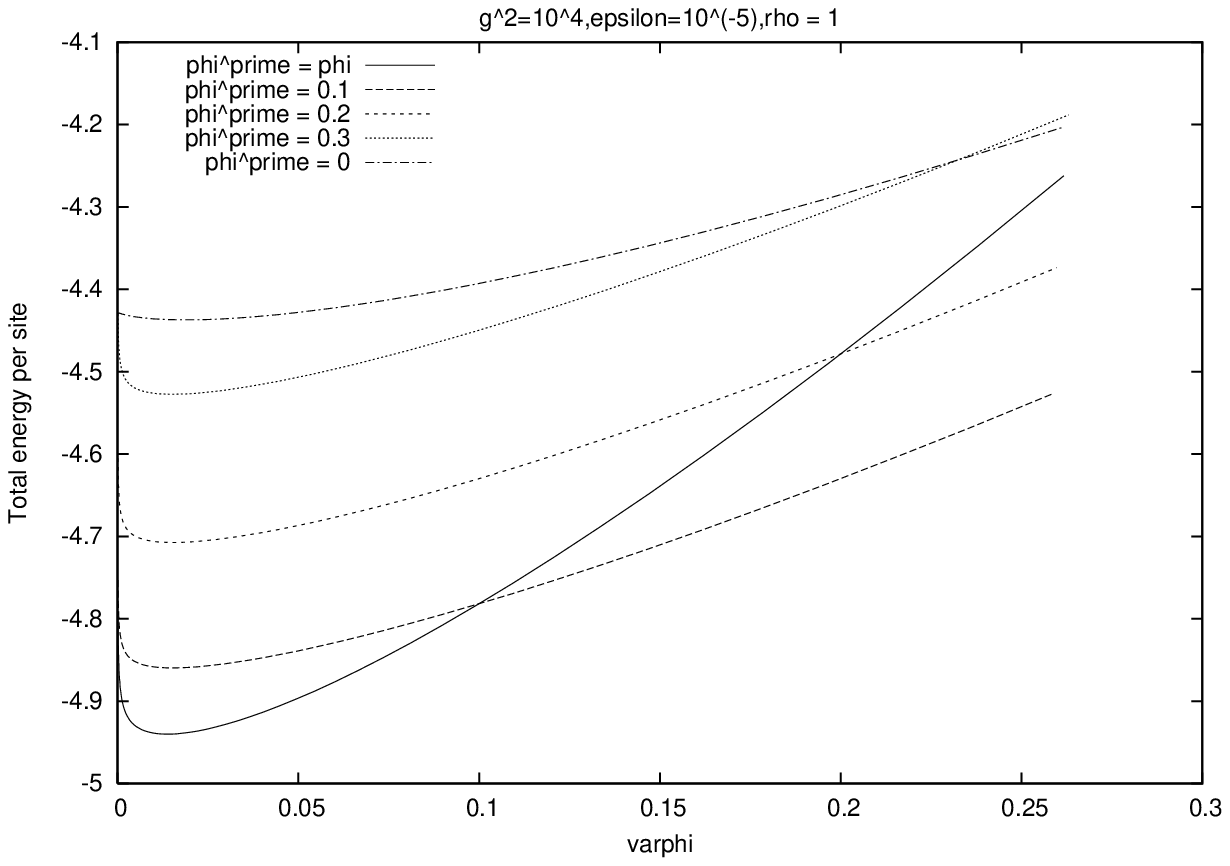}
\caption{The total energy per site for the massive case, 
$\rho=1$, $\epsilon=10^{-5}$.}
\label{totenmass-2pcase}
\end{figure}
In summary,
although the different variable choice has caused some 
dramatic differences in the energy curves between the two
versions of the mean field approximation, the actual
physics contained in the two approximation schemes is 
actually quite similar. And the basic conclusions we
have drawn are robust at least against different styles of
mean field approximation.

\section{Conclusion}
The mean field study undertaken in this article goes
some way toward illuminating the role played by the
fishnet diagrams in the sum of planar diagrams.
For the $\Phi^3$ theory studied, we identified
a mean field description involving two mean
fields $\phi,\phi^\prime$ such that the fishnet
diagram is singled out by the field values $1,0$
respectively. But, within our mean field approximation, 
over the whole range of
couplings the lowest energy is reached
when $\phi=\phi^\prime$ and the effective string
tension for these field values is 0. This indicates that 
the role of the fishnet diagram is negligible in
the sum of planar diagrams of this theory.
Since the underlying theory is unstable, this negative
conclusion is probably reasonable. 

On a more positive
note, our mean field method gives a concrete derivation of
the kind of string description of field theory envisioned
in the AdS/CFT duality of Maldacena. For generic
values of the fields, we identify an effective
string tension $T_{\rm eff}(\phi,\phi^\prime)$ that
characterizes the excitations close in energy to
the energy associated with those fields. For
$\Phi^3$ theory it happens that $T_{\rm eff}=0$ for
the lowest energy configuration of fields. But
in a theory like QCD which is supposed to confine
quarks, we can easily imagine that the tension won't
vanish. Application of our methods to this case
will clearly be an interesting next step.

But the well-known deficiencies of mean field theory
still cast doubt on the reliability of such studies.
We should regard them only as rough intuitive indicators
of the physics potentially contained in these theories.
It is therefore very important to develop calculational
methods without such defects. One promising
approach is Monte Carlo simulation. The worldsheet
path integral for the $q$ variables and the Ising
spins clearly involves a positive definite integrand.
The Grassmann $bc$ variables, introduced to give
a local description, clearly introduce minus signs
and ensuing complications for Monte Carlo methods.
On the other hand, if one integrates out the ghosts,
positivity is again restored at the expense of some
non-locality. Because
the ghost dynamics is independent on each time
slice, the relevant determinants are only one dimensional,
and there is hope that their evaluation won't 
be prohibitively time consuming. 
Another numerical approach that is in principle exact
is the block spin renormalization group. 
The feasibility of such numerical methods is 
under active investigation. We hope that the
mean field results of the present article will
be useful in interpreting any results that come from them.

Even within the narrow application of mean field
theory to the worldsheet for $\Phi^3$ theory, we have
left some issues to be resolved. First, we left
the alternative method \cite{bardakcitimp} of approximating the matter
variables with mean fields unexplored. In particular
this method leads to a two-dimensional Ising spin
system (\ref{2dising-2pcase}) which we despaired of solving analytically.
However Monte Carlo methods are ideal for such systems
and their application to this spin system is
worth studying. The resulting hybrid of mean
field methods and Monte Carlo methods provides
a viewpoint midway between the intuitive
mean field approach of the current article, and
a more exact numerical approach.
We have also proposed a worldsheet model for
conventionally compactified $\Phi^3$ theory in appendix D.
The methods of the current article have yet to be
applied to this model. The dimensional reduction
of this compactified model gives a worldsheet
model of two dimensional scalar field theory, which
has already been investigated using discrete
light-cone Hamiltonian dynamics (DLCQ) \cite{dalleyk}.
It will certainly be interesting to compare the
results of this work with those of the path history
worldsheet approach described here.

\vskip14pt
\noindent\underline{ Acknowledgments}: 
We are grateful to K. Bardakci, J. Greensite,
M. Peskin, C. Rebbi, S. Shabanov,
and M. Weinstein for valuable discussions. This research
was supported in part by the Department
of Energy under Grant No. DE-FG02-97ER-41029.
\appendix
\section{Spin Sums in Mean Fields}
\subsection{Mean field is linear in $P$}
We turn first to the
case where the mean field is taken linear in $P$. Then after
replacing various monomials of $P$'s in the action by 
their corresponding expectation values,
we are left with the spin sums
\bea
\sum_{s=\pm1}\exp\left\{\sum_{kj}\left[\kappa_k^jP_k^j
+{1-s_k^js_k^{j-1}\over2}\ln{\hat g}\right]\right\}.
\eea
One option is to also replace the $s_k^j=2P_k^j-1$ by $2\phi_k^j-1$
as discussed in the text. But when a simple pattern is assumed
for the $\kappa_k^j$, the sum is just that for the partition
function of a one dimensional Ising model with uniform
magnetic field and can be done exactly.

When $\kappa_k^j\equiv \kappa$ is uniform, the spin sum is just the
$N$th power of the $2\times2$ matrix
\bea
{\cal T}=\pmatrix{e^\kappa&{\hat g}e^\kappa\cr{\hat g}&1\cr}
\eea
whose eigenvalues can be immediately written
down
\bea
t_{\pm}={1+e^\kappa\pm\sqrt{(1-e^\kappa)^2+4{\hat g}^2e^\kappa}\over2}
=e^{\kappa/2}\left(\cosh{\kappa\over2}
\pm\sqrt{\sinh^2{\kappa\over2}+{\hat g}^2}\right)
\eea
Thus there are two branches for the spin contribution to the
energy per site $-\ln t_\pm(\kappa)$, the branch with $t_+$ clearly
being the lowest energy. Altogether the $\kappa$ dependent terms
in the exponent of the Boltzmann factor are $MN[\ln t_+-\phi\kappa]$ 
so the saddle point in $\kappa$ determines its relation to $\phi$
\bea
\phi&=&{\partial\over\partial\kappa}\ln t_+={1\over2}\left[1+{\sinh(\kappa/2)
\over \sqrt{{\hat g}^2+\sinh^2(\kappa/2)}}\right]\label{phiofkappa}\\
\kappa&=&-2\sinh^{-1}{{\hat g}(1-2\phi)\over2\sqrt{\phi(1-\phi)}}
=2\ln\left\{{{\hat g}(2\phi-1)\over2\sqrt{\phi(1-\phi)}}
+\sqrt{1+{{\hat g}^2(1-2\phi)^2\over4{\phi(1-\phi)}}}\right\}.
\eea
The second equation enables expressing the spin energy per site
as a function of $\phi$,
\bea
{\cal E}_s&=&\kappa\phi-\ln t_+\\
&=& \left(2\phi-1\right)\ln\left\{{{\hat g}(2\phi-1)\over2\sqrt{\phi(1-\phi)}}
+\sqrt{1+{{\hat g}^2(1-2\phi)^2\over4{\phi(1-\phi)}}}\right\}
-\ln\left\{{{\hat g}\over2\sqrt{\phi(1-\phi)}}
+\sqrt{1+{{\hat g}^2(1-2\phi)^2\over4{\phi(1-\phi)}}}\right\}
\eea

For the fishnet spin pattern, which repeats only after
four time steps, we have a more complicated 
matrix to diagonalize. To force the $\phi^\prime,\phi,\phi,\phi$
pattern for the $\phi_k^j$'s, we need the $\kappa_k^j$ in the
pattern $\kappa^\prime,\kappa,K,\kappa$ so the transfer
matrix is a power of the 4 step matrix
\bea
{\cal T}_4=\pmatrix{e^\kappa&{\hat g}e^\kappa\cr{\hat g}&1\cr}
\pmatrix{e^K&{\hat g}e^K\cr{\hat g}&1\cr}
\pmatrix{e^\kappa&{\hat g}e^\kappa\cr{\hat g}&1\cr}
\pmatrix{e^{\kappa^\prime}&{\hat g}e^{\kappa^\prime}\cr{\hat g}&1\cr}.
\eea
The eigenvalues of a general $2\times2$ matrix can be written
\bea
t^4_\pm={\Tr{\cal T}\pm\sqrt{(\Tr{\cal T})^2-4\det{\cal T}}\over2}.
\eea
It is straightforward to work out the trace and determinant:
\bea
\det{\cal T}&=&e^{2\kappa+K+\kappa^\prime}({\hat g}^2-1)^4\\
\Tr{\cal T}&=&e^{2\kappa+K+\kappa^\prime}+{\hat g}^2e^{2\kappa+\kappa^\prime}
+2{\hat g}^2e^{\kappa+K+\kappa^\prime}
+{\hat g}^2e^{2\kappa+K}+{\hat g}^4(e^{2\kappa}
+e^{K+\kappa^\prime})\nonumber\\
&&+2{\hat g}^2(e^{K+\kappa}
+e^{\kappa+\kappa^\prime})+2{\hat g}^2e^{\kappa}+{\hat g}^2e^{K}+
{\hat g}^2e^{\kappa^\prime}+1.
\eea
After a bit of algebraic manipulation the eigenvalues can be
written
\bea
t^4_\pm&=&{1\over4}\left[\sqrt{(1+e^{( 2\kappa+K+\kappa^\prime)/2})^2
+2{\hat g}^2e^\kappa(1-e^{(K+\kappa^\prime)/2})^2
+{\hat g}^4(e^{\kappa}+e^{(K+\kappa^\prime)/2})^2
+{\hat g}^2(e^{K}+e^{\kappa^\prime})(1+e^\kappa)^2}\right.
\nonumber\\
&&\hskip-.5cm\left.\pm\sqrt{(1-e^{( 2\kappa+K+\kappa^\prime)/2})^2
+2{\hat g}^2e^\kappa(1+e^{(K+\kappa^\prime)/2})^2
+{\hat g}^4(e^{\kappa}-e^{(K+\kappa^\prime)/2})^2
+{\hat g}^2(e^{K}+e^{\kappa^\prime})(1+e^\kappa)^2}\right]^2
\eea
Again it is clear that $t_+^4$ is the branch with lower energy.
The exponent of the Boltzmann factor is now $MN(\ln t_+^4 - \phi(K+2\kappa)
-\phi^\prime\kappa^\prime)/4$, so we have
\bea
\phi^\prime&=&\langle P^\prime \rangle=\partial_{\kappa^\prime}\ln t_+^4
={\partial_{\kappa^\prime}\Tr{\cal T}-t_-^4
\over\sqrt{(\Tr{\cal T})^2-4\det{\cal T}}}
={1\over2}+{\partial_{\kappa^\prime}\Tr{\cal T}-\Tr{\cal T}/2
\over\sqrt{(\Tr{\cal T})^2-4\det{\cal T}}}
\nonumber\\
&=&{e^{\kappa^\prime}\left(e^{2\kappa+K}+{\hat g}^2e^{2\kappa}
+2{\hat g}^2e^{\kappa+K}
+{\hat g}^4e^{K}+2{\hat g}^2e^{\kappa}
+{\hat g}^2\right)-t_-^4
\over\sqrt{(\Tr{\cal T})^2-4\det{\cal T}}}\\
\phi&=&\langle P\rangle
={1\over2}\partial_\kappa\ln t_+^4=\partial_K\ln t_+^4 \\
&=&{e^{K}\left(e^{2\kappa+\kappa^\prime}+{\hat g}^2e^{2\kappa}
+2{\hat g}^2e^{\kappa+\kappa^\prime}
+{\hat g}^4e^{\kappa^\prime}+2{\hat g}^2e^{\kappa}
+{\hat g}^2\right)-t_-^4
\over\sqrt{(\Tr{\cal T})^2-4\det{\cal T}}}
\eea
The equality of the two expressions for $\phi$ determines $K$:
\bea
e^K=e^\kappa{e^{\kappa+\kappa^\prime}+{\hat g}^2e^\kappa
+1+e^{\kappa^\prime}\over e^{\kappa+\kappa^\prime}
+{\hat g}^2e^{\kappa^\prime}+1+e^{\kappa}}
\eea

We also list the expectations of the higher monomials
in $P$ that occur in the action. Let 
$(\kappa_1,\kappa_2,\kappa_3,\kappa_4)$ denote 
$(\kappa,K,\kappa,\kappa^\prime)$ in a general cyclic order.
Then, in this notation
\bea
\Tr{\cal T}
&=&e^{\kappa_1+\kappa_2+\kappa_3+\kappa_4}+1+{\hat g}^2\left[e^{\kappa_1+\kappa_2+\kappa_3}
+e^{\kappa_1+\kappa_3+\kappa_4}+e^{\kappa_1+\kappa_2+\kappa_4}
+e^{\kappa_1+\kappa_2}+e^{\kappa_1+\kappa_4}
+e^{\kappa_1}\right.\nonumber\\&&\left.
+e^{\kappa_2+\kappa_3+\kappa_4}+e^{\kappa_2+\kappa_3}
+e^{\kappa_3+\kappa_4}+e^{\kappa_2}+e^{\kappa_3}+e^{\kappa_4}\right]
+{\hat g}^4(e^{\kappa_1+\kappa_3}+e^{\kappa_2+\kappa_4})\\
\partial_{\kappa_1}\Tr{\cal T}
&=&e^{\kappa_1+\kappa_2+\kappa_3+\kappa_4}
+{\hat g}^2\left[e^{\kappa_1+\kappa_2+\kappa_3}
+e^{\kappa_1+\kappa_3+\kappa_4}+e^{\kappa_1+\kappa_2+\kappa_4}
+e^{\kappa_1+\kappa_2}+e^{\kappa_1+\kappa_4}
+e^{\kappa_1}\right]\nonumber\\
&&\hskip2cm+{\hat g}^4e^{\kappa_1+\kappa_3}\\
{\hat g}Y&\equiv&\Tr\pmatrix{0&1\cr0&0\cr}{\cal T}\nonumber\\
&=&{\hat g}\left[e^{\kappa_2+\kappa_3+\kappa_4}+
e^{\kappa_3+\kappa_4}+e^{\kappa_4}+1
+{\hat g}^2(e^{\kappa_2+\kappa_4}+e^{\kappa_2+\kappa_3}
+e^{\kappa_2}+e^{\kappa_3})\right].
\eea
Then we obtain for $P$'s on consecutive temporal sites:
\bea
\langle P^1\rangle&=&
{1\over2}+{\partial_{\kappa_1}\Tr{\cal T}-\Tr{\cal T}/2
\over\sqrt{(\Tr{\cal T})^2-4\det{\cal T}}}\\
\langle P^1P^2\rangle&=&{e^{\kappa_1+\kappa_2}\over t_+^4}
\left[\left(e^{\kappa_3+\kappa_4}+{\hat g}^2(e^{\kappa_3}+e^{\kappa_4}
+1)\right)\langle P^1\rangle
+{{{\hat g}\left(e^{\kappa_3+\kappa_4}+{\hat g}^2e^{\kappa_4}+e^{\kappa_3}
+1\right)Y}\over\sqrt{(\Tr{\cal T})^2-4\det{\cal T}}}\right]\nonumber\\
&=&{e^{\kappa_1+\kappa_2}\over t_+^4}
\left[{1\over2}\left(e^{\kappa_3+\kappa_4}+{\hat g}^2(e^{\kappa_3}+e^{\kappa_4}
+1)\right)
+{N_1\over\sqrt{(\Tr{\cal T})^2-4\det{\cal T}}}\right]\\
N_1&=&-{1\over2}\left(e^{\kappa_3+\kappa_4}
+{\hat g}^2(e^{\kappa_3}+e^{\kappa_4}
+1)\right)\Tr{\cal T}
+e^{\kappa_1+\kappa_2+\kappa_3+\kappa_4}
(1+e^{\kappa_3+\kappa_4})\nonumber\\
&&+{\hat g}^2\left[e^{\kappa_1+\kappa_2}
(1+e^{\kappa_3+\kappa_4})(1+e^{\kappa_3}+e^{\kappa_4})
+e^{\kappa_1+\kappa_2+\kappa_3+\kappa_4}(e^{\kappa_3}+e^{\kappa_4})
\right.\nonumber\\
&&\left.
+(1+e^{\kappa_3+\kappa_4}+e^{\kappa_3})(1
+e^{\kappa_4}+e^{\kappa_3+\kappa_4})\right.\nonumber\\
&&\left.+e^{\kappa_3+\kappa_4}(e^{\kappa_1+\kappa_4}+e^{\kappa_2+\kappa_3})
+e^{\kappa_3+\kappa_4}(1+e^{\kappa_3+\kappa_4})(e^{\kappa_1}+e^{\kappa_2})
\right]\nonumber\\
&&+{\hat g}^4\left[e^{\kappa_1+\kappa_2}
(e^{\kappa_3}+e^{\kappa_4})(1+e^{\kappa_3}+e^{\kappa_4})
+e^{\kappa_1+2\kappa_3+\kappa_4}\right.\nonumber\\
&&\left.+(1+e^{\kappa_3}+e^{\kappa_4})
(e^{\kappa_1+\kappa_3+\kappa_4}
+e^{\kappa_1+\kappa_4}+e^{\kappa_1})
+e^{\kappa_4}(1+e^{\kappa_2+\kappa_3+\kappa_4}
+e^{\kappa_4}+e^{\kappa_3+\kappa_4})\right.\nonumber\\
&&\left.
+(1+e^{\kappa_3+\kappa_4}+e^{\kappa_3})
(e^{\kappa_2+\kappa_4}+e^{\kappa_2+\kappa_3}+e^{\kappa_2}+e^{\kappa_3})
\right]\nonumber\\
&&+{\hat g}^6\left[(e^{\kappa_1+\kappa_3}+e^{\kappa_2+\kappa_4})
(1+e^{\kappa_3}+e^{\kappa_4})
+e^{\kappa_3+\kappa_4}
\right]
\\
\langle P^1P^2P^3\rangle&=&{e^{\kappa_1+\kappa_2+\kappa_3}\over t_+^4}
\left[\left({\hat g}^2+e^{\kappa_4}\right)\langle P^1\rangle
+{{\hat g}\left(e^{\kappa_4}+1\right)Y
\over\sqrt{(\Tr{\cal T})^2-4\det{\cal T}}}\right]\nonumber\\
&=&{e^{\kappa_1+\kappa_2+\kappa_3}\over t_+^4}
\left[{1\over2}\left({\hat g}^2+e^{\kappa_4}\right)
+{N_2
\over\sqrt{(\Tr{\cal T})^2-4\det{\cal T}}}\right]\\
N_2&=&-{1\over2}\left({\hat g}^2+e^{\kappa_4}\right)\Tr{\cal T} 
+e^{\kappa_1+\kappa_2+\kappa_3+\kappa_4}
+{\hat g}^6e^{\kappa_1+\kappa_3+\kappa_4}\nonumber\\
&&+{\hat g}^2\left[2e^{\kappa_1+\kappa_2+\kappa_3+\kappa_4}
+e^{\kappa_1+\kappa_3+\kappa_4}
+(e^{\kappa_1}+e^{\kappa_3})
(e^{\kappa_2+\kappa_4}
+e^{\kappa_2+2\kappa_4}
+e^{2\kappa_4}+e^{\kappa_4})\right]\nonumber\\
&&+{\hat g}^4\left[e^{\kappa_2}(1+e^{\kappa_4})^2
+(1+e^{\kappa_4})(1+e^{\kappa_2})(e^{\kappa_1}+e^{\kappa_3})
+e^{\kappa_1+\kappa_3}(e^{\kappa_2}+2e^{\kappa_4})\right]
\eea
By inspecting $N_1$ and $N_2$, we see that 
the double $P$ expectations $\langle P^1P^2\rangle$
are symmetric under $(\kappa_1,\kappa_3)\leftrightarrow(\kappa_2,\kappa_4)$.
This means that the fishnet patterns $+++-$ and $++-+$ give
identical values, \vpp, of $\langle P^1P^2\rangle$, and
similarly the patterns $+-++$ and $-+++$ give
identical values \vmp. The triple
$P$ expectations have the symmetry $\kappa_1\leftrightarrow\kappa_3$
which implies identical values, \vmpp, for $-+++$ and $++-+$.
But the triple expectations for $+-++$ and $+++-$ are
unrelated in general, and we denote
them by \vpmp and \vppp
respectively. Thus for the fishnet spin pattern
$\langle P^1P^2\rangle$ assumes two distinct values and
$\langle P^1P^2P^3\rangle$ assumes three distinct values.

For $\kappa^\prime=\kappa$, it immediately follows that $K=\kappa$
and $\phi^\prime=\phi$. Therefore this case should reduce to the
uniform mean field results. Indeed, one finds $t_+^4=(t_+)^4$
as expected. Then it is straightforward to show that the
expression for $\phi$ reduces to (\ref{phiofkappa}), and the
expressions for the higher correlators become
\bea
\langle P^1P^2\rangle&\to&{e^{\kappa}\over t_+}{t_+-1\over t_+-t_-}
=\phi{e^{\kappa}\over t_+}\\
\langle P^1P^2P^3\rangle&\to&
{e^{2\kappa}\over t_+^2}{t_+-1\over t_+-t_-}
=\phi\left({e^{\kappa}\over t_+}\right)^2.
\eea

As discussed in the text, the $\epsilon\to0$ limit is
smooth if one of the mean fields also goes to zero. For the
fishnet pattern we achieve this by putting $\phi^\prime=\epsilon f$
and taking the limit at fixed $f,\phi$. We therefore
need to specialize the spin contribution to the energy per
site to this case. We note that $\phi^\prime\to0$ requires
$\kappa^\prime\to-\infty$. Holding $\phi$ away from its
endpoints $0,1$, we find in this limit
\bea
e^K&\to&e^\kappa{1+{\hat g}^2e^\kappa\over 1+e^\kappa }\\
\phi&\to&{{\hat g}^2e^\kappa(1+e^\kappa)
\over1+2{\hat g}^2e^\kappa+{\hat g}^2e^{2\kappa}}\\
t_+^4&\to&\Tr{\cal T}=(1+{\hat g}^2e^\kappa)[1+2{\hat g}^2e^\kappa
+{\hat g}^2e^{2\kappa}]\\
\vpp&\to&{\phi e^\kappa\over1+e^\kappa}\\
\vppp&\to&{\phi e^{2\kappa}\over(1+e^\kappa)^2}\\
\phi^\prime&\to& {\hat g}^2e^{\kappa^\prime}
{1+2(1+{\hat g}^2)e^\kappa+(2+(1+{\hat g}^2)^2)e^{2\kappa}
+2(1+{\hat g}^2)e^{3\kappa}+e^{4\kappa}
\over(1+2{\hat g}^2e^\kappa+{\hat g}^2 e^{2\kappa})^2}\\
\vmp&\to& e^{\kappa^\prime}\phi{1+e^\kappa+{\hat g}^2e^\kappa+e^{2\kappa}
\over1+2{\hat g}^2e^\kappa+{\hat g}^2 e^{2\kappa}}\\
\vmpp&\to&e^{\kappa^\prime}{\phi e^\kappa\over1+e^\kappa}
{1+e^\kappa+{\hat g}^2e^\kappa+e^{2\kappa}
\over1+2{\hat g}^2e^\kappa+{\hat g}^2 e^{2\kappa}}\\
\vpmp&\to&e^{\kappa^\prime}{\phi e^\kappa(1+e^\kappa)
\over1+2{\hat g}^2e^\kappa+{\hat g}^2 e^{2\kappa}}
\eea
Notice that the quantity \vmp, which figures in the
effective string tension is proportional to $\phi^\prime$
in this limit:
\bea
\vmp\to \phi^\prime{e^\kappa(1+e^\kappa)
(1+e^\kappa+{\hat g}^2e^\kappa+e^{2\kappa})\over
1+2(1+{\hat g}^2)e^\kappa+(2+(1+{\hat g}^2)^2)e^{2\kappa}
+2(1+{\hat g}^2)e^{3\kappa}+e^{4\kappa}}\equiv\phi^\prime G .
\eea
For visualization purposes we display graphs of the 
quantities $\langle{\scriptstyle++}\rangle$,
$\langle{\scriptstyle+++}\rangle$, and $G$ in
Fig.~\ref{correlators}.
\begin{figure}[htb]
\psfrag{g}[l][l][.5]{${\hat g}$}
\includegraphics[width=8cm]{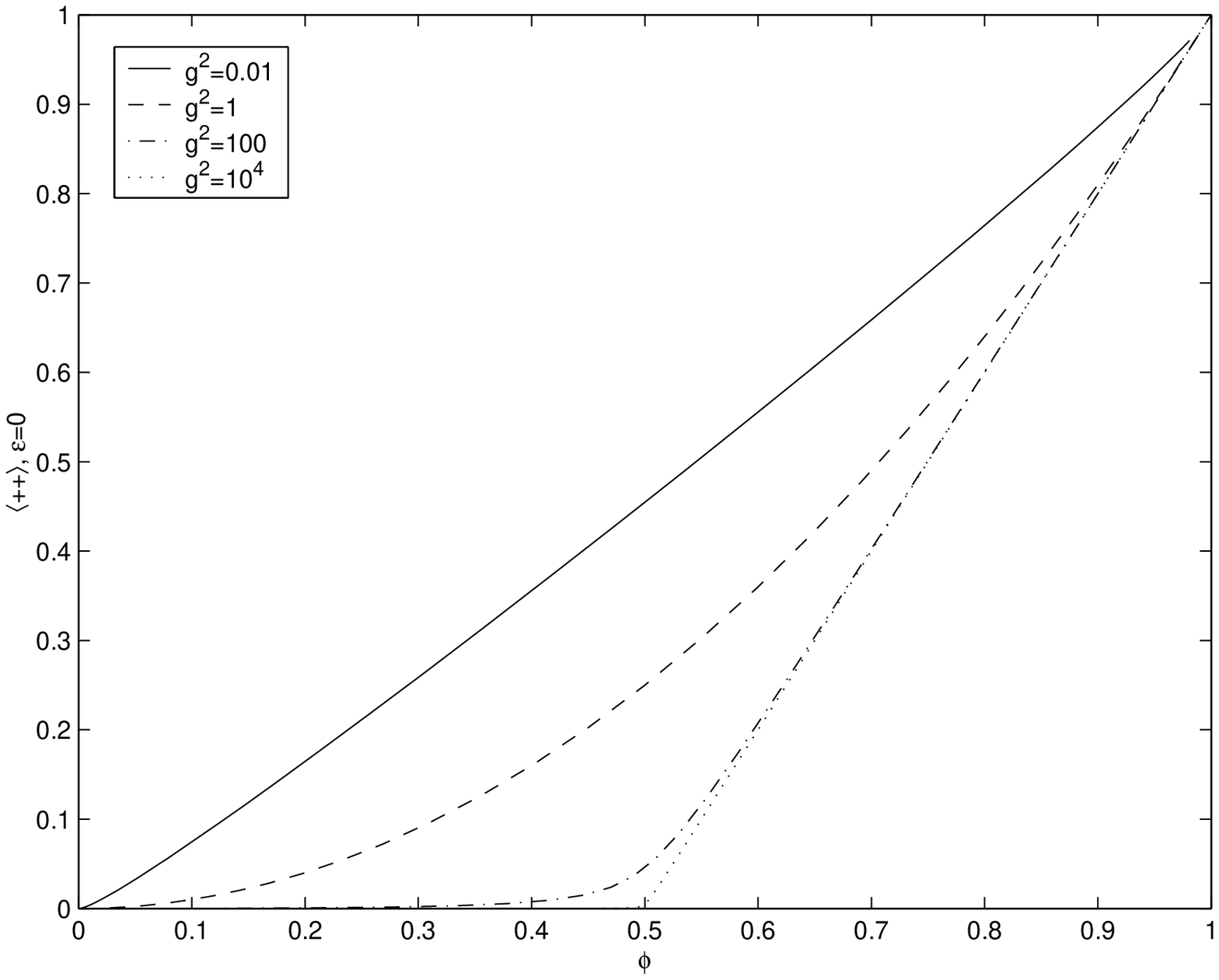}
\includegraphics[width=8cm]{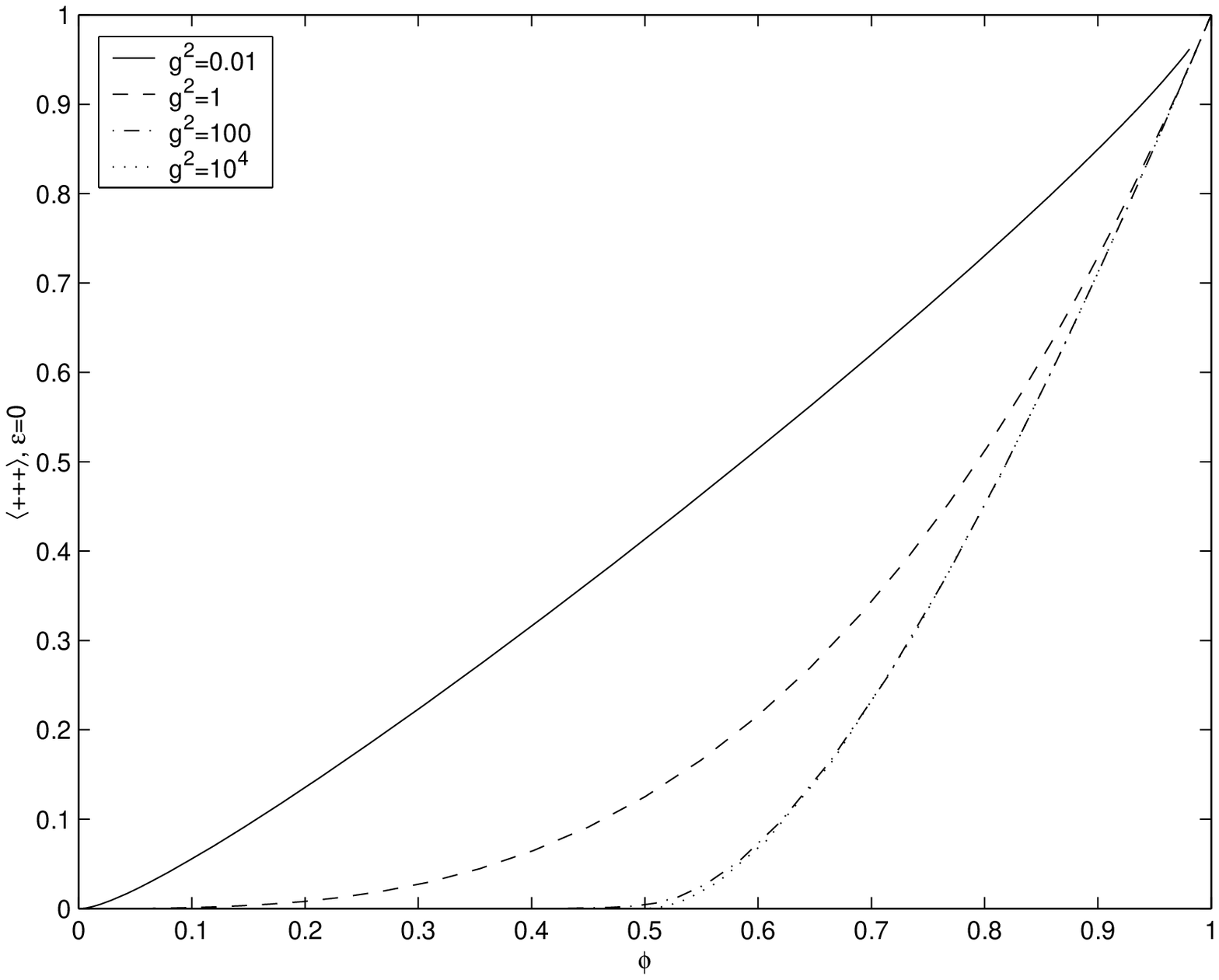}
\centerline{\includegraphics[width=8cm]{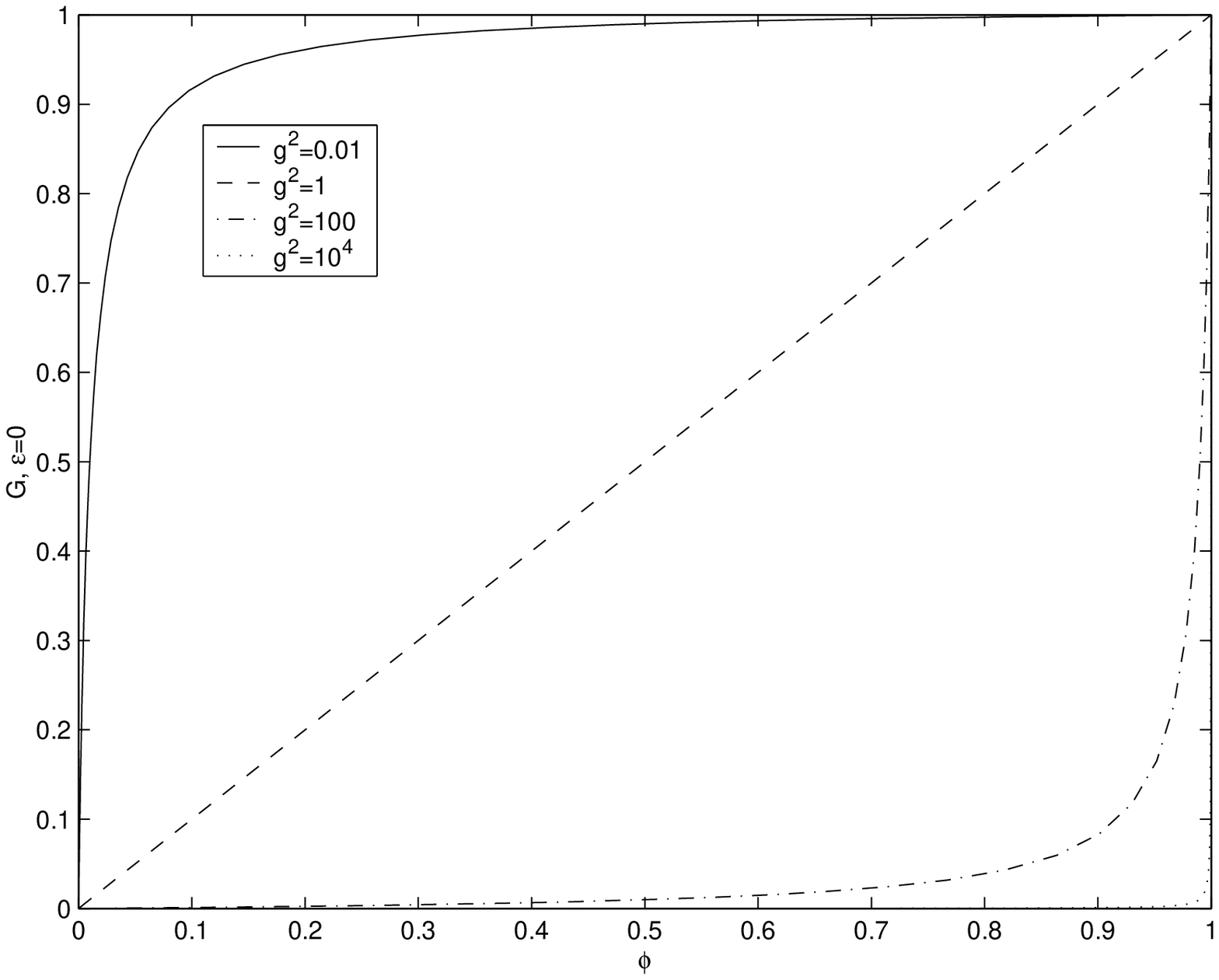}}
\caption{The quantities $\langle{\scriptstyle++}\rangle$,
$\langle{\scriptstyle+++}\rangle$, and $G$ for $\epsilon=0$ and
various ${\hat g}^2$. Note that these quantities are $\phi^2,\phi^3,\phi$
respectively for ${\hat g}^2=1$.}
\label{correlators}
\end{figure}

Moreover, we see that in this case $e^\kappa$ can be eliminated
in favor of $\phi$ by solving a quadratic equation:
\bea
e^{\kappa(\phi)}={\sqrt{(1-2\phi)^2+4\phi(1-\phi)/{\hat g}^2}
-1+2\phi\over2(1-\phi)}.
\eea
Then the spin energy per site,
\bea
{\cal E}_s&=&{1\over4}\left[\phi^\prime\kappa^\prime
+2\kappa\phi+K\phi-\ln t_+^4\right]\\
&\to&{1\over4}\left[
(2\kappa+K)\phi-\ln (1+{\hat g}^2e^\kappa)
-\ln[1+{\hat g}^2e^\kappa+{\hat g}^2e^\kappa(1+e^\kappa)]\right]
\nonumber\\
&=&{1\over4}\left[(1-\phi)\ln(1-\phi)+\phi\ln\phi-\phi\ln{\hat g}^2\right]
\nonumber\\
&&+\phi\ln\left\{\sqrt{1
+{\hat g}^2{(1-2\phi)^2\over4\phi(1-\phi)}}
-{\hat g}{1-2\phi\over2\sqrt{\phi(1-\phi)}}\right\}
-{1\over2}\ln (1+{\hat g}^2e^{\kappa(\phi)})
\eea
can be expressed as an explicit function of $\phi$.
Note that the term $\phi^\prime\kappa^\prime=O(\epsilon\ln\epsilon)$
and therefore drops out for $\epsilon\to0$.
\subsection{Mean field is quadratic in $P$}

We now turn our attention to the case where the mean field is taken
quadratic in $P$. Like the previous subsection, we replace the $P$, $PP$'s
in the action by corresponding $\f$ and $\varphi$, then we are left
with the spin sums
\bea
\sum_{s_i^j=\pm}\exp\left\{\sum_{k,j}\left[\chi_k^jP_k^jP^{j-1}_k
  + \fr{1 - s_k^js_k^{j-1}}{2}\ln\hg\right]\right\}
\eea
For the uniform case in which $\chi_k^j=\chi$, the spin sum as
before is just matrix multiplication of the $N$ $2\times 2$ matrix
\bea
\pmatrix{e^{\chi}&\hg\cr \hg&1\cr}
\eea
which has eigenvalues
\bea
t_\pm ={1+e^{\chi}\pm\sqrt{(1-e^{\chi})^2+4\hg^2}
\over2}
\eea
For large $N$, $t_+$ is the largest eigenvalue of the
transfer matrix. For our purpose, we need to calculate expectation
values of single, double and triple $P$. After some algebraic
manipulation, one gets
\bea
\f &=& \langle P\rangle = 
\fr{\rme^{\chi} - 1 + \sqrt{(1-\rme^{\chi})^2 + 4\hg^2}}
{2\sqrt{(1 - \rme^{\chi})^2 + 4\hg^2}}\label{uni-phi}\\
\varphi &=& \langle PP^\prime\rangle = \fr{
\rme^{\chi}(\rme^{\chi} - 1 + \sqrt{(1-\rme^{\chi})^2 + 4\hg^2})}
{(1+e^{\chi}+\sqrt{(1-e^{\chi})^2+
4\hg^2})\sqrt{(1-\rme^{\chi})^2+ 4\hg^2}}\label{uni-varphi}\\
\langle PP^\prime P^{\prime\prime}\rangle &=&
\fr{2\rme^{2\chi}(\rme^{\chi} - 1 + \sqrt{(1-\rme^{\chi})^2 + 4\hg^2})}
{(1+e^{\chi}+\sqrt{(1-e^{\chi})^2+
4\hg^2})^2\sqrt{(1-\rme^{\chi})^2 + 4\hg^2}}
\label{uni-ppp}
\eea
For the fishnet pattern we have to evaluate the spin sums by diagonalizing the
4 time-step transfer matrix
\bea
T_4=\pmatrix{e^{\chi^\prime}&\hg\cr \hg&1\cr}
\pmatrix{e^{\chi}&\hg\cr \hg&1\cr}
\pmatrix{e^{\chi}&\hg\cr \hg&1\cr}
\pmatrix{e^{\chi^\prime}&\hg\cr \hg&1\cr}.
\eea
To present the results of this analysis in a reasonably compact way it
is useful to rewrite $T_4$ using Pauli matrices, i.e.
\bea
\pmatrix{e^{\chi}&\hat{g}\cr \hat{g} &1\cr}\equiv
{1+e^{\chi}\over2}(I+\boldsymbol{u}\cdot\sigma)
\eea
and a similar definition for $\chi_i^\prime$.
Then
\bea
T_4=\left({1+e^{\chi}\over2}\right)^2
\left({1+e^{\chi^\prime}\over2}\right)^2
(I + \boldsymbol{u}^\prime\cdot\boldsymbol{\sigma})(I +
\boldsymbol{u}\cdot\boldsymbol{\sigma})(I + \boldsymbol{u}\cdot
\boldsymbol{\sigma})
(I + \boldsymbol{u}^\prime\cdot\boldsymbol{\sigma})
\label{transfermatrix-2pcase}
\eea
One can explicitly calculate $\boldsymbol{u}$ and
$\boldsymbol{u}^\prime$
\bea
&&\boldsymbol{u} =\left(\fr{2\hg}{1+\rme^{\chi}},\ 0,\ 
\fr{\rme^{\chi}-1}{\rme^{\chi}+1}\right)\;,\;\;\;
\boldsymbol{u}^\prime =\left(\fr{2\hg}{1+\rme^{\chi^\prime}},\ 0,\ 
\fr{\rme^{\chi^\prime}-1}{\rme^{\chi^\prime}+1}\right)\nn\\ 
&& \boldsymbol{u}^2 = \fr{(1-\rme^{\chi})^2 +
  4\hg^2}{(1+\rme^{\chi})^2}\;,\;\;
\boldsymbol{u}^{\prime\;2} = \fr{(1-\rme^{\chi^\prime})^2 +
4\hg^2}{(1+\rme^{\chi^\prime})^2}
\eea
It is straightforward to show that the eigenvalues of $T_4$
are
\bea
t^4_\pm&=&\left({1+e^{\chi}\over2}\right)^2
\left({1+e^{\chi^\prime}\over2}\right)^2
\left(|\boldsymbol{u}^\prime + \boldsymbol{u}|\pm
\sqrt{1 + \boldsymbol{u}^2\boldsymbol{u}^{\prime2} +
2\boldsymbol{u}\cdot\boldsymbol{u}^\prime
}\right)^2\label{eigen-tpm-2pcase}\\
&=&{1\over4}\left(
\sqrt{\hg^{2}(2+e^{\chi}+e^{\chi^\prime})^2+
(1-e^{\chi^\prime+\chi})^2}\right.\nonumber\\
&&\qquad\qquad\left. \pm\sqrt{\hg^{2}(e^{\chi}+e^{\chi^\prime})^2
+4\hg^{2}(1+\hg^{2})+
(1+e^{\chi+\chi^\prime})^2}\right)^2
\eea
To simplify our formulas we introduce some notations:
\bea
\boldsymbol{\cA} = \boldsymbol{u} + \boldsymbol{u}^\prime,\;\;
\cB = 1 + \boldsymbol{u}^2 \boldsymbol{u}^{\prime\;2} + 
2\boldsymbol{u}\cdot\boldsymbol{u}^\prime,\;\;
\boldsymbol{\cG} = (1-\boldsymbol{u}^{\prime\;2}) \boldsymbol{\cA}
+ \boldsymbol{\cA}^2 \boldsymbol{u}^\prime,\;\;
\boldsymbol{\wtd{\cG}} = (1-\boldsymbol{u}^2) \boldsymbol{\cA}
+ \boldsymbol{\cA}^2 \boldsymbol{u}
\label{notations-2pcase}
\eea
With these notations, we can rewrite the transfer matrix as follows
\bea
T_4 = \left({1+e^{\chi}\over2}\right)^2
\left({1+e^{\chi^\prime}\over2}\right)^2 
\left(\begin{array}{cc}
\boldsymbol{\cA}^2 + \cB + 2 \cG_3 & 2\cG_1\\
2 \cG_1 & \boldsymbol{\cA}^2 + \cB - 2 \cG_3
\end{array}\right) 
\eea
and the eigenvalues can be written as
\bea
t_\pm = \left({1+e^{\chi}\over2}\right)^2
\left({1+e^{\chi^\prime}\over2}\right)^2 \left(\boldsymbol{\cA}^2
+ \cB \pm 2 \sqrt{\boldsymbol{\cG}.\boldsymbol{\cG}} \right)
\eea
Notice that $\boldsymbol{\cA}$, $\cB$ and $t_\pm$ are symmetric under
$\chi\to\chi^\prime$ while
$\boldsymbol{\cG}\to\boldsymbol{\wtd{\cG}}$. However, one can easily
show that $\boldsymbol{\cG}\cdot\boldsymbol{\cG} =
\boldsymbol{\wtd{\cG}}\cdot\boldsymbol{\wtd{\cG}} = \boldsymbol{\cA}^2\cB$
 is symmetric under $\chi\to\chi^\prime$.
For our problem, we need to calculate correlation function for
$\langle P\rangle$, $\langle PP^\prime\rangle$ and 
$\langle PP^\prime P^{\prime\prime}\rangle$. The results
are
\bea
\f^\prime &=& \fr{\cG_3 + 
\sqrt{\boldsymbol{\cG}\cdot\boldsymbol{\cG}}}
{2\sqrt{\boldsymbol{\cG}\cdot\boldsymbol{\cG}}},\;\;\;
\f = \fr{\wtd{\cG}_3 + \sqrt{\boldsymbol{\cG}\cdot\boldsymbol{\cG}}}
{2\sqrt{\boldsymbol{\cG}\cdot\boldsymbol{\cG}}},\;\;\;
\wtd{\f}=
\fr{\cG_3 + \sqrt{\boldsymbol{\cG}\cdot\boldsymbol{\cG}}}
{2\sqrt{\boldsymbol{\cG}\cdot\boldsymbol{\cG}}} +
\fr{\hg\,\cG_1(\rme^{\chi^\prime}-1) - 2\hg^2\,\cG_3}
{(\hg^2-\rme^{\chi^\prime})
\sqrt{\boldsymbol{\cG}\cdot\boldsymbol{\cG}}}\label{phitilde-gen-2pcase}\\
\langle ++\rangle = \varphi &=& 
\fr{\rme^{\chi}(\hg\,\wtd{\cG}_1 - \wtd{\cG}_3 -
\sqrt{\boldsymbol{\cG}\cdot\boldsymbol{\cG}})}{2(\hg^2 -
  \rme^{\chi})\sqrt{\boldsymbol{\cG}\cdot\boldsymbol{\cG}}},\;\;\;
\langle -+\rangle = \varphi^\prime = 
\fr{\rme^{\chi^\prime}(\hg\, \cG_1 - \cG_3 -
\sqrt{\boldsymbol{\cG}\cdot\boldsymbol{\cG}})}{2(\hg^2 -
  \rme^{\chi^\prime})\sqrt{\boldsymbol{\cG}\cdot\boldsymbol{\cG}}}
\label{varphi-gen-2pcase} \\
\vmpp &=& 
\fr{8\rme^{\chi+\chi^\prime}[(\hg^2 + \rme^{\chi+\chi^\prime})(\cG_3 +
\sqrt{\boldsymbol{\cG}\cdot\boldsymbol{\cG}}) + \hg\cG_1 
(1+\rme^{\chi^\prime})]}{(\boldsymbol{\cA}^2+\cB + 
2\sqrt{\boldsymbol{\cG}\cdot\boldsymbol{\cG}})
\sqrt{\boldsymbol{\cG}\cdot\boldsymbol{\cG}}
(1+\rme^{\chi})^2(1+\rme^{\chi^\prime})^2}\label{mpp-2pcase}\\ 
\vppp &=& 
\fr{8\rme^{2\chi}[(\hg^2+\rme^{2\chi^\prime})
\sqrt{\boldsymbol{\cG}\cdot\boldsymbol{\cG}} - (\hg^2 - \rme^{2\chi})
\cG_3 + 2\hg\cG_1\rme^{\chi^\prime}]}{(1+\rme^{\chi})^2(1+
\rme^{\chi^\prime})^2(\boldsymbol{\cA}^2 + \cB
  + 2 \sqrt{\boldsymbol{\cG}\cdot\boldsymbol{\cG}})
\sqrt{\boldsymbol{\cG}\cdot\boldsymbol{\cG}}}\label{ppp-2pcase}\\
\vpmp &=& 
\fr{8\rme^{2\chi^\prime}[(\hg^2+\rme^{2\chi})
\sqrt{\boldsymbol{\cG}\cdot\boldsymbol{\cG}} - (\hg^2 - \rme^{2\chi^\prime})
\wtd{\cG}_3 + 2\hg\wtd{\cG}_1\rme^{\chi}]}{(1+\rme^{\chi})^2(1+
\rme^{\chi^\prime})^2(\boldsymbol{\cA}^2 + \cB
  + 2 \sqrt{\boldsymbol{\cG}\cdot\boldsymbol{\cG}})
\sqrt{\boldsymbol{\cG}\cdot\boldsymbol{\cG}}}\label{pmp-2pcase}
\eea
For large $M,N$ the spin energy per site is
\bea
{\cal E}_s(\chi,\chi^\prime) = -{1\over4}(\ln t^4_+(\chi,\chi^\prime)
- 2\varphi\chi - 2\varphi^\prime\chi^\prime) + 
\frac{d}{8}\ln(1+\rho)(\phi+\phi^\prime + 2\wtd{\phi})
\label{spinenergy-2pcase}
\eea
One can easily show that when $\chi=\chi^\prime$ (uniform case)
\bea
\f &=& \f^\prime = \wtd{\f} \to
\fr{e^\chi -1 + \sqrt{(e^\chi - 1)^2 + 4\hg^2}}
{2\sqrt{(e^\chi - 1)^2 + 4\hg^2}}\label{phi-uniform-2pcase}\\
\varphi &=& \varphi^\prime \to e^\chi\fr{\left[
e^\chi -1 -2\hg^2 + \sqrt{(e^\chi-1)^2 + 4\hg^2}\right]}{
2(e^\chi-\hg^2)\sqrt{(e^\chi-1)^2 + 4\hg^2}}\label{varphi-uniform-2pcase}\\
\vmpp &\to& \fr{2\rme^{2\chi}(\rme^{\chi} - 1 +
  \sqrt{(1-\rme^{\chi})^2 + 4\hg^2})}{(1+e^{\chi}+\sqrt{(1-e^{\chi})^2+
4\hg^2})^2\sqrt{(1-\rme^{\chi})^2 + 4\hg^2}}\label{ppp-uniform-2pcase}\\
\vpmp &=& \vppp = \vmpp
\label{expectations-chiechip}
\eea
as expected in Eqs.~(\ref{uni-phi}), (\ref{uni-varphi}) and (\ref{uni-ppp}).
In this
special case, the spin energy can be written as
\bea
\cE_s = \varphi\chi- \ln\fr{1+e^{\chi} + \sqrt{(1-e^{\chi})^2 + 4\hg^2}}{2} +
\frac{d}{2}\f\ln(1+\rho).
\label{uniformspinen-2pcase}
\eea
Another special case is $\varphi^\prime = 0$ or 
$\chi^\prime\rightarrow -\infty$. In this case
\bea
t_\pm &\to& \fr14\left(\sqrt{\hg^2(2+e^\chi)^2+1} \pm
\sqrt{(1+2\hg^2)^2 + \hg^2 e^{2\chi}}\right)^2,\label{tpm-vpp0-2pcase}\\ 
t_+-t_- &=& \sqrt{\left[1 + \left( 2 + e^\chi \right)^2\,\hg^2\right]
\left[(1 + 2\hg^2)^2+ \hg^2\,e^{2\chi}\right]},\nn\\
\f &\to&\frac{-1 + \left(-3 + (1+e^\chi)^2\right) \,\hg^2 +
t_+-t_-}{2(t_+-t_-)},\label{phiprime-vpp0-2pcase}\\
\f^\prime &\to& \frac{-1 - \left(2 + e^\chi\,\left( 2 + 
e^\chi \right)  \right) \,\hg^2 + t_+-t_-}{2(t_+-t_-)},
\label{phi-vpp0-2pcase}\\
\wtd{\f} &\to& \frac{-1 + \left(-2 + e^{2\,\chi} \right) \,\hg^2 +
t_+-t_-}{2(t_+-t_-)},
\label{phitilde-vpp0-2pcase}\\
\varphi &\to&\frac{e^\chi\,\hg^2\left[t_+\left(e^\chi + 1\right) +
        \hg^2 - e^\chi\right]}{t_+(t_+-t_-)}
\label{varphi-vpp0-2pcase}
\eea
In this limit, $\vmpp = \vpmp = 0$ and $\vppp$ is
\bea
\vppp =\frac{e^{2\chi}\hg^2\left[ \hg^2 + (1 + e^\chi)^2 \hg^2 +
t_+-t_-\right]}
{(t_+-t_-)\left[1 + 2\hg^4(t_+-t_-) + \hg^2 (e^\chi+2)^2 (t_+-t_-)\right]}
\label{ppp-vpp0-2pcase}
\eea
In this case, the spin energy is
\bea
\cE_s = -\fr12\ln\left[\fr{\sqrt{\hg^2(2+e^\chi)^2+1}+
\sqrt{(1+2\hg^2)^2 + \hg^2 e^{2\chi}}}{2}
\right] + \frac{1}{2}\vp\chi 
+ \frac{d}{8}(\phi + \phi^\prime + 2\wtd{\phi})\ln(1 + \rho)
\label{spinen-chipe0-2pcase}
\eea

\section{The coefficients}
\subsection{$\f_i^j = \langle P_i^j\rangle$}
The Gaussian integrals over ghost and $q$ fields for the fishnet
spin pattern requires the extraction of the coefficients of the
various bilinear forms. Let us write the ghost action for a single
$b,c$ pair on time slice $j$ as
\bea
S^j_g&=&\alpha^j\sum_kb_kc_k+\beta^j\sum_k(-)^kb_kc_k
+\gamma^j\sum_k(b_{k+1}-b_k)(c_{k+1}-c_k)\nonumber\\
&&+\delta^j\sum_k(-)^k(b_{k+1}-b_k)(c_{k+1}-c_k).
\eea
Then for four consecutive time slices (and periodic boundary
conditions) we read off:
\bea
\alpha^1&=&\alpha^3={1\over2}\bigg[
\left({1\over\epsilon}+1+\rho(1-\phi^\prime)-2\eta(1-\phi^\prime)
-\eta\phi\right){\vpp}-\vppp-\vpmp
+2(\phi+\phi^\prime)-4\phi\phi^\prime\nonumber\\
&&\hskip2cm +
\left({1\over\epsilon}+1+\rho(1-\phi)-2\eta(1-\phi)
-\eta\phi^\prime\right){\vmp} +2\eta(1-\phi)\vpmp
\nonumber\\&&\hskip2cm
+2\eta(1-\phi^\prime)\vppp+\eta(\phi+\phi^\prime)\vmpp\bigg]\\
\beta^1&=&-\beta^3={1\over2}\bigg[
\left(-{1\over\epsilon}-1+\rho(1-\phi^\prime)-2\eta(1-\phi^\prime)
-\eta\phi\right){\vpp}+\vppp-\vpmp
+2(\phi-\phi^\prime)\nonumber\\
&&\hskip2cm +
\left({1\over\epsilon}+1-\rho(1-\phi)+2\eta(1-\phi)
+\eta\phi^\prime\right){\vmp} -2\eta(1-\phi)\vpmp
\nonumber\\&&\hskip2cm
+2\eta(1-\phi^\prime)\vppp+\eta(\phi-\phi^\prime)\vmpp\bigg]\\
\gamma^1&=&\gamma^3=(1-\phi)(1-\phi^\prime)
-{1\over2}\eta(1-\phi^\prime)(\vmp-\vmpp)
-{1\over2}\eta(1-\phi)(\vpp-\vmpp)\\
\delta^1&=&-\delta^3=\eta(1-\phi^\prime)\vmp-\eta(1-\phi)\vpp
-(\phi-\phi^\prime)\vmpp\\
\alpha^2&=&\alpha^4={1\over2}\bigg[({\vpp+\vmp})
\left({1\over\epsilon}+1+\rho(1-\phi)-\eta(2-\phi)\right)
+4\phi(1-\phi)\nonumber\\
&&\hskip2cm-2\vmpp+4\eta(1-\phi)\vmpp+\eta\phi(\vppp+\vpmp)\bigg]\\
\beta^2&=&-\beta^4={1\over2}\bigg[({\vpp-\vmp})
\left(-{1\over\epsilon}+1+\rho(1-\phi)+\eta(2-\phi)\right)
-\eta\phi(\vppp-\vpmp)\bigg]\\
\gamma^2&=&\gamma^4=(1-\phi)^2-{1\over2}\eta(1-\phi)[\vpp+\vmp-\vppp-\vpmp]\\
\delta^2&=&-\delta^4=\eta(\vpp-\vmp-\vppp+\vpmp)\\
\eea
Similarly we write the bilinear form for the $q$'s as
\bea
S^j_q=\alpha\sum_k q_k^{j2}\pm\beta\sum_k(-)^k q_k^{j2},
\eea
with the $\pm$ sign alternating with time slice according to the
pattern $++--++--\cdots$. Then
\bea
\alpha={\vpp+\vmp\over2\epsilon}, \qquad 
\beta={\vpp-\vmp\over2\epsilon}.
\eea 
\subsection{$\varphi_i^j = \langle P_i^jP_i^{j-1}\rangle$}

Just as the previous subsection we need to extract coefficients of the
various bilinear forms. The ghost action and $q$ action have the same
form as the previous subsection. After reading off, coefficients are
\bea
\wtd{\a}^1 &=& \wtd{\a}^3 = \fr12\left\{\left[\fr1{\epsilon} + 1 +
  \rho (1 - \f^\prime)\right]\varphi + \left[\fr1{\epsilon} + 1 +
  \rho (1 - \f)\right]\varphi^\prime + 2(\f+\f^\prime)\right.\nn\\
&&\hskip1cm\left.-4\f\f^\prime - \vppp - \vpmp \right\}\nn\\
\wtd{\b}^1 &=& -\wtd{\b}^3 = \fr12\left\{\left[\fr1{\epsilon} + 1 -
  \rho (1 - \f^\prime)\right]\varphi - \left[\fr1{\epsilon} + 1 -
  \rho (1 - \f)\right]\varphi^\prime - 2(\f - \f^\prime)\right.\nn\\
&&\hskip1cm\left. - \vppp + \vpmp \right\}\nn\\
\wtd{\g}^1 &=& \wtd{\g}^3 = (1 - \f ) (1 - \f^\prime),\;\;\;\;
\wtd{\d}^1 = -\wtd{\d}^3 = 0\nn\\
\wtd{\a}^2 &=& \wtd{\a}^4 = \fr12\left\{\left[\fr1{\epsilon} + 1 +
  \rho (1 - \tilde{\f})\right]\varphi + \left[\fr1{\epsilon} + 1 +
  \rho (1 - \tilde{\f})\right]\varphi^\prime + 4 \tilde{\f}(1 - 
\tilde{\f}) - 2\vmpp\right\}\nn\\
\wtd{\b}^2 &=& -\wtd{\b}^4 = \fr12\left\{\left[\fr1{\epsilon} - 1 -
  \rho (1 - \tilde{\f})\right]\varphi - \left[\fr1{\epsilon} - 1 -
  \rho (1 - \tilde{\f})\right]\varphi^\prime \right\}\nn\\
\wtd{\g}^2 &=& \wtd{\g}^4 = (1 - \tilde{\f})^2,\;\;\;\;
\wtd{\d}^2 = -\wtd{\d}^4 = 0,\nn\\
\label{ghostcoeff-2pcase}
\eea
for the ghosts and for the $q$'s they are
\bea
\wtd{\a} = \fr{\varphi +\varphi^\prime}{2\epsilon},\;\;\;
\wtd{\b} = \fr{\varphi -\varphi^\prime}{2\epsilon}.
\eea
Note that we denote coefficients correspond to $\langle PP^\prime\rangle$
case with tilde.

\section{Fishnet Determinants}
\subsection{Single $P$ case}

The easiest way to do the $q$ integrals is to diagonalize
the bilinear form in the exponent by going to normal modes.
We describe this procedure by first going to normal modes
in $i$, and then in $j$. We let the $i$ labels range
from $i=0,1,\ldots,M$ and choose Dirichlet conditions at
the boundaries: $q_0^j=q_M^j=0$. Thus our system carries zero
transverse momentum (we are working in the center of
mass system). Then the mode expansion reads:
\bea
q_k^j=\sqrt{2\over M}\sum_{l=1}^{M-1} Q_l^j\sin{{\pi l k\over M}}.
\eea
For completeness, we also mention the mode expansion
for periodic boundary conditions (which will be useful
in our discussion of the ghost determinant.) 
\bea
q_k^j=\sqrt{1\over M}\sum_{l=0}^{M-1} A_l^j\exp\left\{{2i\pi l k\over M}
\right\}.
\qquad{\rm Periodic~b.c.}
\eea 
Defining the frequencies $\omega_l^2\equiv4\sin^2{l\pi/2M}$,
we find 
\bea
\sum_k(q_{k+1}^j-q_k^j)^2=\sum_l Q_l^{j2}\omega_l^2
\qquad\left(\sum_l A_lA_{M-l}\omega_{2l}^2,\quad{\rm Periodic~b.c.}\right)
\eea

For the remaining terms in the exponent the $q_{odd}^{j2}$'s
have a different coefficient that the  $q_{even}^{j2}$'s.
Thus in addition to the usual orthonormality condition
\bea
{2\over M}\sum_{k=1}^{M-1} \sin{\pi l^\prime k\over M}
\sin{\pi lk\over M}=\delta_{l^\prime,l},
\qquad{1\over M}\sum_{k=0}^{M-1} 
\exp\left\{{2\pi i(l^\prime+l) k\over M}\right\}=\delta_{l^\prime,M-l}
\eea
we also need the corresponding formulas when the $k$
sum is restricted to even and odd values. 
A short calculation shows that
\bea
{2\over M}\sum_{k=1}^{M-1} (-)^k\sin{\pi l^\prime k\over M}
\sin{\pi lk\over M}=-\delta_{l^\prime,M-l},
\qquad{1\over M}\sum_{k=0}^{M-1}(-)^k 
\exp\left\{{2\pi i(l^\prime+l) k\over M}\right\}=\delta_{l^\prime,M/2-l}\quad 
{\rm mod}~M,
\eea
where the periodic case assumes that $M$ is {\it even}.
From these we derive the identities
\bea
{2\over M}\sum_{k=odd} \sin{\pi l^\prime k\over M}
\sin{\pi lk\over M}&=&{1\over2}(\delta_{l^\prime,l}+\delta_{l^\prime,M-l})\\
{1\over M}\sum_{k=odd} 
\exp\left\{{2\pi i(l^\prime+l) k\over M}\right\}
&=&{\delta_{l^\prime,M-l}-\delta_{l^\prime,M/2-l}\over2}
\\
{2\over M}\sum_{k=even} \sin{\pi l^\prime k\over M}
\sin{\pi lk\over M}&=&{1\over2}(\delta_{l^\prime,l}-\delta_{l^\prime,M-l})\\
{1\over M}\sum_{k=even} 
\exp\left\{{2\pi i(l^\prime+l) k\over M}\right\}
&=&{\delta_{l^\prime,M-l}+\delta_{l^\prime,M/2-l}\over2}
\eea
The respective sums are then
\bea
\sum_{k=odd} q_k^{j2}={1\over2}\sum_l Q_m^{j2}+{1\over2}\sum_l 
Q_l^{j}Q_{M-l}^j;
\qquad
\sum_{k=even} q_k^{j2}={1\over2}\sum_l Q_l^{j2}-{1\over2}\sum_l 
Q_l^{j}Q_{M-l}^j
\eea
for Dirichlet boundary conditions and
\bea
\sum_{k=odd} q_k^{j2}={1\over2}\sum_l A_l^{j}A_{M-l}^j-{1\over2}\sum_l 
A_l^{j}A_{M/2-l}^j;
\qquad
\sum_{k=even} q_k^{j2}={1\over2}\sum_l A_l^{j}A_{M-l}^j+{1\over2}\sum_l 
A_l^{j}A_{M/2-l}^j
\eea
for periodic boundary conditions.
Depending on the value of $j$, either $\vpp$ multiplies the
even sum and $\vmp$ multiplies the odd sum, or
{\it vice versa}. Recalling that $\alpha=(\vpp+\vmp)/2\epsilon$
and $\beta=(\vpp-\vmp)/2\epsilon$,we find that
the entire exponent for fixed $j$ for the
$q$ integral can be written as $-a/2m$ times
\bea
\sum_l \omega_l^2 Q_l^{j2}+\alpha
\sum_l(Q_l^j-Q_l^{j-1})^2\pm
\beta\sum_l(Q_l^j-Q_l^{j-1})(Q_{M-l}^j-Q_{M-l}^{j-1}).
\eea
for Dirichlet conditions and
\bea
\sum_l \omega_{2l}^2 A_l^{j}A_{M-l}^j+\alpha
\sum_l(A_l^j-A_l^{j-1})(A_{M-l}^j-A_{M-l}^{j-1})\pm
\beta
\sum_l(A_l^j-A_l^{j-1})(A_{M/2-l}^j-A_{M/2-l}^{j-1}).
\eea
for periodic conditions.
The $\pm$ stays fixed for two time steps and then switches.
For example the simple function $\sqrt{2}\cos[\pi(2j-1)/4]$
produces the signs $+--++--+\cdots$ as $j=1,2,3,4,\cdots$,
so we can write the complete exponent as $-a/2m$ times:
\bea
\sum_{l,j} \omega_l^2Q_l^{j2}+\alpha
\sum_{l,j}(Q_l^j-Q_l^{j-1})^2+
\beta\sum_{l,j}\sqrt{2}\cos{\pi\over4}(2j-1)
(Q_l^j-Q_l^{j-1})(Q_{M-l}^j-Q_{M-l}^{j-1}).
\eea
with a similar expression for the periodic case.

Now we are in a position to choose normal modes in $j$. Let us
let the $j$ values range from $0$ to $N$, and impose Dirichlet conditions
at $j=0,N$: $Q_l^0=Q_l^N=0$, so that we have for the Dirichlet case
 \bea
Q_l^j=\sqrt{2\over N}\sum_{n=1}^{N-1} A_l^n\sin{{\pi n j\over N}}.
\eea 
Defining the frequencies $\omega_n^{\prime2}\equiv4\sin^2{n\pi/2N}$,
we easily find 
\bea
\sum_j Q_l^{j2}=\sum_n A_l^{n2};\qquad
\sum_j(Q_l^j-Q_l^{j-1})^2=\sum_n A_l^{n2}\omega_n^{\prime2}.
\eea
But we also need the more complicated sums in the last
term of the exponent. To do that we need
\bea
{2\over N}\sum_{j=1}^N\cos{\pi\over2}\left(j-{1\over2}\right)
\left[\sin{n\pi j\over N}-\sin{n\pi(j-1)\over N}\right]
\left[\sin{n^\prime\pi j\over N}-\sin{n\pi(j-1)\over N}\right]&=&\\
&&\hskip-4in 4\sin{n\pi\over2N}\sin{n^\prime\pi\over2N}
{2\over N}\sum_{j=1}^N\cos{\pi\over2}\left(j-{1\over2}\right)
\cos{n\pi\over N}\left(j-{1\over2}\right)
\cos{n^\prime\pi\over N}\left(j-{1\over2}\right). 
\eea
We shall restrict $N$ to even values, in which case we find that
this expression is essentially block diagonal with the
blocks being at most $4\times4$:
\bea
&&\hskip-1in 
4\sin{n\pi\over2N}\sin{n^\prime\pi\over2N}{2\over N}
\sum_{j=1}^N\cos{\pi\over2}\left(j-{1\over2}\right)
\cos{n\pi\over N}\left(j-{1\over2}\right)
\cos{n^\prime\pi\over N}\left(j-{1\over2}\right)\nonumber\\
&&=2\sin{n\pi\over2N}\sin{n^\prime\pi\over2N}
\left[\delta_{n^\prime,{N\over2}-n}+\delta_{n^\prime,{N\over2}+n}
+\delta_{n^\prime,n-{N\over2}}-\delta_{n^\prime,{3N\over2}-n}\right] . 
\eea
The exponent now becomes $-a/2m$ times
\bea
&&\hskip-1in \sum_{l,n} (\omega_l^2+\alpha\omega_n^{\prime2})A_l^{n2} 
+ \beta{\sqrt2\over2}
\sum_{l,n}A_l^nA_{M-l}^{n^\prime}
\omega^\prime_n\omega^\prime_{n^\prime}\left[\delta_{n^\prime,{N\over2}-n}
+\delta_{n^\prime,{N\over2}+n}+\delta_{n^\prime,n-{N\over2}}
-\delta_{n^\prime,{3N\over2}-n}\right]\nonumber\\
&&\hskip-1in =\sum_{l,n} (\omega_l^2+\alpha\omega_n^{\prime 2})A_l^{n2} 
+ \beta{\sqrt2}
\sum_{l<M/2,n}A_l^nA_{M-l}^{n^\prime}
\omega^\prime_n\omega^\prime_{n^\prime}\left[\delta_{n^\prime,{N\over2}-n}
+\delta_{n^\prime,{N\over2}+n}+\delta_{n^\prime,n-{N\over2}}
-\delta_{n^\prime,{3N\over2}-n}\right]. 
\eea

For $l<M/2$ the mode $A_l^n$ couples to itself as well as
$A_{M-l}^{N/2-n},A_{M-l}^{N/2+n}$ for $n<N/2$ and
to $A_{M-l}^{N-n},A_{M-l}^{3N/2-n}$ for $n>N/2$. For
clarity, give a different letter name to each of the 
four possibilities:
\bea
A_l^n =\cases{A_l^n&for $l<M/2$, $n<N/2$\cr
B_l^{N-n}&for $l<M/2$, $n>N/2$\cr
C_{M-l}^{N/2-n}&for $l>M/2$, $n<N/2$\cr
D_{M-l}^{n-N/2}&for $l>M/2$, $n>N/2$\cr}
\eea
Then the bilinear form can be written
\bea
&&\sum_{l<M/2,n<N/2} 
\left[(\omega_l^2+\alpha\omega_n^{\prime2})A_l^{n2}
+(\omega_l^2+\alpha\omega_{N-n}^{\prime2})B_l^{n2}
+(\omega_{M-l}^2+\alpha\omega_{N/2-n}^{\prime2})C_l^{n2}
+(\omega_{M-l}^2+\alpha\omega_{N/2+n}^{\prime2})D_l^{n2}
\right]\nonumber\\ 
&&+ \beta{\sqrt2}\hskip-.25in
\sum_{l<M/2,n<N/2}\left[A_l^n\omega^\prime_n(C_l^{n}
\omega^\prime_{N/2-n}
+D_l^{n}\omega^\prime_{N/2+n})
+B_l^{n}\omega^\prime_{N-n}(C_l^{n}\omega^\prime_{N/2-n}
-D_l^{n}\omega^\prime_{N/2+n})
\right] .
\eea
The Gaussian integral then involves the determinant of a
block diagonal matrix, the size of each
block at most $4\times4$. The computation of each 
determinant is not hard: Each $4\times4$ block has the
structure and determinant
\bea
M&=&\pmatrix{a&0&ef&eh\cr
0&b&gf&-gh\cr ef&gf&c&0\cr eh&-gh&0&d}\\
\nonumber\\
\det M&=& abcd+4(efgh)^2-ac(gh)^2-ad(gf)^2-bd(ef)^2-bc(eh)^2.
\eea
For the $4\times4$ matrix for mode $(l,n)$ with $l<M/2$ and
$n<N/2$, we have the following values for $a,b,\ldots,h$:
\bea
a=\omega_l^2+\alpha\omega_n^{\prime2};\quad
b=\omega_l^2+\alpha\omega_{N-n}^{\prime2};\quad
c=\omega_{M-l}^2+\alpha\omega_{N/2-n}^{\prime2};\quad
d=\omega_{M-l}^2+\alpha\omega_{N/2+n}^{\prime2}\\
ef={\beta\omega^\prime_n\omega^\prime_{N/2-n}\over\sqrt{2}};
\quad eh={\beta\omega^\prime_n\omega^\prime_{N/2+n}\over\sqrt{2}};
\quad gf={\beta\omega^\prime_{N-n}\omega^\prime_{N/2-n}\over\sqrt{2}};
\quad gh={\beta\omega^\prime_{N-n}\omega^\prime_{N/2+n}\over\sqrt{2}}
\eea
Plugging these values into $\det M$ we get
\bea
\det M_{l,n}&=&16(\alpha^2-\beta^2)^2\sin^2{n\pi\over N}\cos^2{n\pi\over N}
+16\alpha(\alpha^2-\beta^2)
[\omega_l^2\cos^2{n\pi\over N}+
\omega_{M-l}^2\sin^2{n\pi\over N}]\nonumber\\
&&
+4\alpha^2\left(\omega_l^4\cos^2{n\pi\over N}
+\omega_{M-l}^4\sin^2{n\pi\over N}\right)
+64\left(\alpha^2-{\beta^2\over2}+\alpha\right)\sin^2{l\pi\over M}
+16\sin^4{l\pi\over M}
\label{qdeterminant}
\eea
In obtaining this result
we took advantage of the explicit forms for the 
$\omega$'s:
\bea
\omega_l^2&=&4\sin^2{l\pi\over2M},\quad\omega_{M-l}^2=4\cos^2{l\pi\over2M},
\quad \omega_n^{\prime2}=4\sin^2{n\pi\over2N},
\quad \omega_{N-n}^{\prime2}=4\cos^2{n\pi\over2N}\\
\omega_{N/2-n}^{\prime2}&=&4\sin^2\left({\pi\over4}-{n\pi\over2N}\right),
\qquad\qquad \omega_{N/2+n}^{\prime2}=4\cos^2\left({\pi\over4}-
{n\pi\over2N}\right)
\eea
In particular note that 
\bea
\omega_l^2+\omega_{M-l}^2&=&\omega_n^{\prime2}+\omega_{N-n}^{\prime2}
=\omega_{N/2-n}^{\prime2}+\omega_{N/2+n}^{\prime2}=4\\
\omega_l^2\omega_{M-l}^2&=&4\sin^2{l\pi\over M},
\quad \omega_n^{\prime2}\omega_{N-n}^{\prime2}=4\sin^2{n\pi\over N},
\quad\omega_{N/2-n}^{\prime2}\omega_{N/2+n}^{\prime2}=4\cos^2{n\pi\over N}.
\eea
Note that $\det M_{l,n}$ is a quadratic polynomial in $\sin^2(n\pi/N)$
which we can factorize by finding the roots $r_\pm$:
\bea
\det M_{l,n}&=&16(\alpha^2-\beta^2)^2\left(r_+-\sin^2{n\pi\over N}\right) 
\left(\sin^2{n\pi\over N}-r_-\right).
\eea
We can conveniently parametrize the roots as
\bea
r_+&=&R_l\cosh^2\kappa_l,\qquad\qquad r_-=-R_l\sinh^2\kappa_l\\
R_l&=&1+4{\alpha^3-\alpha\beta^2+\alpha^2\over
(\alpha^2-\beta^2)^2}\cos{l\pi\over M}\\
R_l\sinh^2\kappa_l&=&{1\over2}\left[\sqrt{
R_l^2+{\omega_l^2(\omega_{M-l}^2+2\alpha)
[(\omega_l^2+4\alpha)(\omega_{M-l}^2+2\alpha)-8\beta^2]
\over 4(\alpha^2-\beta^2)^2}}-R_l\right].
\eea

The result of doing the integral over all the $q$'s is then
simply
\bea
e^{W_q}=\left({2\pi m\over a}\right)^{dMN/2}\prod_{l<M/2,n<N/2} 
\det{}^{-d/2}M_{l,n}.
\eea
Note that for $M$ odd, which we can take for convenience, there
is no special unpaired $l=M/2$ mode. However, since we must restrict
$N$ to be even there is a special $n=N/2$ mode which must be handled
separately. However, as far as the energy is concerned, it's 
presence or absence doesn't matter because the energy is
extracted from the coefficient of $N$ in $W$ as $N\to\infty$.

Using the above factorization, we can perform the product over
$n$ by using one of the identities:
\bea
\prod_{n=1}^{N-1}\left(4\sin^2{n\pi\over N}+4\sinh^2\xi\right)
&=&4\cosh^2\xi\prod_{n<N/2}\left(4\sin^2{n\pi\over N}+4\sinh^2\xi\right)^2 
={\sinh^2 N\xi\over\sinh^2\xi}\\
\prod_{n=1}^{N-1}\left(4\cosh^2\xi-4\sin^2{n\pi\over N}\right)
&=&4\sinh^2\xi\prod_{n<N/2}\left(4\cosh^2\xi-4\sin^2{n\pi\over N}\right)^2
={\sinh^2 N\xi\over\cosh^2\xi},~~ {\rm for}~N~{\rm even}\\
\prod_{n=1}^{N-1}\left(4\cosh^2\xi-4\sin^2{n\pi\over N}\right)
&=&\prod_{n<N/2}\left(4\cosh^2\xi-4\sin^2{n\pi\over N}\right)^2
={\cosh^2 N\xi\over\cosh^2\xi},\qquad {\rm for}~N~{\rm odd}
\eea
Since we want to product over only the modes with $n<N/2$,
we will require the square roots of these identities:
\bea
\prod_{n<N/2}\left(4\sin^2{n\pi\over N}+4\sinh^2\xi\right) 
&=&{\sinh N\xi\over\sinh 2\xi}\\
\prod_{n<N/2}\left(4\cosh^2\xi-4\sin^2{n\pi\over N}\right)
&=&{\sinh N\xi\over\sinh2\xi},
\eea
which are now specialized to even $N$. the appropriate $\xi$'s
to use are given by
\bea
\sinh\xi^l_-=\sqrt{-r_-}=\sqrt{R_l}\sinh\kappa_l,\qquad 
\cosh\xi^l_+=\sqrt{r_+}=\sqrt{R_l}\cosh\kappa_l,
\eea
which can be solved:
\bea
\xi^l_-&=&\ln\left(\sqrt{R_l}\sinh\kappa_l+\sqrt{1+R_l\sinh^2\kappa_l}\right)\\
\xi^l_+&=&\ln\left(\sqrt{R_l}\cosh\kappa_l+\sqrt{R_l\cosh^2\kappa_l-1}\right).
\eea
Thus we have
\bea
\left({2\pi m/a}\right)^{-dMN/2}e^{W_q}=\prod_{l<M/2}\prod_{n<N/2}\det M_{l,n}
=\prod_{l<M/2}\left[(\alpha^2-\beta^2)^{N-2}{\sinh N\xi^l_+
\over\sinh 2\xi^l_+}{\sinh N\xi^l_-\over\sinh 2\xi^l_-}\right]^{-d/2}.
\label{qint}
\eea
From this we extract the $q$ contribution to the energy per site
(${\cal E}_q=-W_q/MN$):
\bea
{\cal E}_q =
{d\over2M}\sum_{l<M/2}[\x^l_++\xi^l_-+\ln(\alpha^2-\beta^2)].
\label{bulkenergyq}
\eea

We will eventually need the contribution of
the general ghost action which will have 
the coefficients of the terms $(b_{k+1}-b_k)(c_{k+1}-c_k)$
different for $k$ even and odd. Unfortunately, for Dirichlet
boundary conditions this translates to non-local expressions
in mode space. However, the bulk energy, which we have been extracting
is the same for Dirichlet and periodic boundary conditions, and
for even $M$ this more general anti-ferromagnetic pattern
of couplings is local in mode space:
\bea
&&\hskip-0.5in\alpha\sum_kb_kc_k+\beta\sum_k(-)^kb_kc_k
+\gamma\sum_k(b_{k+1}-b_k)(c_{k+1}-c_k)
+\delta\sum_k(-)^k(b_{k+1}-b_k)(c_{k+1}-c_k) = \nonumber\\
&&\alpha B_0C_0+(\alpha+4\gamma)B_{M/2}C_{M/2}
+\sum_{l\neq0,M/2}B_lC_{M-l}\left[\alpha+4\gamma\sin^2{l\pi\over M}\right]
+\beta(B_0C_{M/2}+B_{M/2}C_0)\\
&&+\sum_{l=1}^{M/2-1}B_lC_{M/2-l}
\left(\beta-4i\delta\sin{l\pi\over M}\cos{l\pi\over M}\right)
+\sum_{l=1}^{M/2-1}B_{M/2+l}C_{M-l}
\left(\beta+4i\delta\sin{l\pi\over M}\cos{l\pi\over M}\right).\nonumber
\eea
The corresponding ghost determinant for a single
time slice is easily read off in this periodic case:
\bea
{\rm Det}_j^P&=&(\alpha^2+4\alpha\gamma-\beta^2)\prod_{l=1}^{M/2-1}
\left[\alpha^2-\beta^2+4\alpha\gamma+16(\gamma^2-\delta^2)\sin^2{l\pi\over M}
\cos^2{l\pi\over M}\right]\\
&=&(\alpha^2+4\alpha\gamma
-\beta^2)(\gamma^2-\delta^2)^{M/2-1}\left({\sinh M\zeta_P/2
\over\sinh\zeta_P}\right)^2\\
\zeta_P&=&\ln\left(\sqrt{{\alpha^2-\beta^2+4\alpha\gamma\over
4(\gamma^2-\delta^2)}}+\sqrt{1+{\alpha^2-\beta^2+4\alpha\gamma\over
4(\gamma^2-\delta^2)}}\right).
\eea
The prefactor $(\alpha^2+4\alpha\gamma-\beta^2)$ 
shows that the determinant is zero
for $\beta^2=\alpha^2+4\alpha\gamma$ 
which is due to a zero mode that occurs in that
case. But this is insignificant for the bulk energy per site:
\bea
{\cal E}_g^P&=&-{d\over2N}\sum_j
\ln\left(\sqrt{{\alpha_j^2-\beta_j^2\over4}
+\alpha_j\gamma_j}+\sqrt{\gamma_j^2-\delta_j^2+{\alpha_j^2-\beta_j^2\over4}
+\alpha_j\gamma_j}\right)\equiv-{d\over2N}\sum_j\xi^j_g.
\eea
Here $j$ labels the time slice. For the fishnet spin pattern,
there are two distinct types of time slice and two different
values for $\xi^j_g$, which we simply call $\xi_1$ and $\xi_2$.
Then we write the total matter contribution ($q$ plus ghosts) 
to the energy per site, including the boundary term for
the massive case, as
\bea
{\cal E}_m(\phi,\phi^\prime)={d\over2}{3\phi+\phi^\prime\over4}
\ln(1+\rho)-{d\over4}(\xi_1+\xi_2)+{d\over2M}\sum_{l<M/2}
[\x^l_++\xi^l_-+\ln(\alpha^2-\beta^2)].
\label{bulkenergy}
\eea

In view of the complexity of this formula, it is useful to
explicitly display simplifying special cases. First 
take the uniform case $\phi^\prime=\phi$. Then $\beta=\beta^j_g
=\delta^j_g=0$, and 
\bea
\alpha_g^j&=&{\vpp}\left({1\over\epsilon}+1+\rho(1-\phi)
-\eta(2-\phi)\right)+2\phi(1-\phi)-\vppp(1-\eta(2-\phi))\nonumber\\
&\equiv&{\vpp\over\epsilon}{\hat\alpha}\\
\gamma_g^j&=&\gamma=(1-\phi)^2-\eta(1-\phi)(\vpp-\vppp) , \hskip1in
\alpha={\vpp\over\epsilon}.
\eea
Then the matter energy per site simplifies to
\bea
{\cal E}_m&=&{d\over2}{\phi}
\ln(1+\rho)+{d\over M}\sum_{l=1}^{M-1}\ln\left[{\sqrt{\epsilon\over\vpp}
\sin{l\pi\over2M}}+\sqrt{1+{\epsilon\over\vpp}\sin^2{l\pi\over2M}}\right]
\nonumber\\
&&\quad-{d\over2}
\ln\left(\sqrt{{{\hat\alpha}\gamma\epsilon\over\vpp}
+{{\hat\alpha}^2\over4}}
+\sqrt{\gamma^2{\epsilon^2\over\vpp^2}+{\epsilon{\hat\alpha}\gamma\over
\vpp}+{{\hat\alpha}^2\over4}}\right)\nonumber\\
&\to&{d\over2}{\phi}\ln(1+\rho)+{2d\over\pi}{\rm Re}\left\{i{\rm Li}_2
\left({\sqrt{\epsilon\over\vpp}}\right)\right\}\nonumber\\
&&\quad-{d\over2}
\ln\left(\sqrt{{{\hat\alpha}\gamma\epsilon\over\vpp}
+{{\hat\alpha}^2\over4}}
+\sqrt{\gamma^2{\epsilon^2\over\vpp^2}+{\epsilon{\hat\alpha}\gamma\over
\vpp}+{{\hat\alpha}^2\over4}}\right)
\eea

The second special case is one in which we take $\epsilon\to0$
holding $f=\phi^\prime/\epsilon$ fixed. In this limit 
$\alpha^2-\beta^2=\vpp\vmp/\epsilon^2=fG\vpp/\epsilon$ 
is of $O(1/\epsilon)$
and so are the $\alpha_g^2-\beta_g^2$. Thus the $\xi_g$
simplify to
\bea
\xi_g\to\ln 2\sqrt{\alpha_g\gamma_g+{\alpha_g^2-\beta_g^2\over4}}.
\eea 
Keeping only the leading terms as $\epsilon\to0$ we find
\bea
\xi_1+\xi_2&\to&\ln{\vpp\over\epsilon}
+\ln\sqrt{2+fG+\vpp(\rho-\eta)-2\eta(\vpp-\vppp)}\nonumber\\
&&+\ln\sqrt{\vpp+fG+(1-\phi)
(2+\vpp(\rho-\eta)+\eta\vppp)}
\eea
The simplification of the $q$ determinant is less dramatic.
We list the limiting forms of the various ingredients:
\bea
R_l&\to& 1+{1+2fG\over f^2G^2}\cos{l\pi\over M}\\
R_l\sinh^2\kappa_l&\to&{1\over2}\left[\sqrt{R_l^2+\sin^2{l\pi\over2M}
{8(1+fG)-4\sin^2{l\pi/2M}\over f^2G^2}}-R_l\right]\\
\ln(\alpha^2-\beta^2)&=&\ln{fG\vpp\over\epsilon}
\eea
Combining these ingredients we end up with total energy per site
for $\phi^\prime=\epsilon f$:
\bea
{\cal E}&\to&{\cal E}_f(\phi)
={3d\phi\over8}\ln(1+\rho)
+{d\over 2M}\sum_{l<M/2}[\ln(fG)+\xi_+^l+\xi_-^l]\nonumber\\
&&\qquad-{d\over4}\ln\sqrt{2+fG+\vpp(\rho-\eta)-2\eta(\vpp-\vppp)}\nonumber\\
&&\qquad-{d\over4}\ln\sqrt{\vpp+fG+(1-\phi)
(2+\vpp(\rho-\eta)+\eta\vppp)}
\eea  
We note that the $f\to0$ limit of the right side is finite
because $\ln(fG)+\xi_+^l\to \ln(2\sqrt{\cos(\pi l/M)})$
in that limit.
\bea
{\cal E}_{f=0}(\phi)
&=&{3d\phi\over8}\ln(1+\rho)+{d\over 2M}\sum_{l<M/2}
\ln\left[\sin{l\pi\over2M}\sqrt{1+\cos^2{l\pi\over2M}}
+\sqrt{\cos{l\pi\over M}+\sin^2{l\pi\over2M}\left(
1+\cos^2{l\pi\over2M}\right)}\right]\nonumber\\
&&+{d\over4}\ln{2}-{d\over4}\ln\sqrt{2+\vpp(\rho-\eta)-2\eta(\vpp-\vppp)}
\nonumber\\&&
-{d\over4}\ln\sqrt{\vpp+(1-\phi)
(2+\vpp(\rho-\eta)+\eta\vppp)}
\eea  

\subsection{Double $P$ case}

The above computations are general so one can obtain the results for
the double $P$ correlator by replacing $\a \rightarrow \wtd{\a}$,
etc. Instead of repeating the same lengthy procedure as before we just
quote some important results. The $q$ contribution to the energy per
site is
\bea
\cE_q = \fr{d}{2M}\sum_{l<M/2}[\wtd{\xi}_+^l + \wtd{\xi}_-^l + 
\ln(\wtd{\a}^2 - \wtd{\b}^2)],
\label{qen-2pcase}
\eea
where
\bea
\wtd{\xi}^l_-&=&\ln\left(\sqrt{R_l}\sinh\kappa_l+
\sqrt{1+R_l\sinh^2\kappa_l}\right)\\
\wtd{\xi}^l_+&=&\ln\left(\sqrt{R_l}\cosh\kappa_l+
\sqrt{R_l\cosh^2\kappa_l-1}\right)\\
\wtd{R}_l&=&1+4{\wtd{\alpha}^3-\wtd{\alpha}\wtd{\beta}^2+\wtd{\alpha}^2\over
(\wtd{\alpha}^2-\wtd{\beta}^2)^2}\cos{l\pi\over M}\\
\wtd{R}_l\sinh^2\kappa_l&=&{1\over2}\left[\sqrt{
\wtd{R}_l^2+{\omega_l^2(\omega_{M-l}^2+2\wtd{\alpha})
[(\omega_l^2+4\wtd{\alpha})(\omega_{M-l}^2+2\wtd{\alpha})-8\wtd{\beta}^2]
\over 4(\wtd{\alpha}^2-\wtd{\beta}^2)^2}}-\wtd{R}_l\right],\\
\wtd{R}_l\cosh^2\kappa_l&=&{1\over2}\left[\sqrt{
\wtd{R}_l^2+{\omega_l^2(\omega_{M-l}^2+2\wtd{\alpha})
[(\omega_l^2+4\wtd{\alpha})(\omega_{M-l}^2+2\wtd{\alpha})-8\wtd{\beta}^2]
\over 4(\wtd{\alpha}^2-\wtd{\beta}^2)^2}}+\wtd{R}_l\right].
\eea
The ghost energy per site is
\bea
\cE_g = -\fr{d}{2N}\sum_j \ln\left(\sqrt{\fr{\wtd{\a}_j^2 -
    \wtd{\b}_j^2}{4} + \wtd{\a}_j\wtd{\g}_j} + \sqrt{\wtd{\g}_j^2 +
\fr{\wtd{\a}_j^2 -\wtd{\b}_j^2}{4} + \wtd{\a}_j\wtd{\g}_j}
\right) \equiv -\fr{d}{4}(\wtd{\xi}_1 + \wtd{\xi}_2)
\label{ghosten-2pcase}
\eea
here we have use the fishnet spin pattern where there two only two
distinct types of time slice and two different values for
$\wtd{\xi}_g^j$ and we will call them $\wtd{\xi}_1$ and
$\wtd{\xi}_2$. The total matter contribution to the energy per site
as
\bea
\cE_m(\varphi,\varphi^\prime) = -\fr{d}{4}(\wtd{\xi}_1 +
\wtd{\xi}_2) + \fr{d}{2M}\sum_{l<M/2}[\wtd{\xi}_+^l + \wtd{\xi}_-^l
  + \ln(\wtd{\a}^2-\wtd{\b}^2)]
\label{maten-2pcase}
\eea
There are two special cases. First
we take the uniform case in which $\varphi=\varphi^\prime$. In this
case, $\wtd{\b}^j_g = 0$, $\wtd{\b}=0$, $\wtd{\a}=\varphi/\epsilon$ and 
\bea
\wtd{\a}^j_g &=& \left[\fr1{\epsilon} + 1 + \rho (1-\f)\right]\varphi -
\langle +++\rangle + 2(1-\f)\f,\\
\wtd{\g}^j_g &=& (1-\f)^2
\eea 
where $\f, \vp$ and $\vppp$ are those in
Eqs.~(\ref{phi-uniform-2pcase}), 
(\ref{varphi-uniform-2pcase}) and (\ref{ppp-uniform-2pcase}). 
The matter energy is
\bea
\cE_m^{uniform} = \fr{2d}{\pi}{\rm Re}\left\{i{\rm
  Li}_2\left(\fr1{i}\sqrt{\fr{\epsilon}{\varphi}}\right)\right\} +
\fr{d}{2}\ln\left(\fr{\varphi}{\epsilon}\right) -
\fr{d}{2}\ln\left[\fr{\wtd{\a}_g}{2} + \wtd{\g}_g + 
\sqrt{\fr{\wtd{\a}_g^2}{4} + \wtd{\a}_g\wtd{\g}_g}\right]
\label{maten-fiefip-2pcase}
\eea
Another special case is $\varphi^\prime = 0$ corresponding
$\chi^\prime\to -\infty$. In this limit, $\wtd{\a}=\wtd{\b}$ and the
$q$ determinant Eq.~(\ref{qdeterminant}) can be written in a simpler
form
\bea
{\rm det}M_{l,n} = 16\wtd{\a}^2\cos\left(\fr{l\pi}{M}\right)
\left[4\sin^2\left(\fr{n\pi}{N}\right) + 
\fr{4\sin^2\left(\fr{l\pi}{M}\right) + \wtd{\a}^2\w_l^4 +
    8(\wtd{\a}^2+2\wtd{\a})
    \sin^2\left(\fr{l\pi}{M}\right)}{4\wtd{\a}^2\cos\left( \fr{l\pi}{M}\right)
}
\right]
\eea
and the $q$ integral can be perform and the result of $q$ energy
contribution in the large $M,N$ limit is
\bea
\cE_q^{\vp^\prime=0} &=&
\fr{d}{2M}\sum_{l<M/2}\ln\left[16\wtd{\a}^2\cos\left(\fr{l\pi}{M}\right)\right]
+ \fr{d}{M}\sum_{l<M/2}\ln\left(\bar{R}_l +
\sqrt{\bar{R}_l^2+1}\right)\nn\\
&\to& d\,\int_0^{1/2}dz\,\ln\left[
\sqrt{4\sin^4(\pi z) + 16\wtd{\a}^2\sin^4(\pi z/2) +8(\wtd{\a}^2 +
  2\wtd{\a})\sin^2(\pi z)}\right.\nn\\
&&+\left.\sqrt{16\wtd{\a}^2\cos(\pi z) + 
4\sin^4(\pi z) + 16\wtd{\a}^2\sin^4(\pi z/2) + 8(\wtd{\a}^2 +
  2\wtd{\a})\sin^2(\pi z)}\right]
\eea
where
\bea
\bar{R}_l = \sqrt{\fr{4\sin^2\left(\fr{l\pi}{M}\right) + \wtd{\a}^2\w_l^4 +
    8(\wtd{\a}^2+2\wtd{\a})
    \sin^2\left(\fr{l\pi}{M}\right)}{16\wtd{\a}^2\cos\left( \fr{l\pi}{M}\right)
}}
\eea
The matter energy in this case is
\bea
\cE_m^{\vp^\prime=0} &=&
-\fr{d}{4}(\wtd{\xi}_1|_{\vp^\prime=0} + \wtd{\xi}_2|_{\vp^\prime=0}) +
\fr{d}{2}\,\int_0^{1/2}dz\,\ln\left[
\sqrt{4\sin^4(\pi z) + 16\wtd{\a}^2\sin^4(\pi z/2) + 8(\wtd{\a}^2 +
  2\wtd{\a})\sin^2(\pi z)}\right.\nn\\
&&+\left.\sqrt{16\wtd{\a}^2\cos(\pi z) + 
4 \sin^4(\pi z) + 16\wtd{\a}^2\sin^4(\pi z/2) + 8(\wtd{\a}^2 +
  2\wtd{\a})\sin^2(\pi z)}\right]
\label{maten-vpp0-2pcase}
\eea

\section{Compactification and dimensional Reduction}
A final noteworthy simplification occurs when one employs
dimensional reduction on some of the $q$ components by imposing
exact Dirichlet conditions, $q=0$, on the boundaries for those components.
Then instead of the $\epsilon$ trick,
one can use the spin projectors to rearrange
the bond structure in the way already used for the ghosts which
do obey exact Dirichlet conditions. Let's rewrite the path integral in
the case where this is done for {\it all} components of the $q$'s:
\begin{eqnarray}
T_{fi}&=&
\sum_{s_i^j=\pm1}\int DcDbD{\boldsymbol q}
\exp\left\{\sum_{i,j}
({P}_i^j+{P}_i^{j-1}-2{P}_i^j{P}_i^{j-1})
\left(\ln{\hat g}+{d\over4}\ln{a\over2\pi mR^2}\right)
\right\}
\nonumber\\
&&\exp\left\{{a\over m}
\sum_{i,j}
P_i^j\left[{\boldsymbol b}^j_{i}{\boldsymbol c}^j_{i}
-{1\over2}{\boldsymbol q}_i^{j2}\right]+
\rho{a\over m}\sum_{i,j}(1-P_i^j)
P_{i-1}^{j-1}P_{i-1}^j{\boldsymbol b}_i^j{\boldsymbol c}_i^j
+{d\over2}\sum_{i,j}P_i^j\ln\left(1+\rho\right)\right\}\nonumber\\
&&\exp\left\{{a\over m}\sum_{i,j}(1-P_i^j)\left[(P_{i+1}^j+
P_{i-1}^j){\boldsymbol b}_{i}^{j}{\boldsymbol c}_{i}^{j}
-{1\over2}P_{i+1}^j\left({{\boldsymbol n}_{i+1}^j\over R}
-{\boldsymbol q}_i^{j}\right)^2
-{1\over2}P_{i-1}^j\left({\boldsymbol q}_i^{j}
-{{\boldsymbol n}_{i-1}^j\over R}
\right)^2\right]\right\}\nonumber\\
&&\exp\left\{{a\over m}\sum_{i,j}\left[({\boldsymbol b}_{i+1}^j
-{\boldsymbol b}_i^j)({\boldsymbol c}_{i+1}^j-{\boldsymbol c}_i^j)
-{1\over2}({\boldsymbol q}_{i+1}^j
-{\boldsymbol q}_{i}^{j})^2\right](1-P_i^j)(1-P_{i+1}^j)\right\}
\nonumber\\
&&\exp\left\{-{a\over m}
\sum_{i,j}(1-P_i^j)P_i^{j-1}
\left[(1-P_{i+1}^j)({b}_{i+1}^j-{b}_i^j)({c}_{i+1}^j-{c}_i^j)
+P_{i+1}^j{b}_i^j{c}_i^j\right]        \right\}
\nonumber\\
&&\exp\left\{-{a\over m}\sum_{i,j}(1-P_i^j)\left(P_{i+1}^j
(1-P_{i+1}^{j-1})+
P_{i-1}^j{(1-P_{i-1}^{j-1})}\right)
{b}_{i}^{j}{c}_{i}^{j}\right\}\label{isingsumdir}
\end{eqnarray}
To understand this formula, focus on a solid line crossing
$n>1$ up spins. Then in the original formula there are $n-1$
Gaussian approximated delta functions:
\bea
\epsilon^{-d/2}\exp\left\{-{a\over2m\epsilon}\boldsymbol{q}^2\right\}
\to \left({2\pi m\over a}\right)^{d/2}\delta(\boldsymbol{q}).
\eea
Integrating over $n-1$ $q$'s leaves a single $q$ on the solid line to
be integrated:
\bea
\left({2\pi m\over a}\right)^{d(n-1)/2}\int d\boldsymbol{q}
\to\left({a\over2\pi mR^2}\right)^{d/2}\sum_{\boldsymbol{k}}
\int d\boldsymbol{q}_1\cdots d\boldsymbol{q}_n
\exp\left\{-{a\over2m}\sum_{i=1}^n\boldsymbol{q}_i^2\right\},
\eea
where we have discretized the single integral, $\boldsymbol{q}\to
\boldsymbol{k}/R$, where the vector $\boldsymbol{k}$ 
has integer components and $R$ is 
the corresponding compactification radius 
(note that one could have a different $R$ for each component).
The process of dimensional reduction is to keep {\it only} the
contributions with all $\boldsymbol{k}=0$ ($R\to0$ is
one way to effect this), absorbing the
factor $(a/2\pi mR^2)^{d/2}$ in the coupling constant:
\bea
{\hat g}_{d=0}^2\equiv 
\left({a\over2\pi mR^2}\right)^{d/2}{\hat g}_{d}^2.
\eea
More generally, one could stop before this last step
and have a worldsheet formalism for the compactified
quantum field theory. Then one would need to explicitly
implement the constraints that all $\boldsymbol{k}_i^j$'s
on a solid line are equal and all $\boldsymbol{k}_i^j=0$
on dotted lines (i.e. those that cross dotted lines).
These constraints are easily imposed by introducing
angular variables $\boldsymbol{\theta}_i^j$ on each
temporal link $(j-1,j)$. Then the constraints follow
by inserting
\bea
\prod_{ij}\int{d\boldsymbol{\theta}\over(2\pi)^d}
\exp\left\{i\boldsymbol{\theta}_i^j\cdot
[\boldsymbol{k}_i^j(1-P_i^j)+P_i^jP_i^{j-1}
(\boldsymbol{k}_i^j-\boldsymbol{k}_i^{j-1})]\right\}.
\eea
Taking the limit $R\to\infty$ returns to the uncompactified
theory. The explicit infra-red cutoff $R$ then serves the
purpose of $1/\epsilon$ which need not be introduced. This
gives an alternative infra-red cutoff in which conservation of
transverse momentum is maintained throughout, and hence is
superior (at least conceptually) to the $\epsilon$ cutoff.
Whether it is computationally superior remains to be seen.

\end{document}